\documentclass[aps,prl,twocolumn,nofootinbib,superscriptaddress,reprint,floatfix]{revtex4-2}
\usepackage[utf8]{inputenc}
\usepackage[a4paper, margin = 2cm]{geometry}
\usepackage{pgf,tikz}
\usetikzlibrary{arrows}
\usepackage{mathrsfs}
\usepackage{amssymb}
\usepackage{amsmath}
\usepackage{subfigure}
\usepackage{hyperref}
\usepackage{braket}
\usepackage{MnSymbol}
\usepackage{float}
\usepackage{pgfplots}
\usepackage[normalem]{ulem}

\newcommand\ballsize{0.15}
\newcommand\ax{0.8}
\newcommand\ay{0.6}

\definecolor{red}{rgb}{1., 0.0, 0.0}
\definecolor{grey}{rgb}{0.7, 0.75, 0.71}
\definecolor{blue}{rgb}{0.01, 0.28, 1.0}

\hypersetup{
    colorlinks = True,
    linkcolor = {teal},
    linkbordercolor = {teal},
    citecolor = {magenta},
    citebordercolor = {magenta},
    urlcolor = {magenta},
}

\usetikzlibrary{arrows}
\begin{document}

\title{Scar-induced imbalance in staggered Rydberg ladders}

\author{Mainak Pal}
\affiliation{School of Physical Sciences, Indian Association for the Cultivation of Science, Kolkata 700032, India}

\author{Madhumita Sarkar}
\affiliation{Department of Physics and Astronomy, University of Exeter, Stocker Road, Exeter EX4 4QL, United Kingdom}

\author{K. Sengupta}
\affiliation{School of Physical Sciences, Indian Association for the Cultivation of Science, Kolkata 700032, India}

\author{Arnab Sen}
\affiliation{School of Physical Sciences, Indian Association for the Cultivation of Science, Kolkata 700032, India}

\begin{abstract}
  We demonstrate that the kinematically-constrained model of Rydberg atoms on a two-leg ladder with staggered detuning, $\Delta \in [0,1]$, has quantum many-body scars (QMBS) in its spectrum and represents a non-perturbative generalization of the paradigmatic PXP model defined on a chain. We show that these QMBS result in coherent many-body revivals and site-dependent magnetization dynamics for both N\'eel and Rydberg vacuum initial states around $\Delta=1$. The latter feature leads to eigenstate thermalization hypothesis (ETH)-violating finite imbalance at long times in a disorder-free system. This is further demonstrated by constructing appropriate local imbalance operators that display nonzero long-time averages for N\'eel and vacuum initial states. We also study the fidelity and Shannon entropy for such dynamics which, along with the presence of long-time finite imbalance,
  brings out the qualitatively different nature of QMBS in PXP ladders with $\Delta \sim 1$ from those in the PXP chain. Finally, we identify additional exact mid-spectrum zero modes that stay unchanged as a function of $\Delta$ and violate ETH. 
\end{abstract}

\date{\today}

\maketitle

%\tableofcontents

{\bf{Introduction}}:--
ETH proposes that thermalization of local observables under purely unitary dynamics happens at the level of individual eigenstates of the Hamiltonian for many-body quantum systems~\cite{Deutsch1991, Srednicki1994, Rigol2008, Alessio2016, Dymarsky2018}. Evasion of ETH can occur due to many-body localization (MBL)~\cite{Basko1, Basko2, Huse1, Pal2010, Nandkishore2015, Abanin2019, Aletmultifractal,SerbynMBL2013, HuseMBL2014} in interacting systems with strong disorder. Such systems exhibit anomalous temporal relaxation of imbalance to a nonzero value starting from inhomogeneous initial states. This feature is often treated as an experimental signature of MBL~\cite{ImbalanceExp2015, Smith2016}.

 Recent quench experiments with a kinematically-constrained chain of $51$ Rydberg atoms~\cite{Bernien2017} showed non-trivial persistent revivals when initialized in a N\'eel state while other high-energy initial states showed rapid thermalization. This unusual behavior was shown, using a so-called PXP model~\cite{Turner2018scars1, Turner2018scars2, Sachdev2002, Fendley2004, Lesanovsky2012}, to be due to the substantial overlap of the N\'eel state with certain highly athermal ETH-violating states, dubbed QMBS, embedded in an otherwise ETH-satisfying spectrum~\cite{Turner2018scars1, Turner2018scars2}. Other models with kinematic constraints in one and higher dimensions have already given several realizations for QMBS~\cite{Lin2019, Ok2019, Iadecola2020, Lin2020, Surace2021, Banerjee2021, Karle2021, Mukherjee2021a, Langlett2022, Biswas2022, Aramthottil2022, Zhang2023, Desaules2023a, Desaules2023b, Sau2024, Budde2024} (including in driven quantum matter~\cite{Mizuta2020, Mukherjee2020b, Mukherjee2020c, Mukherjee2020d, Mukherjee2022e, Hudomal2022}) as well as other anomalous dynamical features such as Hilbert space fragmentation~\cite{Mukherjee2021a, Langlett2021, Mukherjee2021f, Chattopadhyay2023, QuantumEast1, QuantumEast2} and superdiffusive energy transport~\cite{Marko2023}.

QMBS, in the PXP model, can be described within a forward scattering approximation (FSA)~\cite{Turner2018scars1, Turner2018scars2} in terms of an {{\it emergent}} free spin whose magnitude scales linearly with system size~\cite{Ho2019, Turner2021c, Omiya2023a, Omiya2023b}. Within FSA, the maximally up and down states of this ``giant'' spin are given by the period-$2$ N\'eel state and its translated partner. The persistent many-body revivals can then be interpreted as a coherent spin precession connecting these maximally up and down states~\cite{Turner2018scars1, Turner2018scars2}. However, unlike in MBL, no long-time average imbalance survives dynamically for a quench from the inhomogeneous N\'eel state.

In this Letter, we consider a two-leg Rydberg ladder in the presence of finite $\Delta$ \cite{Sarkar2023} which show several features of QMBS that are absent in the paradigmatic PXP chain. Mapping the presence (absence) of a Rydberg excitation to a $S=1/2$ degree of freedom, $\sigma^z=+1$ ($-1$), we show that a quench from either the N\'eel or the vacuum state leads to site-dependent magnetization dynamics and to nonzero long-time average values for appropriately defined imbalance operators for $\Delta \gtrapprox 0.5$ on finite-sized ladders. In contrast, these imbalances vanish for unitary dynamics from a generic initial state as per ETH. These constitute imbalance as a signature of scar-induced ETH violation; to our knowledge, such an effect has not been reported earlier in disorder-free systems. The PXP ladder studied here leads to persistent many-body revivals from both the N\'eel and the vacuum states (only N\'eel states) around $\Delta=1$ ($\Delta=0$). The analysis of the fidelity and Shannon entropy dynamics starting from these initial states indicates absence of a FSA picture in an unentangled Fock basis for QMBS in ladders around $\Delta \sim 1$; this feature, along with the presence of ETH violating imbalance as well as other anomalous mid-spectrum zero modes that stay unchanged~\cite{Sau2024, Udupa2023} as a function of $\Delta$, sets these QMBS apart from their counterparts in PXP chains.    

\begin{figure}[!tpb]
    \begin{tikzpicture}[x=1.7cm,y=1.7cm]
          
        \draw [dashed] (0,0)-- (0,\ay);
        \draw [dashed] (\ax,0)-- (\ax,\ay);
        \draw [dashed] (2*\ax,0)-- (2*\ax,\ay);
        \draw [dashed] (3*\ax,0)-- (3*\ax,\ay);
        \draw [dashed] (4*\ax,0)-- (4*\ax,\ay);
        \draw [dashed] (5*\ax,0)-- (5*\ax,\ay);
        \draw [dashed] (0,0)-- (5*\ax,0);
        \draw [dashed] (0,\ay)-- (5*\ax,\ay);
        \begin{scriptsize}          
            \shade [ball color=gray] (0,0)   circle (\ballsize);
            \shade [ball color=gray] (0,\ay) circle (\ballsize);
            \shade [ball color=blue] (\ax,0)   circle (\ballsize);
            \shade [ball color=gray] (\ax,\ay) circle (\ballsize);
            \shade [ball color=red] (2*\ax,0)   circle (\ballsize);
            \shade [ball color=blue] (2*\ax,\ay) circle (\ballsize);
            \shade [ball color=blue] (3*\ax,0)   circle (\ballsize);
            \shade [ball color=gray] (3*\ax,\ay) circle (\ballsize);   
            \shade [ball color=gray] (4*\ax,0)   circle (\ballsize);
            \shade [ball color=gray] (4*\ax,\ay) circle (\ballsize);   
            \shade [ball color=gray] (5*\ax,0) circle (\ballsize);
            \shade [ball color=gray] (5*\ax,\ay) circle (\ballsize);            
          \end{scriptsize}
    \end{tikzpicture}    
    \caption{A depiction of the Hilbert space constraint for the Hamiltonian in Eq.\ \eqref{eq:hamiltonian_ladder} for $N=12$.  A site with 
    a Rydberg excited atom (red circle) implies that its nearest-neighbor sites (blue circles) cannot have another Rydberg excitation.}
    \label{fig:model-cartoon}
   \end{figure}
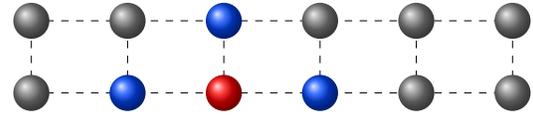      
    
{\bf{Model Hamiltonian}}:--
We consider $N$ $S=1/2$ spins on the sites of a two-leg ladder, with $L$ sites on
each leg ($N=2L$), described by the Hamiltonian:
    \begin{equation}
        \hat{\mathcal{H}}= -\Delta \sum_{j=1}^{L}\sum_{a=1}^2 (-1)^j \hat{\sigma}_{j,a}^z -w\sum_{j=1}^L\sum_{a=1}^2 \hat{\tilde{\sigma}}^{x}_{j,a},
        \label{eq:hamiltonian_ladder}
    \end{equation}
    where $\Delta\ge0$ denotes a staggered detuning while $w>0$ is the coupling strength between the ground and excited states of Rydberg atoms. The site index $(j,a)$ denotes the bottom [top] of the $j$-th rung with $(j,1)$ [$(j,2)$] (Fig.\ \ref{fig:model-cartoon}), $\hat{\tilde{\sigma}}^{x}_{j,a}:=\hat{P}^{\downarrow}_{j,\overline{a}}\hat{P}^{\downarrow}_{j-1,a}\hat{P}^{\downarrow}_{j+1,a}\hat{\sigma}_{j,a}^{x}$ where $(j,\overline{a})$ denotes the other site that belongs to the same rung as $(j,a)$, with $\hat{P}^{\downarrow}_{j,a}=(1-\hat{\sigma}^z_{j,a})/2$. The string of $\hat{P}^\downarrow$ operators ensures that no two nearest neighbor spins are simultaneously in the $\sigma^z=+1$ state for this ladder geometry (Fig.\ \ref{fig:model-cartoon}). We take periodic (open) boundary conditions along each leg (rung) of the ladder with even $L$, fix $w=\hbar=1$ and $\Delta \in [0,1]$ to stay well away from the integrable $\Delta \gg 1$ limit, and measure $t$ in units of $\hbar w^{-1}$. The dimension of this constrained Hilbert space is $\mathcal{D}_{H} = (1+\sqrt{2})^L + (1-\sqrt{2})^L + (-1)^L$ (see Ref.~\cite{SImat}).

%%%%%%%%%%%%%%%%%%%%%%%%%%%current Fig 1 
\begin{figure*}[!htpb]
        \centering
        \rotatebox{0}{\includegraphics*[width= 0.45\textwidth]{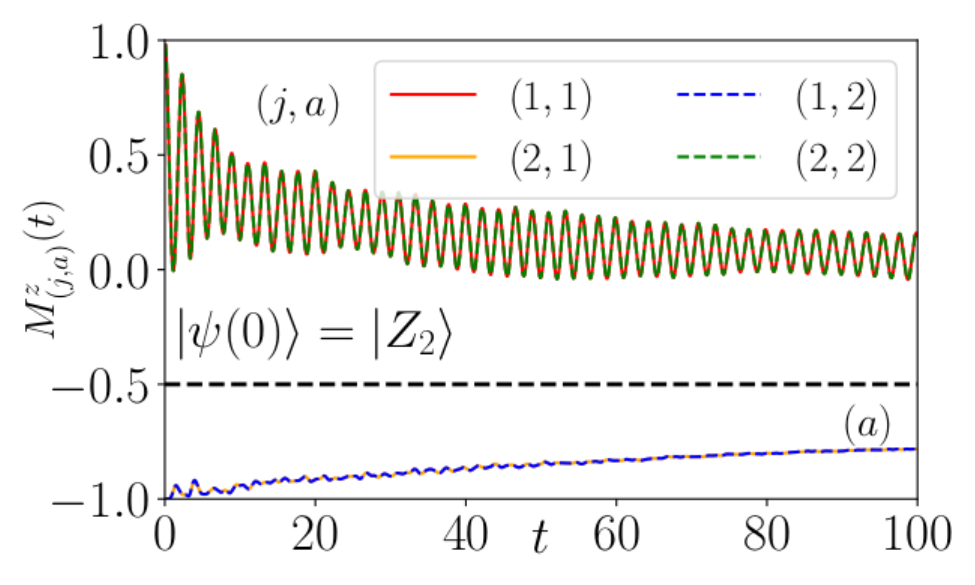}}
        %\rotatebox{0}{\includegraphics*[width= 0.49 \textwidth]{suppl_figs/quench_L16_Mz_idelta_00050.pdf}}
        \rotatebox{0}{\includegraphics*[width= 0.45\textwidth]{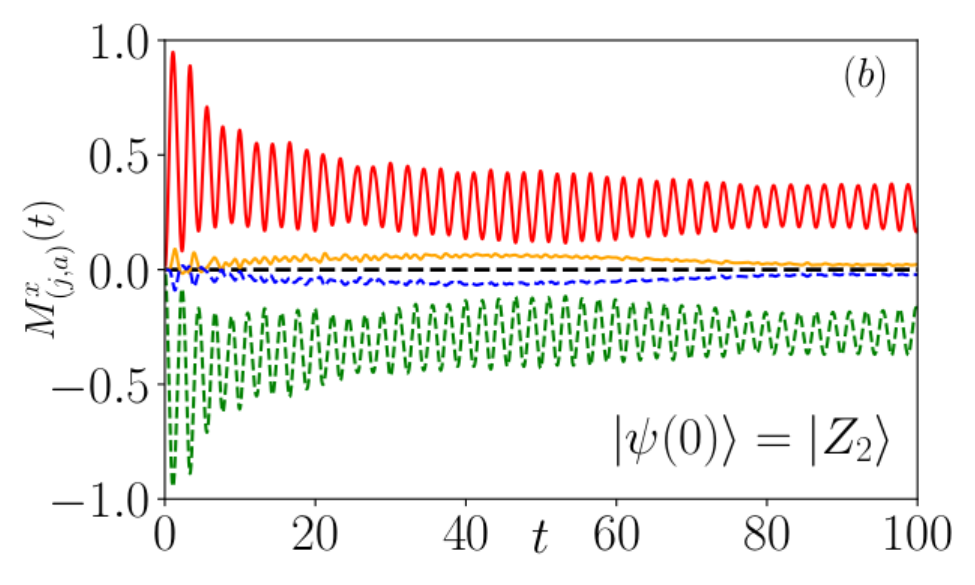}}
        \rotatebox{0}{\includegraphics*[width= 0.45\textwidth]{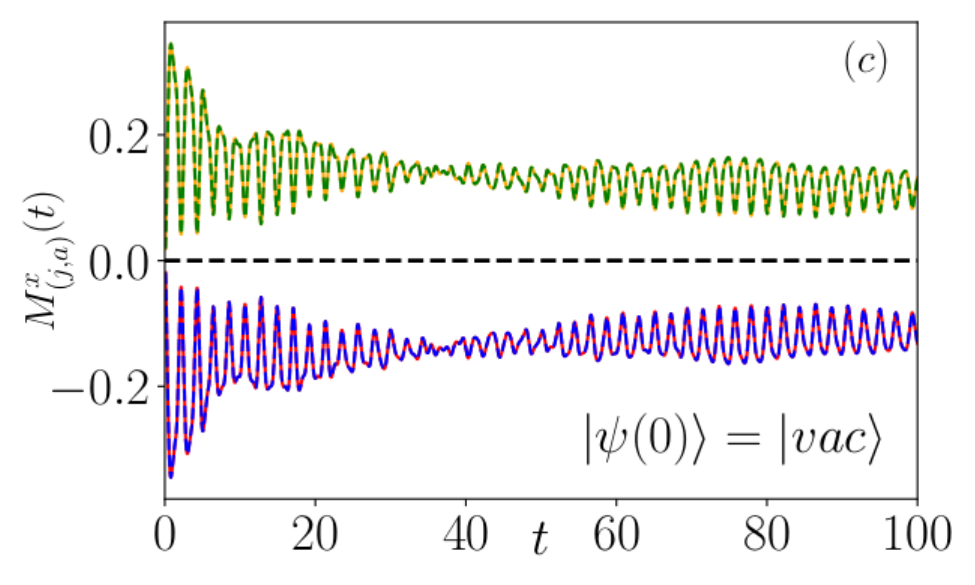}}
        \rotatebox{0}{\includegraphics*[width= 0.45\textwidth]{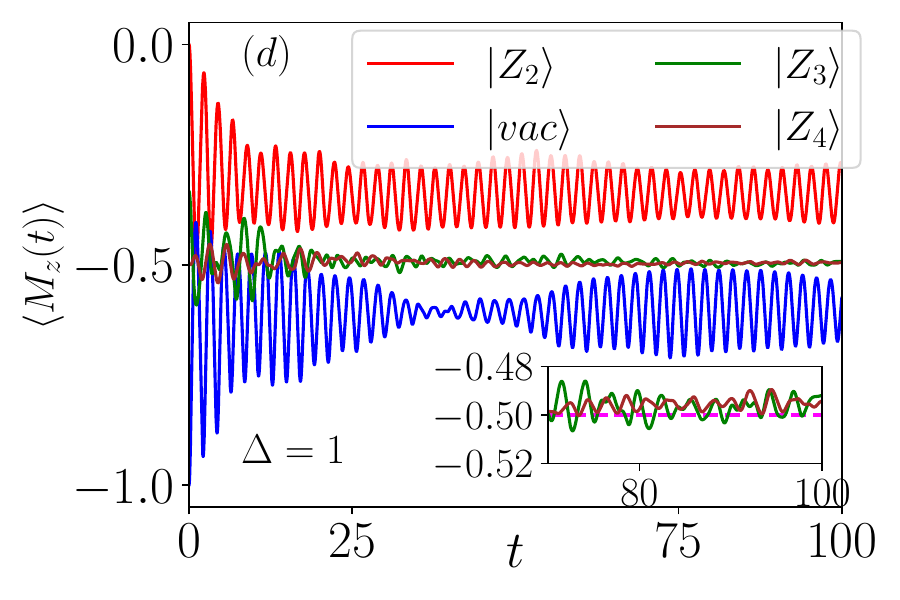}}
        %\rotatebox{0}{\includegraphics*[width= 0.49 \textwidth]{main_figs/quench_L16_Mz_idelta_00100.pdf}}
        \caption{Time evolution of (a) $M^z_{j,a}(t)$ and (b) $M^x_{j,a}(t)$ starting from $\ket{\mathbb{Z}_2}$ state for 
        $j=1,2$ and $ a=1,2$.  (c) Plot of $M^x_{j,a}(t)$ starting from the $\ket{{\rm vac}}$ state. For (a), (b), (c), the color scheme follows panel (a), $\Delta=1$ and $N=32$. Time evolution of $\langle \psi(t) | \hat{M}_z | \psi(t) \rangle $ starting from $\ket{\mathbb{Z}_2}$ (red), $\ket{{\rm vac}}$ (blue) , $\ket{\mathbb{Z}_3}$ (green) and $\ket{\mathbb{Z}_4}$ (brown) for $\Delta=1$ and $N=24$ shown in (d). The inset of (d) display the late time behaviour more clearly; the horizontal dotted line in black indicates  $\llangle \hat{M}_z \rrangle_{\beta=0}$.}
        \label{fig:quench_dynamics}
\end{figure*} 
%%%%%%%%%%%%%%%%%%%%%%%%%%%%%%%%%%%%%%%%%%%%%%%%

%\section{Quench Dynamics\label{sec:quench}}
{\bf{Local magnetization dynamics}}:- Similar to the PXP chain with a staggered field~\cite{Turner2018scars2, Iadecolazeromodes2018}, for every eigenstate $|E\rangle$ of $\hat{\mathcal{H}}$ (Eq.~\eqref{eq:hamiltonian_ladder}) with eigenvalue $E$, there exists a chiral partner $|-E\rangle$
for any $\Delta$ (see Ref.~\cite{SImat}). The $E \rightarrow -E$ symmetry of the spectrum fixes $\llangle \hat{\mathcal{H}} \rrangle_{\beta=0}=0$ ($\llangle \hat{\mathcal{H}} \rrangle_\beta$ denotes a thermal average at inverse temperature $\beta$); thus typical initial states have a zero average energy density.
%The $E \rightarrow -E$ symmetry of the spectrum fixes the thermal average at inverse temperature $\beta$, i.e., $\llangle \hat{\mathcal{H}} \rrangle_{\beta=0}=0$ ; thus typical initial states have a zero average energy density.
Performing a quantum quench from an initial state $|\psi(0)\rangle$ such that $\langle \psi(0)|\hat{\mathcal{H}}|\psi(0)\rangle=0$ implies that $\langle\psi(t)|\hat{\mathcal{A}}|\psi(t)\rangle=\llangle \hat{\mathcal{A}} \rrangle_{\beta=0}$ for any local operator $\hat{\mathcal{A}}$ at late time from ETH if $N=2L \gg 1$. 

%While all classical Fock states in the $\sigma^z$ basis satisfy $\langle \psi(0)|\hat{\mathcal{H}}|\psi(0)\rangle=0$ for $\Delta=0$, this is not true for %$\Delta \neq 0$.

We probe the ergodicity of the model (Eq.~\eqref{eq:hamiltonian_ladder}) by studying quench dynamics from  the N\'eel $\ket{\mathbb{Z}_2}=\ket{\substack{{\circ\bullet\circ\bullet...}\\{\bullet\circ\bullet\circ...}}}$ and vacuum $\ket{{\rm vac}}=\ket{\substack{{\circ\circ\circ\circ...}\\{\circ\circ\circ\circ...}}}$ states at $\Delta=1$; these product states satisfy $\langle \psi(0)|\hat{\mathcal{H}}|\psi(0)\rangle=0$ for any $\Delta$. Here, the filled (open) circles denote $\sigma^z=+1 (-1)$ on the corresponding sites

%(see Ref.~\cite{SImat} for dynamics from other initial states with  $\langle \psi(0)|\hat{\mathcal{H}}|\psi(0)\rangle=0$). 

In Fig.~\ref{fig:quench_dynamics}(a),(b),(c), we show the quench dynamics of such states for $N=2L=32$ by computing the dynamics of the site-resolved magnetization 
\begin{eqnarray}
M_{j, a}^{z[x]}(t)= \langle \psi(t)|\hat{\sigma}^{z}_{j,a} [\tilde \sigma^x_{j,a}]|\psi(t)\rangle \label{sitemag} 
\end{eqnarray}
starting from $\ket{\mathbb{Z}_2}$ and $\ket{{\rm vac}}$ initial states at $\Delta=1$ by evolving $\ket{\psi(0)}$ by integrating the Schr\"odinger equation as an initial value problem using the fourth-order Runge-Kutta integration scheme for $t \in [0,100]$. For both initial states, we find $M^{z[x]}_{j,a}=M_{j\pm 2,a}^{z[x]}$; this ensures that Fig.\ \ref{fig:quench_dynamics}, plotted for $j=1,2$  represents dynamics for any site $j$. 

Starting the dynamics from $\ket{\mathbb{Z}_2}$ results in all spins that were initially $\sigma^z=+1$ (e.g., $M_{1,1}^z(t)$ and $M^z_{2,2}(t)$) oscillating around a superthermal value (i.e., higher than $\llangle \hat{\sigma}^z_{j,a} \rrangle_{\beta=0}$). In contrast, the spins that were initially $\sigma^z=-1$ (e.g., $M_{1,2}^z(t)$ and $M^z_{2,1}(t)$) seem to approach a subthermal value without any oscillations (Fig.\ \ref{fig:quench_dynamics}(a)). In general, we find $M_{2j-1,1}^z(t)=M_{2j,2}^z(t) \ne M_{2j,1}^z(t)=M_{2j-1,2}^z(t)$; this generates a longitudinal imbalance at $\Delta=1$ for $t \in [0,100]$. The corresponding $M^x_{j,a} (t)$ (Fig.\ \ref{fig:quench_dynamics}(b)) for the sites that show oscillating longitudinal magnetization is sizeable in magnitude compared to the rest of the spins (for which it is close to $0$ for these $t$): $M_{2j-1,1}^x(t)=- M_{2j,2}^x(t)$ and  $M_{2j,1}^x(t)=- M_{2j-1,2}^x(t)$  with $|M_{2j,2}^x(t)| \ll |M_{2j,1}^x(t)|$. This generates a transverse imbalance.  

In contrast, starting the dynamics from the vacuum state, one finds oscillating transverse magnetization (Fig.\ \ref{fig:quench_dynamics}(c)) which satisfies $M_{2j-1,1}^x(t)=M_{2j-1,2}^x(t)=-M_{2j,1}^x(t)=- M_{2j,2}^x(t)$ for $t \in [0,100]$ at $\Delta=1$. This generates an imbalance in the transverse magnetization with half of the spins being superthermal (subthermal) for $t \in [0,100]$. Some of these features can be analytically, albeit qualitatively, understood using a trial wavefunction (see Ref.~\cite{SImat}). In Fig.~\ref{fig:quench_dynamics}(d), we show the evolution of $\displaystyle \hat{M}_z=\frac{1}{N}\sum_{j=1}^L\sum_{a=1}^2 \hat{\sigma}^z_{j,a}$ from both $|\text{vac}\rangle$ and $|\mathbb{Z}_2\rangle$, as well as $|\mathbb{Z}_3\rangle$ and $|\mathbb{Z}_4\rangle$ initial states for $N=24$ sites and $\Delta=1$ (see Ref.~\cite{SImat} for more details). While both $|\text{vac}\rangle$ and $|\mathbb{Z}_2\rangle$ lead to oscillations, $|\mathbb{Z}_3\rangle$ and $|\mathbb{Z}_4\rangle$ relax locally to $\beta=0$, as expected from ETH. 
%, while the longitudinal magnetization does not show oscillation.

%%%%%%%%%%%%%%%%%%%%%%%%%%%%%%%%%current fig 3 
\begin{figure*}
        \centering
        \rotatebox{0}{\includegraphics*[width= 0.29\textwidth]{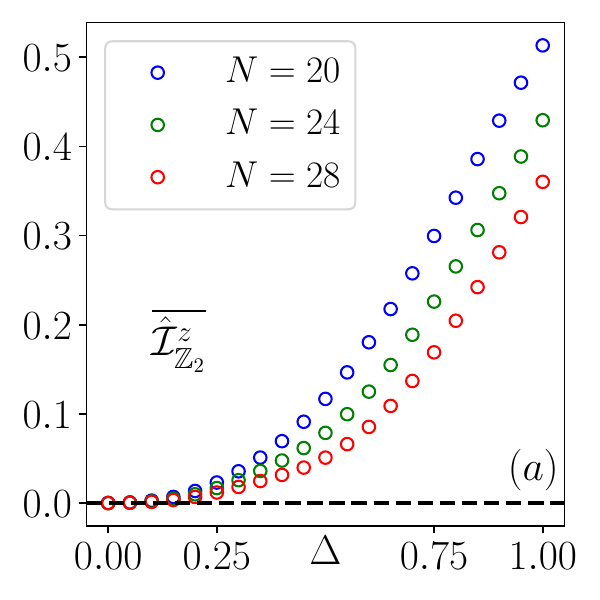}}
        \rotatebox{0}{\includegraphics*[width= 0.29\textwidth]{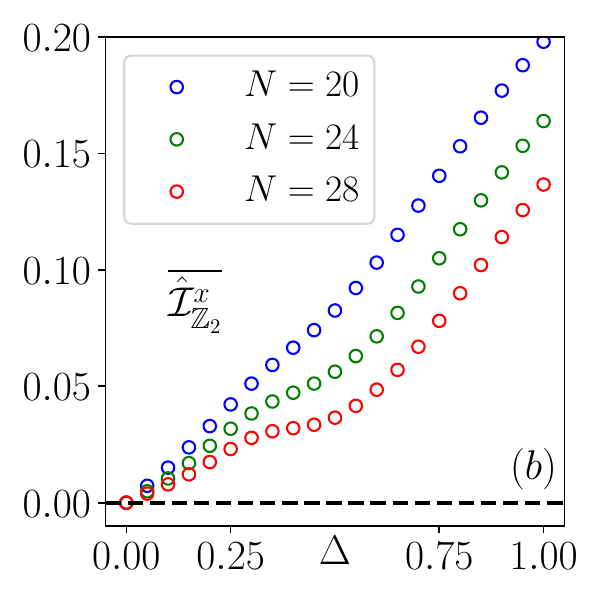}}
        \rotatebox{0}{\includegraphics*[width= 0.29\textwidth]{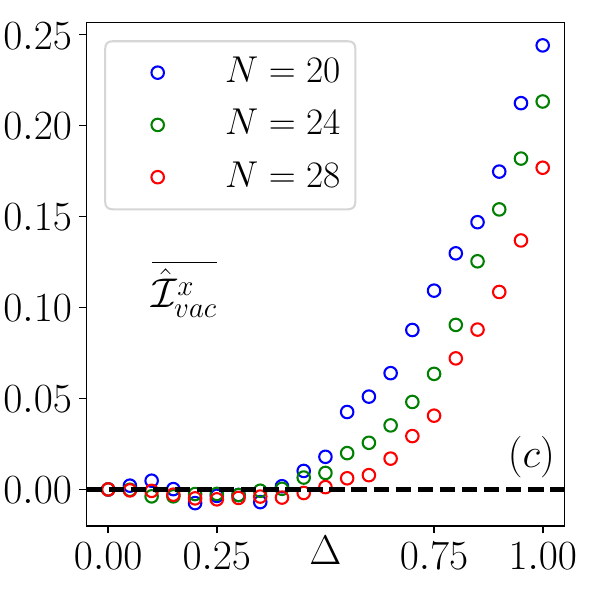}}
        \caption{Infinite-time average of local imbalances (a) $\hat{\mathcal{I}}^z_{\mathbb{Z}_2}$, (b) $\hat{\mathcal{I}}^x_{\mathbb{Z}_2}$ and (c) $\hat{\mathcal{I}}^x_{{\rm vac}.}$ for increasing values of $\Delta$ for $N=20,24,28$.} 
        \label{fig:imb-diag-ensemble}
    \end{figure*}   
%%%%%%%%%%%%%%%%%%%%%%%%%%%%%%%%%%%%%%%%%%%%%%%%%%%%%

These results indicate the following key differences of the local magnetization dynamics at $\Delta=1$ from their counterparts in a PXP chain or a ladder at $\Delta=0$. First, we find ETH violating local magnetization oscillations starting from the $\ket{{\rm vac}}$ state which has no analogue in the latter systems. Second, the plots in  Fig.\ \ref{fig:quench_dynamics} indicate a clear presence of imbalance for $t \gg 1$. We note that whereas the imbalance for short time dynamics may be expected due to the staggered detuning, its persistence at long times, away from the integrable limit $\Delta \gg 1$, constitutes a clear indication of scar-induced imbalance in systems without disorder and provides a novel signature of weak ETH violation. %This is verified by noting the absence of such imbalance starting from other initial states for which ETH is obeyed (see SI).}
Such an imbalance does not occur at $\Delta \simeq 0$ where the scar-induced oscillations lead to zero average imbalance. Moreover, such imbalance is present around $\Delta=0.5$, at least from the N\'eel state, even though no persistent scar-induced oscillations occur around $\Delta=0.5$ (see Ref.~\cite{SImat}).%for which the system conforms to ETH predictions for all initial states(see SI). 

{\bf{Behavior of Imbalance}}:- To further probe the nature of the imbalance, we define local imbalance
operators for both longitudinal and transverse magnetizations as 
\begin{eqnarray}
\hat{\mathcal{I}}^{z}_{\mathbb{Z}_2} &=& \frac{1}{L}\;\sum_{j=1}^L\sum_{a=1}^2 (-1)^{j+a}\;\hat{\sigma}_{j,a}^z, \,\,
        \hat{\mathcal{I}}^{x}_{\mathbb{Z}_2}=\frac{1}{L}\;\sum_{j=1}^L\sum_{a=1}^2 
        (-1)^{a}\;\hat{\tilde{\sigma}}_{j,a}^x, \nonumber\\
        \hat{\mathcal{I}}^{x}_{{\rm vac}} &=& \frac{1}{L}\;\sum_{j=1}^L\sum_{a=1}^2 (-1)^{j}\;\hat{\tilde{\sigma}}_{j,a}^x. \label{eq:imbalance_defn}
    \label{eq:imbalance_defn}
    \end{eqnarray}
    
This particular sign structure is motivated by exact relations between various site resolved magnetizations (see Ref.~\cite{SImat}). Quenching the system from either $\ket{{\rm vac}}$ or $\ket{\mathbb{Z}_2}$ leaves an imprint in long-time average values of these imbalance operators. %which we study.

From ETH, it follows that the long-time average value of any local operator obeys
 \begin{eqnarray}
        \overline{\hat{\mathcal{A}}}= \lim_{T\rightarrow\infty} \frac{1}{T}\int_0^T\;\langle \psi(t) | \hat{\mathcal{A}} | \psi(t) \rangle \; dt =  \llangle \hat{\mathcal{A}} \rrangle_{\beta},
        \label{eq:observable_thermalization_meaning}
    \end{eqnarray}
 where $\beta$ is set by the average energy density of the initial state. For $\ket{\mathbb{Z}_2}$ and $\ket{{\rm vac}}$ initial states $|\psi(0)\rangle$, $\beta=0$ as discussed before. For such initial states, $\llangle \hat{\mathcal{I}}^{x/z}_{\mathbb{Z}_2} \rrangle_{\beta=0} = \llangle \hat{\mathcal{I}}^{x}_{\rm{vac}} \rrangle_{\beta=0}=0$ since $\llangle \hat{\sigma}^z_{j,a}\rrangle_{\beta=0}$ is independent of $(j,a)$ and
 $\llangle \hat{\sigma}^x_{j,a}\rrangle_{\beta=0}=0$. Thus, these imbalances should relax to zero from ETH for such initial states. 

 However, we find that quenching from either $\ket{{\rm vac}}$ or $\ket{\mathbb{Z}_2}$ initial states leaves nonzero long-time average imbalances in finite-sized ladders which reflects a hitherto unexplored, signature of QMBS from specific initial states. To understand this, we express $\overline{\hat{\mathcal{A}}}$ (Eq.~\ref{eq:observable_thermalization_meaning}) using the ``diagonal ensemble''~\cite{Rigol2008}
 \begin{eqnarray}
   \overline{\hat{\mathcal{A}}} &=&  \sum_{\mu \notin \mathscr{H}_0} | \langle \psi(0) | E_\mu \rangle |^2 \; \langle E_\mu |\hat{\mathcal{A}}|E_\mu\rangle \nonumber\\
   && + \sum_{\mu\in\mathscr{H}_0} |\langle \psi(0) | E^{\mathcal{A}}_\mu \rangle|^2 \langle E^{\mathcal{A}}_\mu |\hat{\mathcal{A}}| E^{\mathcal{A}}_\mu\rangle  
   \label{eq:operator_rotated_zero_modes}
    \end{eqnarray}
 In Eq.~\eqref{eq:operator_rotated_zero_modes}, the notation $\mu\in \mathscr{H}_0$ ($\mu \notin \mathscr{H}_0$) refer to the eigenstates of $\hat{\mathcal{H}}$ with $E=0$ ($E \neq 0$). Furthermore, $| E^{\mathcal{A}}_\mu \rangle$ refers to the basis of zero modes that diagonalizes $\hat{\mathcal{A}}$ in the zero mode subspace~\cite{Iadecolazeromodes2018}.  We note here that while an interacting non-integrable model (see Ref. \cite{SImat} for spectral statistics) such as the one considered here is expected to have a non-degenerate spectrum (apart from degeneracies induced by global symmetries), the PXP ladder contains an exponentially large number (in system size) of exact mid-spectrum zero modes ($E=0$) that are protected by an index theorem~\cite{Turner2018scars2, Iadecolazeromodes2018, Udupa2023} for any $\Delta$ (see Ref.~\cite{SImat}). 
 
 We compute the long-time averages of the three imbalances using Eq.~\eqref{eq:operator_rotated_zero_modes} for finite ladders upto $N=28$ for various $\Delta \in [0,1]$ using exact diagonalization (ED) and the results are displayed in Fig.~\ref{fig:imb-diag-ensemble}. We find that $\overline{\hat{\mathcal{I}}^{z}_{\mathbb{Z}_2}}$, $\overline{\hat{\mathcal{I}}^{x}_{\mathbb{Z}_2}}$ and $\overline{\hat{\mathcal{I}}^{x}_{{\rm vac}}}$, where the first two (last) imbalances (imbalance) are (is) calculated from the $\ket{\mathbb{Z}_2}$ ($\ket{{\rm vac}}$) initial state, are all nonzero only when $\Delta \neq 0$. While $\overline{\hat{\mathcal{I}}^{z}_{\mathbb{Z}_2}}$ is monotonic as a function of $\Delta$ for a fixed $N$ (Fig.~\ref{fig:imb-diag-ensemble}(a)), $\overline{\hat{\mathcal{I}}^{x}_{\mathbb{Z}_2}}$ shows a non-convex behavior around $\Delta \approx 0.5$ that sharpens with increasing $N$ (Fig.~\ref{fig:imb-diag-ensemble}(b)). The imbalance generated from the homogeneous vacuum state is even more nontrivial. The behavior of $\overline{\hat{\mathcal{I}}^{x}_{{\rm vac}}}$ (Fig.~\ref{fig:imb-diag-ensemble}(c)) shows that it becomes sizeable only when $\Delta \gtrapprox 0.5$.  
 
 Explicit ED computations of Eq.~\eqref{eq:operator_rotated_zero_modes} for both imbalances starting from the N\'eel state as well as analytic arguments \cite{SImat} shows that nonzero contribution
 %to the imbalances
 solely comes from the zero mode subspace. This clarifies how imbalance can be present around $\Delta =0.5$ even though scar-induced oscillations are absent. Since both imbalances have a thermal expectation value of $0$ for $\beta=0$, this implies that $\sum_{\mu\in\mathscr{H}_0}\hat{\mathcal{A}}_{\mu \mu}=0$ for both from ETH. To have $O(1)$ imbalances, $|\langle \psi(0) | E^{\mathcal{A}}_\mu \rangle|^2$ in Eq.~\eqref{eq:operator_rotated_zero_modes} should deviate strongly from a flat function in $\mu$ to one which gets sharply peaked when $\hat{\mathcal{A}}_{\mu \mu} \sim O(1)$. This immediately implies that the $O(1)$ values of the long-time average imbalances starting from the N\'eel state are due to ``anomalous'' zero modes that have a strong overlap with this initial state. These modes are anomalous because a typical zero mode, that mimics a featureless infinite-temperature state from ETH, will only have $O(1/\mathcal{D}_{\text{H}})$ overlap with the N\'eel state. The long-time average of the transverse imbalance from the vacuum state receives contributions both from zero and nonzero modes. However, an $O(1)$ imbalance again implies QMBS that have a high overlap with the vacuum state.

The features of the magnetization dynamics and imbalance seen above may be tied to the presence of two chirality operators $\hat{\mathcal C}_1 = \hat{T_x} \hat{\mathcal C}$  and $\hat{\mathcal C}_2 = \hat{T_x} \hat{T_y} \hat{\mathcal C}$, where $\hat{T}_{x(y)}$ are translation operator along (perpendicular) to the chains and $\hat{\mathcal C}=\prod_{j} \prod_{a=1,2} \hat{\sigma}_{j,a}^z$, that satisfy $\{\hat{\mathcal{H}}, \hat{\mathcal C}_{1,2}\}=0$. Since there are only two sites along the $y$ direction, the operator $\hat{T_y}$ simply exchanges the two sites of each rung. Using their properties, it is possible to show that \cite{SImat} for all $E \neq 0$ eigenstates $\langle E_{\mu}|\hat{\sigma}_{j,a}^z|E_{\nu}\rangle = \langle -E_{\mu}|\hat{\sigma}_{j+1,a}^z|-E_{\nu}\rangle = \langle -E_{\mu}|\hat{\sigma}_{j+1,\bar a}^z|-E_{\nu}\rangle$; moreover, $|\langle \psi(0)|E_{\mu}\rangle|= |\langle \psi(0)|-E_{\mu}\rangle|$ for $|\psi(0)\rangle =|\mathbb{Z}_2 \rangle$. Thus the difference in dynamics of $M^z_{1,1}(t)$ and $M^z_{1,2}(t)$ and hence ${\mathcal I}^z_{{\mathbb Z}_2}$ do not receive any contribution from the $E \neq 0$ states. Rather, it is entirely due to the contribution of states from the zero energy sector with which $\ket{\mathbb{Z}_2}$ has a large overlap. A similar analysis \cite{SImat} shows $\langle E_{\mu}|\hat{\sigma}_{j,a}^x|E_{\nu}\rangle = -\langle -E_{\mu}|\hat{\sigma}_{j+1,\bar a}^x|-E_{\nu}\rangle$ for the transverse magnetization and leads to $M_{2j-1,1}^x(t)=- M_{2j,2}^x(t)$ and a net transverse imbalance. Finally, an identical analysis for the $|{\rm vac}\rangle$ state yields $M^z_{j,a}(t)= M_{j+1,a}^z(t)= M_{j+1,\bar a}^z (t)$ leading to zero imbalance; in contrast, the transverse magnetization satisfies $M^x_{j,a}(t)= -M^x_{j+1,\bar a}(t)$ and shows a finite ${\mathcal I}^x_{{\rm vac}}$.

%%%%%%%%%%%%%%%%%%%%%%%%%%%%%%current fig. 2
\begin{figure}
        \rotatebox{0}{\includegraphics*[width= 0.4\textwidth]{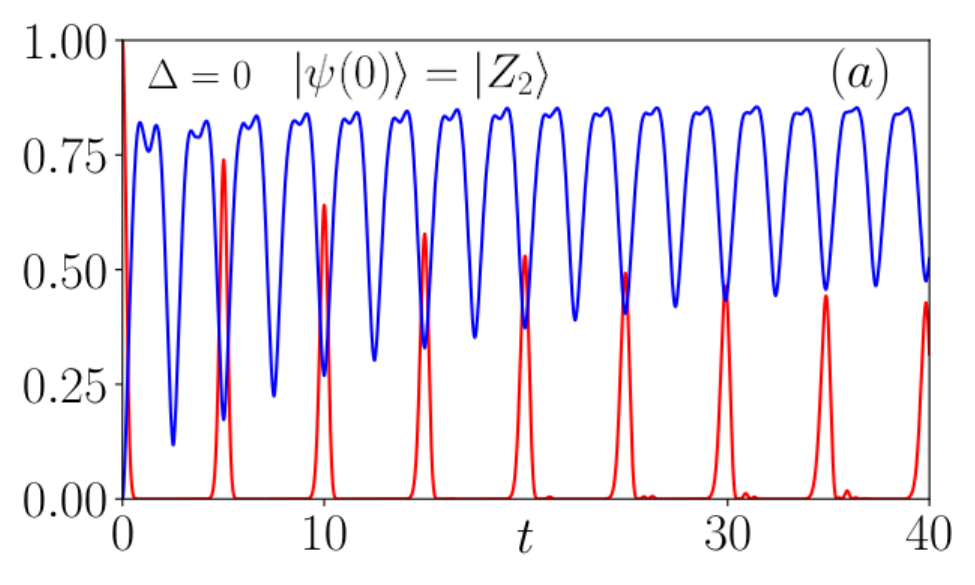}}
        \rotatebox{0}{\includegraphics*[width= 0.4\textwidth]{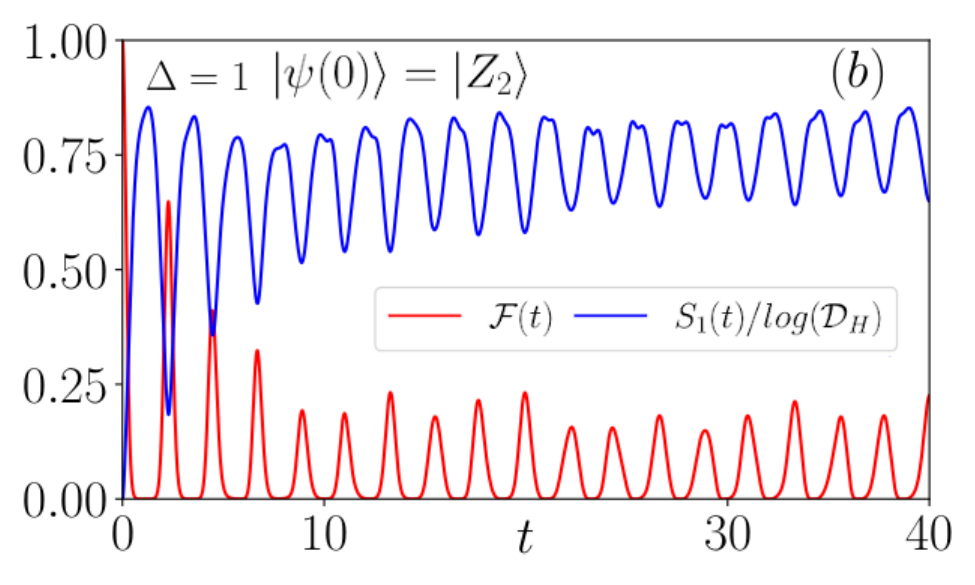}}
        \rotatebox{0}{\includegraphics*[width= 0.4\textwidth]{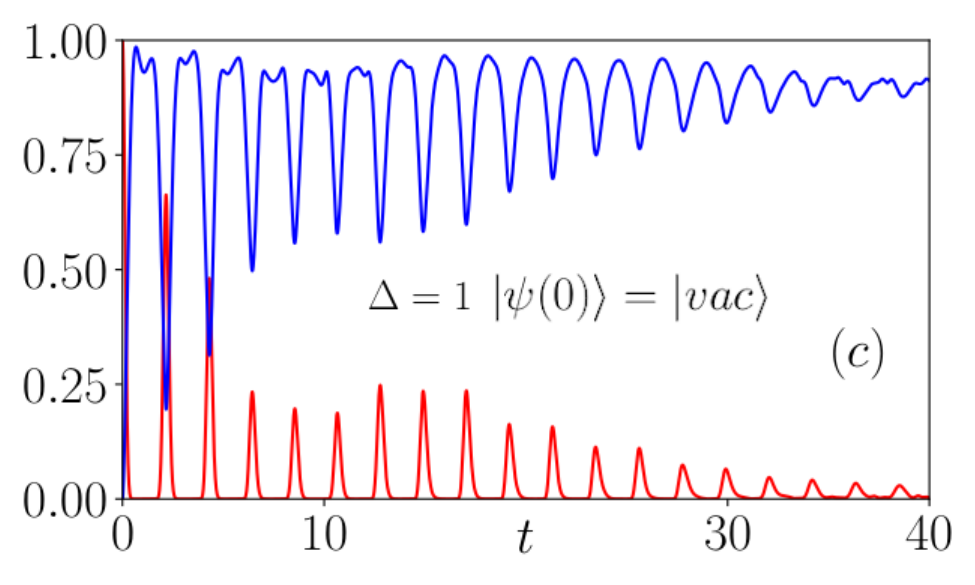}}
        \caption{Evolution of fidelity $\mathcal{F}(t)$ (in red) and (normalized) Shannon entropy $S_1/\log(\mathcal{D}_H)$ (in blue) of $N=32$ ladder for $(a)\;\Delta=0,\ket{\psi(0)}=\ket{\mathbb{Z}_2}$ , $(b)\;\Delta=1,\ket{\psi(0)}=\ket{\mathbb{Z}_2}$ and $(c)\;\Delta=1,\ket{\psi(0)}=\ket{{\rm vac}}$. }
        \label{fig:fid_Shannon_t}
    \end{figure}
%%%%%%%%%%%%%%%%%%%%%%%%%%%%%%%%%%%%%%%%%%%%%%    
    
{\bf Fidelity and Shannon entropy dynamics}:-- To understand the qualitatively different nature of the scar-induced oscillations around $\Delta=1$ and $\Delta=0$, it is instructive to monitor the time evolution of both fidelity $\mathcal{F} (t)=|\langle \psi(t)|\psi(0)\rangle|^2$ and Shannon entropy $S_1 (t) = -\sum_{\alpha=1}^{\mathcal{D}_{\text{H}}} |\psi_{\alpha}(t)|^2 \log(|\psi_{\alpha}(t)|^2)$ where $|\psi(t)\rangle=\sum_{\alpha=1}^{\mathcal{D}_{\text{H}}}\psi_{\alpha}(t) |\alpha\rangle$ when $|\psi(t)\rangle$ is expressed in the classical Fock basis $|\alpha\rangle$ diagonal in $\hat{\sigma}^z$. $S_1(t)$ directly probes the spread of $|\psi(t)\rangle$ in this Fock basis.

At $\Delta=0$ [Fig.\ \ref{fig:fid_Shannon_t}(a)] ($\Delta=1$ [Fig.\ \ref{fig:fid_Shannon_t}(b)]), there are pronounced periodic revivals in $\mathcal{F}(t)$ when the system is initialized in the $\ket{\mathbb{Z}_2}$ state unlike the rapid decay to zero expected from ETH.  However, the behavior of $S_1(t)$ shows a striking difference between $\Delta=0$ and $\Delta=1$. While $S_1(t)$ shows local minima at the same $(t^*, 2t^*,3t^*,\cdots)$ as the revivals in $\mathcal{F}(t)$ in both cases,
%as can be expected since $|\psi(t)\rangle$ closely resembles $\ket{\mathbb{Z}_2}$ at these $t$ and thus has a small spread in the Hilbert space,
it shows additional pronounced local minima at $(t^*/2, 3t^*/2, 5t^*/2,\cdots)$ for $\Delta=0$ which are absent for $\Delta=1$. Such additional local minima in $S_1(t)$ are also present starting from both period-$2$ and period-$3$ initial states for the PXP chain~\cite{SImat}.

These additional minima in $S_1(t)$ at $\Delta=0$ are due to local maxima in $|\bra{\overline{\mathbb{Z}}_2}\exp(-i\hat{\mathcal{H}}t)\ket{\mathbb{Z}_2}|^2$ at $(t^*/2, 3t^*/2, 5t^*/2,\cdots)$ where the state $|\psi(t)\rangle$ can again be approximated by $\ket{\overline{\mathbb{Z}}_2}$ which is related to $\ket{\mathbb{Z}_2}$ by $\hat{\sigma}^z \rightarrow -\hat{\sigma}^z$ on all sites. Thus, the persistent many-body revivals can be interpreted as a coherent spin precession of an emergent giant spin whose maximally polarized states may be approximated by $\ket{\mathbb{Z}_2}$ and $\ket{\overline{\mathbb{Z}}_2}$ respectively. This PXP-like scarring regime from $\ket{\mathbb{Z}_2}$ persists from $\Delta=0$ upto $\Delta \approx 0.2$ from the $N=32$ data \cite{SImat}. Interestingly, the timescale of fidelity revivals is remarkably close to that obtained from two {\it independent} PXP chains, where both chains are individually initialized to N\'eel states, with one being a translated partner of the other to maintain the Hilbert space constraints on a ladder.

On the other hand, the many-body revivals from  $\ket{\mathbb{Z}_2}$ for $\Delta=1$ cannot be interpreted in terms of two end states where the other either appears classical or can be approximated by a superposition of a small number of Fock states.   This non-PXP-like scarring regime from $\ket{\mathbb{Z}_2}$ persists from $\Delta\approx 0.75$ onward from the $N=32$ data \cite{SImat}. The many-body revivals in $\mathcal{F}(t)$ starting from the $\ket{{\rm vac}}$ state at $\Delta=1$ (Fig.~\ref{fig:fid_Shannon_t} (c)) behaves similarly with the local minima of $S_1(t)$ coinciding with the maxima of $\mathcal{F}(t)$. The revivals from the $\ket{{\rm vac}}$ state does not have any analog in the PXP chain and this non-PXP-like scarring regime from $\ket{{\rm vac}}$ persists from $\Delta\approx 0.7$ onward from the $N=32$ data \cite{SImat}. The corresponding entanglement entropy dynamics~\cite{SImat} in both cases suggests a more natural description in a basis where entanglement of spin pairs along rungs plays a crucial role.
%, leading to a breakdown of the {\it independent} chain picture.

In all cases with such periodic revivals, monitoring $|\bra{\mathbb{Z}_2}E\rangle|^2$ and $|\bra{{\rm vac}}E\rangle\rangle|^2$ from ED data shows towers of states with anomalously high overlap with these initial states which are approximately equidistant in energy with a spacing of $\delta E$ near the middle of the spectrum. The revival timescale of $\mathcal{F}(t)$ are related to $\delta E$ by $t^* \approx 2\pi/\delta E$ \cite{SImat}. However, the timescale for oscillations in local magnetization is $t^*/2$ ($t^*$) from the initial $\ket{\mathbb{Z}_2}$ state in the PXP-scarring (non-PXP scarring) regime (Fig.~\ref{fig:quench_dynamics} and Ref.\ \cite{SImat}).

{\bf{Additional anomalous zero modes}}:--The spectrum of this model possesses other anomalous zero modes apart from the ones that contribute to the long-time average imbalances (Eq.~\ref{eq:operator_rotated_zero_modes}) or oscillations 
from $|\text{vac}\rangle$ and $|\mathbb{Z}_2\rangle$. These are {\it simultaneous} zero modes of both the non-commuting terms in $\hat{\mathcal{H}}$ (Eq.~\ref{eq:hamiltonian_ladder}), i.e., $\sum_{j=1}^{L}\sum_{a=1}^2 (-1)^j \hat{\sigma}_{j,a}^z$ and $\sum_{j=1}^L\sum_{a=1}^2 \hat{\tilde{\sigma}}^{x}_{j,a}$. Since such eigenstates stay unchanged as $\Delta$ is varied in spite of the exponentially small (in system size) energy level spacing in their neighborhood, these necessarily violate ETH~\cite{Banerjee2021}. While the existence of such anomalous zero modes was shown even in the thermodynamic limit for certain interacting models~\cite{Udupa2023, Sau2024}, here we demonstrate their presence using ED on finite-sized ladders. Since a typical zero mode is expected to satisfy ETH and coexists with such anomalous zero modes at the same energy, a further numerical procedure is required to identify them from ED~\cite{Biswas2022}. This gives $\mathcal{N}_0=1, 3, 6, 8, 9, 8$ for ladders with $N=4, 8, 12, 16, 20, 24$, where $\mathcal{N}_0$ is the number of such anomalous zero modes.  
   
    \begin{figure}
        \centering
        \rotatebox{0}{\includegraphics*[width= 0.35\textwidth]{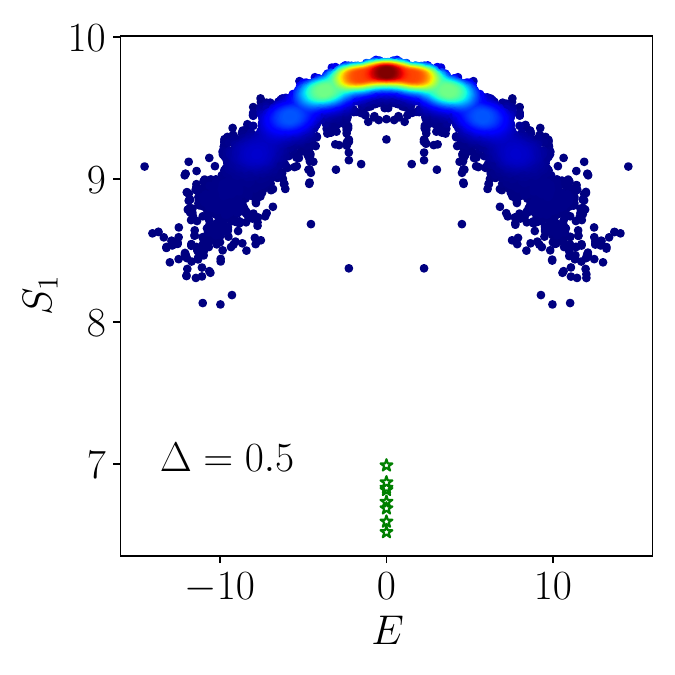}}
        \caption{Shanon entropy $S_1$ of all eigenvectors (where warm colors indicate a higher density of states in the colormap) versus the anomalous zero modes (green star points) that stay unchanged as a function of $\Delta$ for $N=24$ and $\Delta=0.5$.}
        \label{fig:S1N24}
    \end{figure}

We show the data for the Shannon entropy $S_1$ for all eigenstates obtained for a ladder with $N=24$ at $\Delta=0.5$ from ED in Fig.~\ref{fig:S1N24}. While such anomalous modes are much more localized in the Hilbert space than any typical zero mode, these do not necessarily have low entanglement entropy~\cite{Udupa2023, Motrunich_volumescars} (see Ref~\onlinecite{SImat}). To our knowledge, simultaneous zero modes have not been demonstrated in the context of PXP chains.
    
{\bf{Conclusions and outlook}}:--
We have considered a kinematically-constrained model of Rydberg atoms on finite two-leg ladders in the presence of a staggered detuning and demonstrated several new features of QMBS in its spectrum that are absent in the paradigmatic PXP chain (or ladder at $\Delta=0$). Most strikingly, quantum quenches from either the N\'eel or the vacuum state generate nonzero long-time imbalance averages for finite $\Delta$. These imbalances can be probed dynamically by monitoring the site-resolved magnetizations; they decay to zero for dynamics from generic initial states for which ETH is obeyed. We have also demonstrated the co-existence of QMBS that have large overlaps with both the N\'eel and the vacuum initial states, something which has no analog in PXP chains.  Anomalous mid-spectrum zero modes that stay unchanged as a function of the staggered detuning strength have also been shown.

Several open questions arise from our study. While the imbalance operators certainly acquire a nonzero long-time average value, these do not necessarily converge to a time-independent value at late times, unlike the case of MBL; this issue deserves further study. Also, while a PXP-like scarring regime exists for these ladders at small enough $\Delta$, it seems that the regime of co-existing QMBS with large overlap with both N\'eel and vacuum states might not have any simple FSA description in an unentangled basis and may require locally entangled units. A better understanding using ideas similar to Refs.~\onlinecite{Omiya2023a, Omiya2023b} is left as a subject of future study.  Lastly, understanding the simultaneous zero modes that stay unchanged with the staggered detuning and possibly showing their existence in the thermodynamic limit are interesting open questions.  

{\it Acknowledgement}: KS thanks DST for support through JCB/2021/000030. AS thanks Ajit C.~Balram and MP thanks Tista Banerjee for discussions. MS acknowledges the support from UK EPSRC award under the Agreement EP/Y005090/1. 
 
\bibliography{main_v4}

\end{document}

%%%%%%%%%% Merge with supplemental materials %%%%%%%%%%
\pagebreak
\widetext
\begin{center}
\textbf{Supplementary Information for ``Scar-induced imbalance in staggered Rydberg ladders''}
\end{center}
\author{Mainak Pal}
\affiliation{School of Physical Sciences, Indian Association for the Cultivation of Science, Kolkata 700032, India}

\author{Madhumita Sarkar}
\affiliation{Department of Physics and Astronomy, University of Exeter, Stocker Road, Exeter EX4 4QL, United Kingdom}

\author{K. Sengupta}
\affiliation{School of Physical Sciences, Indian Association for the Cultivation of Science, Kolkata 700032, India}

\author{Arnab Sen}
\affiliation{School of Physical Sciences, Indian Association for the Cultivation of Science, Kolkata 700032, India}

%\date{\today}
\maketitle

\onecolumngrid

%%%%%%%%%% Merge with supplemental materials %%%%%%%%%%
%%%%%%%%%% Prefix a "S" to all equations, figures, tables and reset the counter %%%%%%%%%%
\setcounter{equation}{0}
\setcounter{figure}{0}
\setcounter{section}{0}
\setcounter{page}{1}
\makeatletter
\renewcommand{\theequation}{S\arabic{equation}}
\renewcommand{\thefigure}{S\arabic{figure}}
\renewcommand{\thesection}{S\arabic{section}}
\renewcommand{\bibnumfmt}[1]{[S#1]}
\renewcommand{\citenumfont}[1]{S#1}
%%%%%%%%%% Prefix a "S" to all equations, figures, tables and reset the counter %%%%%%%%%%

% \section{Magnetization Dynamics for $N=24$}

%     \begin{figure}[!htpb]
%         \centering
%         \rotatebox{0}{\includegraphics*[width=0.68\textwidth]{suppl_figs/quench_figs/quench_L12_Mz_idelta_00000.pdf}}
%         \rotatebox{0}{\includegraphics*[width=0.68\textwidth]{suppl_figs/quench_figs/quench_L12_Mz_idelta_00050.pdf}}
%         \rotatebox{0}{\includegraphics*[width=0.68\textwidth]{suppl_figs/quench_figs/quench_L12_Mz_idelta_00100.pdf}}
%         \caption{Time evolution of site averaged longitudinal magnetization $\langle M_z(t) \rangle = \langle \psi(t) | \hat{M}_z | \psi(t) \rangle $ starting from simple initial states $\ket{\mathbb{Z}_2}$ (red), $\ket{vac}$ (blue), $\ket{\mathbb{Z}_3}$ (green)  and $\ket{\mathbb{Z}_4}$ (brown) for (a)$\Delta=0$, (b)$\Delta=0.5$ and (c)$\Delta=1$ in $N=24$ ladder.}
%         \label{fig:quench_dynamics_suppl}
%     \end{figure} 

\tableofcontents
%\newpage

\section{Hilbert Space Dimension and spectral statistics \label{app:hsd_count}}

 At inverse temperature $\beta$, the partition function of a quantum theory with Hamiltonian $\hat{\mathcal{H}}$ reads $\mathcal{Z}(\beta)=\text{Tr}\left(e^{-\beta\hat{\mathcal{H}}}\right)$ which implies $\mathcal{Z}(\beta\rightarrow 0 )=\text{Tr} ( 1 ) = \mathcal{D}_H$. The quantity $\mathcal{Z}(\beta\rightarrow 0)$ for Eq.~(1) of main text can be found from the eigenvalues of the corresponding transfer matrix $T$, since $\mathcal{Z}(\beta\rightarrow 0) = \text{Tr}(T^L)$. We note that the state $\ket{\substack{\bullet \\ \bullet}}$ is prohibited due to Rydberg blockade. Using this, in the basis $\{\ket{\substack{\bullet \\ \circ}},\ket{\substack{\circ \\ \bullet}},\ket{\substack{\circ \\ \circ}} \}$, the 
 transfer matrix $T$ can be written as  
\begin{eqnarray}
T &=& \left( \begin{array}{ccc}
                0 & 1 & 1 \\
                1 & 0 & 1 \\
                1 & 1 & 1
                \end{array}  \right).        
        \label{eq:transfer_matrix}
    \end{eqnarray}  

The eigenvalues of the matrix $T$ are $\lambda=-1,1\pm\sqrt{2}$ which implies $\mathcal{D}_H=(1+\sqrt{2})^L + (1-\sqrt{2})^L + (-1)^L$ as given in the main text. Furthermore, since both the product states $\ket{\mathbb{Z}_2}=\ket{\substack{{\circ\bullet\circ\bullet...}\\{\bullet\circ\bullet\circ...}}}$ and $\ket{\text{vac}}=\ket{\substack{{\circ\circ\circ\circ...}\\{\circ\circ\circ\circ...}}}$ lie in the sector with $k_x=0$ (with respect to translations $(j,a) \rightarrow (j+2,a)$ which is also a symmetry of the Hamiltonian), we give the relevant Hilbert space dimension for $k_x=0$ below.

 \begin{table}[!htpb]
        \centering
        \begin{tabular}{||c | c | c ||} 
         \hline
         $N\;(=2L)$ & $\mathcal{D}_H,\: \text{full}$ & $\mathcal{D}_H,k_x=0$\\ [0.5ex] 
         \hline\hline
         4  & 7 & 7 \\ 
         \hline
         8 & 35 & 21 \\
         \hline
         12 & 199 & 71 \\
         \hline
         16 & 1155 & 301 \\
         \hline
         20 & 6727 & 1351 \\ 
         \hline
          24 & 39203 & 6581 \\
          \hline 
          28 & 228487 & 32647 \\
          \hline 
          32 & 1331715 & 166621 \\
         [1ex] 
         \hline
        \end{tabular}
        \caption{Hilbert space dimensions of the model (Eq.~(1) of main text) for various system sizes and $k_x=0$ translational symmetry sector.}
        \label{tab:hsd_table}
    \end{table}

\iffalse
   \begin{table}[!htpb]
        \centering
        \begin{tabular}{||c | c | c | c | c | c ||} 
         \hline
         $N\;(=2L)$ & No Symmetry & $m_{k_x}=0$ & $m_{k_x}=0,m_{k_y}=0$ & $m_{k_x}=0,m_{k_y}=1$ & $m_{k_x}=1,m_{k_y}=1$ \\ [0.8ex] 
         \hline\hline
         4  & 7 & 7 & 5 & 2 & 0 \\
         \hline
         8 & 35 & 21 & 12 & 9 & 8 \\
         \hline
         12 & 199 & 71 & 36 & 35 & 32 \\
         \hline
         16 & 1,155 & 301 & 156 & 145 & 144 \\
         \hline
         20 & 6,727 & 1,351 & 676 & 675 & 672 \\ 
         \hline
          24 & 39,203 & 6,581 & 3,308 & 3,273 & 3,264 \\
          \hline 
          28 & 228,487 & 32,647 & 16,324 & 16,323 & 16,320 \\
          \hline 
          32 & 1,331,715 & 166,621 & 83,388 & 83,283 & 83,232 \\
         [1ex] 
         \hline
        \end{tabular}
        \caption{Hilbert space dimensions of the model Eq.~(1) for various symmetry sectors at different system sizes $N=2L$. The unit of translation along $x$-axis $a_x=2$ and that of $y$-axis $a_y=1$}
        \label{tab:hsd_table}
   \end{table}
   \fi

In order to verify that the model given by Eq.~(1) of the main text is indeed non-integrable, we present numerical results for the distribution ($P(r)$) of the consecutive level spacing ratios ($r$) of eigenvalues of the Hamiltonian.

\begin{eqnarray}
    \delta_\mu = E_{\mu+1}-E_{\mu} , 0 \le r_\mu = \frac{\text{min}\left(\delta_\mu,\delta_{\mu+1}\right)}{\text{max}\left(\delta_\mu,\delta_{\mu+1}\right)} \le 1
\end{eqnarray}

\begin{figure}
    \centering
    \includegraphics[width=0.5\textwidth]{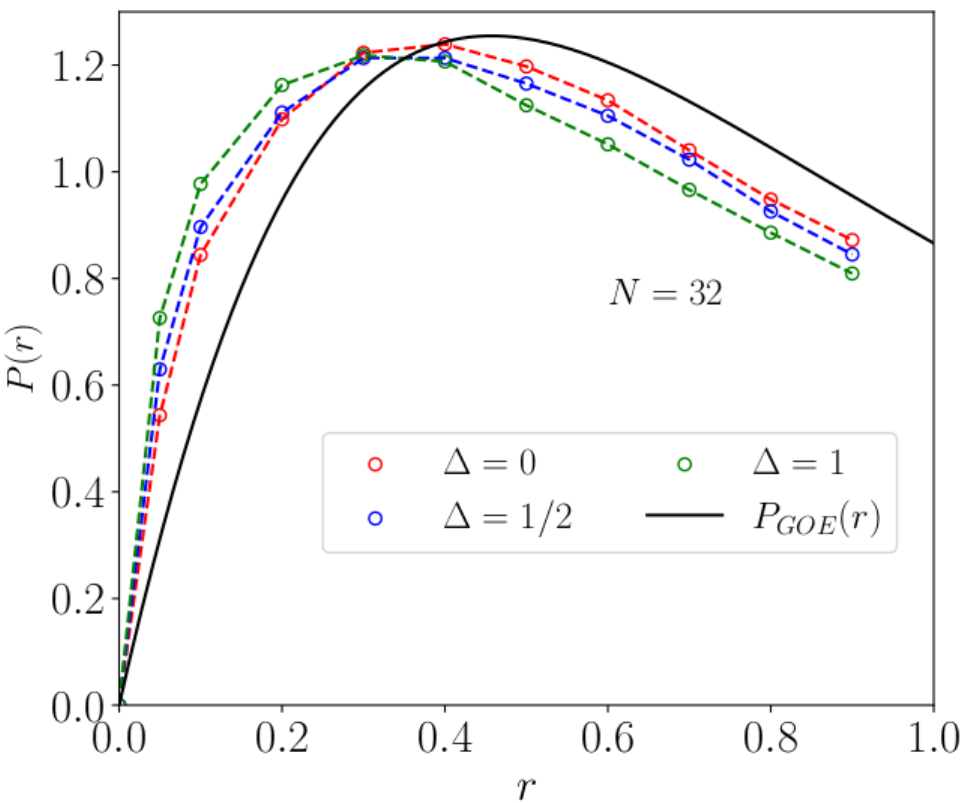}
    \caption{Distribution of consecutive level spacing ratios of model (1) of main text at $\Delta=0,1/2,1$ for a $N=32$ ladder and the RMT prediction of $P_{\text{GOE}}(r)=27(r+r^2)/(4(1+r+r^2)^{5/2})$ (black solid line). The data is generated by averaging over $k_x\ne0,\pi$ sectors.}
    \label{fig:nonint}
\end{figure}

The results shown in Fig.~\ref{fig:nonint} have been generated by averaging over $k_x\ne0,\pi$ symmetry sectors for a $N=32$ ladder for $\Delta=0,1/2,1$ to avoid resolving certain additional discrete symmetries for these two sectors. The results clearly demonstrate that the energy levels exhibits level repulsion, characteristic of non-integrable many-body quantum systems, although the distribution $P(r)$ itself shows a quantitative deviation from the Gaussian orthogonal ensemble (GOE) answer expected from random matrix theory (RMT) considerations at this system size. 

\section{Magnetization Dynamics from $\ket{\mathbb{Z}_2},\ket{\mathbb{Z}_3},\ket{\mathbb{Z}_4}$ and $\ket{{\rm vac}}$ initial states}

\begin{figure}[!h]
    \centering
    \rotatebox{0}{\includegraphics*[width= 0.48\textwidth]{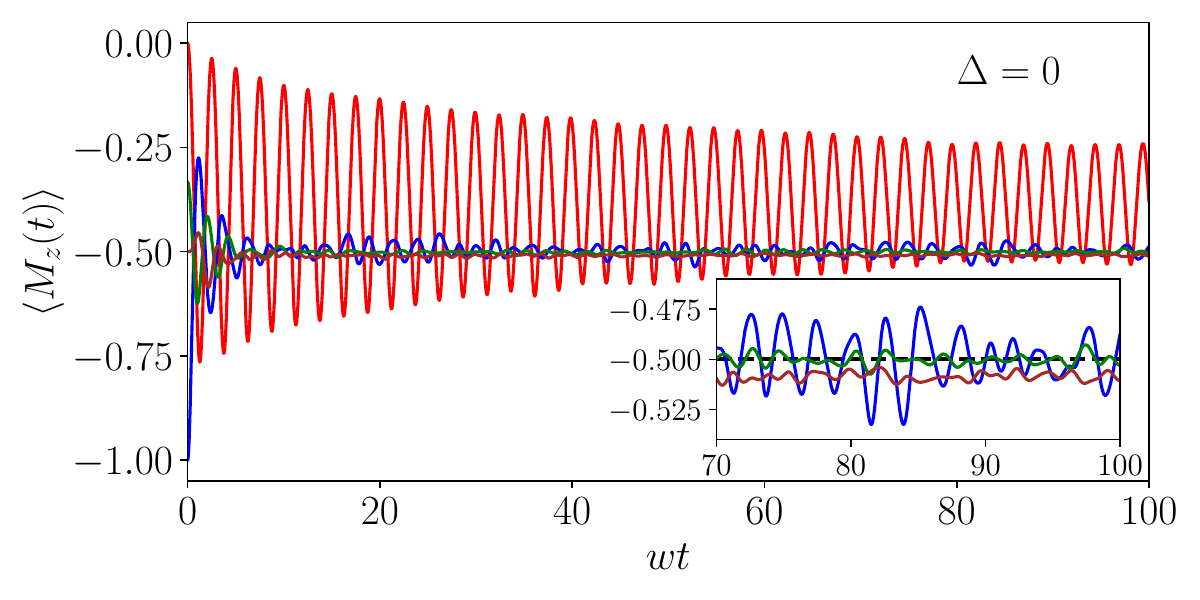}}
    \rotatebox{0}{\includegraphics*[width= 0.48\textwidth]{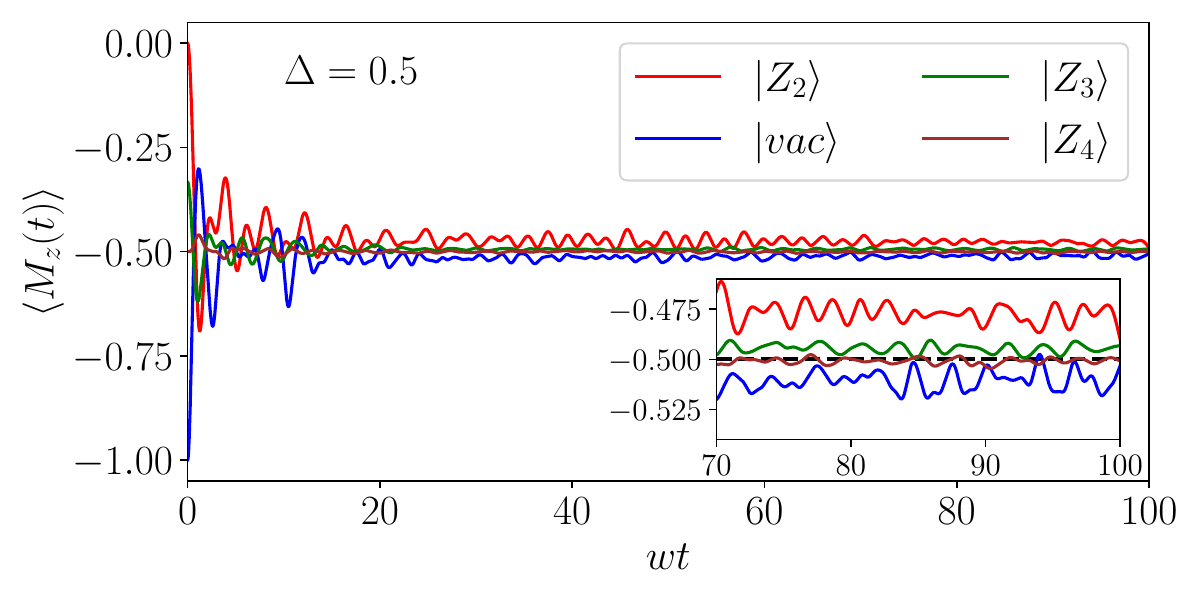}}
    \rotatebox{0}{\includegraphics*[width= 0.48\textwidth]{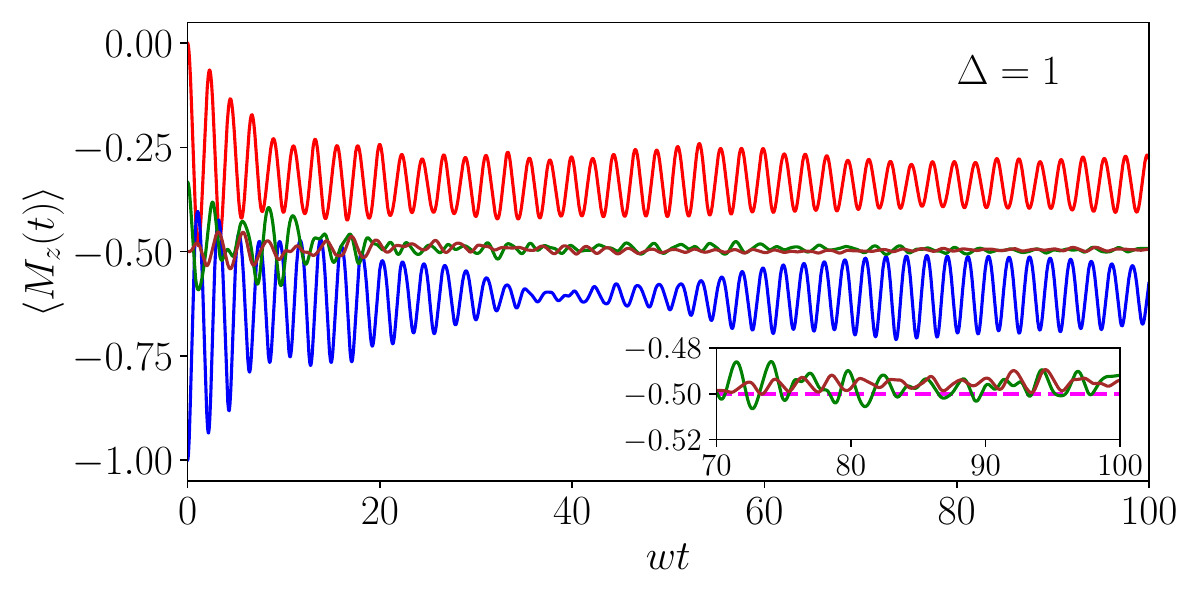}}
    \caption{Time evolution of longitudinal magnetization density $\langle \psi(t) | \hat{M}_z | \psi(t) \rangle $ starting from simple initial states $\ket{\mathbb{Z}_2}$ (red), $\ket{{\rm vac}}$ (blue) , $\ket{\mathbb{Z}_3}$ (green) and $\ket{\mathbb{Z}_4}$ (brown) for (a) $\Delta=0$, (b) $\Delta=0.5$ and (c) $\Delta=1$ for a $N=24$ ladder. The insets display the late time behaviour more clearly; the horizontal dotted line in black indicates  $\llangle \hat{M}_z \rrangle_{\beta=0}$.}
    \label{fig:quench_dynamics}
\end{figure}

We probe the ergodicity of the model by studying quench dynamics from the following product states (i) $\ket{\mathbb{Z}_2}=\ket{\substack{{\circ\bullet\circ\bullet...}\\{\bullet\circ\bullet\circ...}}}$, (ii) $\ket{\mathbb{Z}_3}=\ket{\substack{{\bullet\circ\circ\bullet...}\\{\circ\circ\bullet\circ...}}}$, (iii) $\ket{\mathbb{Z}_4}=\ket{\substack{{\circ\circ\circ\bullet...}\\{\circ\circ\bullet\circ...}}}$, and (iv) $\ket{\text{vac}}=\ket{\substack{{\circ\circ\circ\circ...}\\{\circ\circ\circ\circ...}}}$, all of which satisfy $\langle \psi(0)|\hat{\mathcal{H}}|\psi(0)\rangle=0$ for any $\Delta$. Here, the filled (open) circles denote $\sigma^z=+1 (-1)$ on the corresponding sites and  $\ket{\mathbb{Z}_2}$ ($\ket{\text{vac}}$) denotes a Rydberg N\'eel (vacuum) state.
While the states $\ket{\mathbb{Z}_2}$ and $\ket{\text{vac}}$ are compatible with any $L$ which is even, to accommodate $\ket{\mathbb{Z}_3}$ ($\ket{\mathbb{Z}_4}$) one needs $L$ to be a multiple of $3$ ($4$). In Fig.~\ref{fig:quench_dynamics}, we show the quench dynamics of such states for $N=2L=24$ for three different values of $\Delta$ and observe the time evolution of $\displaystyle \hat{M}_z=\frac{1}{N}\sum_{j=1}^L\sum_{a=1}^2 \hat{\sigma}^z_{j,a}$ (longitudinal magnetization density). 

In all cases ($\Delta=0.0, 0.5, 1.0$) displayed in Fig.~\ref{fig:quench_dynamics}, we see that starting from  $\ket{\mathbb{Z}_3}$ and  $\ket{\mathbb{Z}_4}$ initial states, the local operator $\hat{M}_z$  relaxes close to $\llangle \hat{M}_z \rrangle_{\beta=0}$ (see insets of the panels, where $\llangle \hat{M}_z \rrangle_{\beta=0}$ is denoted by a dotted horizontal black line for clarity) for $t \gg 1$ as can be expected from ETH. For $\Delta=0$ (Fig.~\ref{fig:quench_dynamics} (a)), even the $\ket{{\rm vac}}$ state seems to relax locally to $\beta=0$, while there are clear oscillations from the $\ket{\mathbb{Z}_2}$ state reminiscent of the PXP chain. Increasing $\Delta$ to $0.5$ (Fig.~\ref{fig:quench_dynamics} (b)) leads to all $4$ initial states relaxing locally to $\beta=0$. Remarkably, the oscillations from $\ket{\mathbb{Z}_2}$ reappear as $\Delta$ is increased further to $1$ (Fig.~\ref{fig:quench_dynamics} (c)). In fact, even the $\ket{{\rm vac}}$ state leads to oscillations in the longitudinal magnetization density at $\Delta=1$ (Fig.~\ref{fig:quench_dynamics} (c)) which has no analog in the PXP chain.

\section{Analytical Results} 

In this section, we provide an analysis which may be useful for understanding certain aspects of the imbalance that are numerically obtained in the main text. 
To this end, we first derive analytic expressions for the eigenstates and eigenfunctions of the Hamiltonian corresponding to a single plaquette in Sec.\ \ref{sing1}. This is followed by construction of a trial wavefunction in Sec.\ \ref{var1} which is used to analyze the behavior of imbalance and site resolved magnetization for several initial states. 

\subsection{Single plaquette eigenstates}
\label{sing1} 

We begin with analyzing the Fock states of a single plaquette of the ladder. Denoting $\circ$ as a site with a down spin and $\bullet$ as one with spin-up, we schematically chart the allowed Fock states of a plaquette as 
\begin{eqnarray}
|\mathcal{F}_0\rangle &=& \ket{\substack{{\circ \circ}\\{\circ \circ}}}, \quad  | \mathcal{F}_1\rangle = \ket{\substack{{\bullet  \circ}\\{\circ \circ}}}, \quad |\mathcal{F}_2\rangle = \ket{\substack{{\circ \bullet }\\{\circ \circ}}} 
\quad |\mathcal{F}_3\rangle = \ket{\substack{{\circ \circ}\\{\bullet  \circ}}}, \quad |\mathcal{F}_4\rangle = \ket{\substack{{\circ \circ}\\{\circ \bullet }}}, \quad |\mathcal{F}_5\rangle = \ket{\substack{{\bullet \circ}\\{\circ \bullet}}}, \quad
|\mathcal{F}_6\rangle = \ket{\substack{{\circ \bullet}\\{\bullet \circ}}} 
\label{wavfn1} 
\end{eqnarray} 
The Hamiltonian matrix, in the basis of these Fock states, can be written in terms of a single dimensionless parameter $r=w/(2\Delta)$ as 

\begin{equation}
    H = \begin{pmatrix}
            0 & r & r & r & r & 0 & 0    \\
            r & -1 & 0 & 0 & 0 & r & 0   \\
            r & 0 & 1 & 0 & 0 & 0 & r    \\
            r & 0 & 0 & -1 & 0 & 0 & r   \\
            r & 0 & 0 & 0 & 1 & r & 0    \\
            0 & r & 0 & 0 & r & 0 & 0    \\
            0 & 0 & r & r & 0 & 0 & 0
        \end{pmatrix}.	\label{hmat}
\end{equation}
where we have scaled all energies by $2\Delta$. The corresponding eigenvalues and eigenvectors $\epsilon_n \equiv \epsilon_n/(2\Delta)$ and corresponding eigenfunctions $\ket{\psi_n}=\left(c_{0,n},c_{1,n},c_{2,n},c_{3,n},c_{4,n},c_{5,n},c_{6,n}\right)^T$ can be obtained by diagonalizing $H$. These are given by 
\begin{eqnarray}
    \epsilon_0 &=& 0, \quad \ket{\psi_0} = \frac{1}{\sqrt{1+2r^2}}\left( 0,0,-r,r,0,0,1\right)^T, \\
    \epsilon_1 &=& 0, \quad\ket{\psi_1} = \frac{1}{\sqrt{2+2r^2}}\left( -1,0,r,-r,0,1,0\right)^T, \\
    \epsilon_2 &=& 0, \quad\ket{\psi_2} = \frac{1}{\sqrt{4+1/r^{2}}}\left( -1/r,-1,1,-1,1,0,0\right)^T, \\
     \epsilon_3 &=& -\sqrt{1+2r^2}, \quad \ket{\psi_3} = \frac{1}{\mathcal{N}}\left( 0,\beta,-\alpha,-\beta,\alpha,-1,1\right)^T, \\
     \epsilon_4 &=& +\sqrt{1+2r^2}, \quad \ket{\psi_4} = \frac{1}{\mathcal{N}}\left( 0,-\alpha,\beta,\alpha,-\beta,-1,1\right)^T, \\
     \epsilon_5 &=& -\sqrt{1+6r^2}, \quad \ket{\psi_5} = \frac{1}{\mathcal{N}'}\left( 2,-\beta',-\alpha',-\beta',-\alpha',1,1\right)^T, \\
     \epsilon_6 &=& +\sqrt{1+6r^2}, \quad \ket{\psi_6} = \frac{1}{\mathcal{N}'}\left( 2,\alpha',\beta',\alpha',\beta',1,1\right)^T,  \label{eigen1} 
\end{eqnarray}
where we have introduced the following definitions
\begin{eqnarray}
    \alpha = \frac{1}{2r} \left(\sqrt{1+2r^2}-1\right), \:\: 
    \beta = \frac{1}{2r} \left(\sqrt{1+2r^2}+1\right), \:\: 
    \mathcal{N} = \sqrt{2} \sqrt{2+1/r^2} \\
    \alpha' = \frac{1}{2r}\left(\sqrt{1+6r^2}-1\right), \:\: 
    \beta' = \frac{1}{2r}\left(\sqrt{1+6r^2}+1\right), \:\: 
    \mathcal{N}' = \sqrt{2}\sqrt{1+6/r^2}.
\end{eqnarray}
We note that there are three zero energy modes and the spectrum is symmetric around $ \epsilon=0$ as expected from general arguments given 
using the presence of chirality operator ${\mathcal C}$ in the main text. In the next subsection, we shall these eigenvalues and eigenvectors to study the magnetization dynamics of the system using a trial wavefunction.

\subsection{Trial wavefunction}
\label{var1} 

In this subsection, we shall study the site-resolved magnetization dynamics of the ladder using a trial wavefunction. To this end,  
we first consider the site-resolved magnetization dynamics for a single plaquette using the results of Sec.\ \ref{sing1}. We begin by analyzing the dynamics of the transverse magnetization starting from the $\ket{\mathbb{Z}_2}$ state. We note that $\hat{\tilde \sigma}_j^x$ for sites $j=1,2,3,4$, can be written in terms of the Fock states $\{\ket{\mathcal{F}_\alpha},\alpha=0,1,...,6\}$ (Eq.\ \ref{wavfn1}) as  

\begin{eqnarray}
    \hat{\tilde \sigma}^x_{1} = \left( \ket{\mathcal{F}_0} \, \bra{\mathcal{F}_1}+\ket{\mathcal{F}_4}\, \bra{\mathcal{F}_5} +\text{h.c.}\right), \quad
    \hat{\tilde \sigma}^x_{2} = \left( \ket{\mathcal{F}_0}\, \bra{\mathcal{F}_2}+\ket{\mathcal{F}_3}\, \bra{\mathcal{F}_6} +\text{h.c.}\right)\\
    \hat{\tilde \sigma}^x_{3} = \left( \ket{\mathcal{F}_0}\, \bra{\mathcal{F}_3}+\ket{\mathcal{F}_2}\, \bra{\mathcal{F}_6} +\text{h.c.}\right), \quad
    \hat{\tilde \sigma}^x_{4} = \left( \ket{\mathcal{F}_0}\, \bra{\mathcal{F}_4}+\ket{\mathcal{F}_5}\, \bra{\mathcal{F}_1} +\text{h.c.}\right). \label{opexp1} 
\end{eqnarray} 
To study the magnetization dynamics of a single plaquette, we now obtain the 
time dependent state $|\psi(t)\rangle = \sum_{\alpha=0..6} c_\alpha(t) \ket{{\mathcal{F}}_\alpha}$ by solving the Schr\"odinger equation 
\begin{eqnarray} 
 i \frac{d c_\alpha(t)}{dt} &=& \sum_{\beta=0..6} \langle \mathcal{F}_\alpha|H|\mathcal{F}_{\beta} \rangle c_{\beta}(t), \quad c_{5}(0)=1, \quad c_{\alpha\ne 5}=0, \label{sch1}
\end{eqnarray} 
where we have scaled $t$ in units of $\hbar/(2\Delta)$. Using $E_1=\sqrt{1+2 r^2}$, $E_2=\sqrt{1+ 6r^2}$, we find
\begin{eqnarray} 
c_0(t) &=& 2 r^2 (-1 + \cos E_2 t)/E_2^2  \quad c_1(t) = \frac{r}{2} \left(\sum_{n=1,2} (1- \cos E_n t)/E_n^2 - i \sin(E_n t)/E_n \right) = -c_5^{\ast}(t), \nonumber\\
c_3(t) &=&  \frac{r}{2} \sum_{n=1,2} \left[ 2 r^2/(E_1^2E_2^2) +(-1)^n \left(\cos (E_1 t)/E_n^2 - i \sin (E_n t)/E_n \right) \right], = - c_4^{\ast}(t), \nonumber\\
c_5(t) &=& 1-\frac{2r^2(1+4 r^2)}{E_1^2 E_2^2} + r^2 \sum_{n=1,2} \cos (E_n t)/E_n^2, \quad c_6 (t) = \frac{-4r^2}{E_1^2 E_2^2} + r^2 \sum_{n=1,2} (-1)^n \cos (E_n t)/E_n^2. \label{dyncoeff} 
\end{eqnarray}

\begin{figure}[!htpb]
    \rotatebox{0}{\includegraphics*[width= 0.49\textwidth]{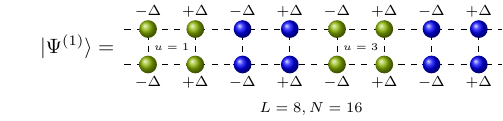}}
    \rotatebox{0}{\includegraphics*[width= 0.49\textwidth]{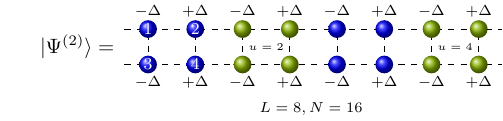}}
    \caption{Schematic representation of $|\psi^{(1,2)}\rangle$ for $N=2L=16$. 
    The green sites represent dynamic spins while the blue sites represent spins frozen to $|\downarrow\rangle$.}
    \label{figsuppa}
\end{figure}

Next, we construct a trial time-dependent many-body wavefunction using $|\psi(t)\rangle$ given by (Fig.\ \ref{figsuppa}) 
\begin{eqnarray} 
        & \ket{\Psi(\{c^{\left(1\right)}\},\{c^{\left(2\right)}\})}(t) = \ket{\Psi^{(1)}(c^{(1)})}(t)+\ket{\Psi^{(2)}(c^{(2)})}(t), \label{eq:variational_state}\\
        & \ket{\Psi^{(1)}(c^{(1)})}(t) = \displaystyle \bigotimes_{u=1,3,5,..} \ket{\psi^{(1)}(c^{(1)})}_u(t), 
        \quad \displaystyle \ket{\Psi^{(2)}(c^{(2)})} (t) = \bigotimes_{u=2,4,6,..} \ket{\psi^{(2)}(c^{(2)})}_u(t), \\
        &  \displaystyle \ket{\psi^{(1,2)}(c^{(1,2)})}_u(t) = \sum_{\alpha=0}^{6} c^{(1,2)}_\alpha(t) \ket{\mathcal{F}_\alpha}_u. \label{varw1} 
\end{eqnarray}
In the above, the single plaquette Fock states, $\{\ket{\mathcal{F}_\alpha},\alpha=0,...,6\}$ are replaced by the corresponding Fock states of the $u$-th plaquette (or unit cell) denoted as $\{\ket{\mathcal{F}_\alpha}_u,\alpha=0,...,6\}$. Eq.~ \eqref{varw1} represents the simplest wavefunction in the total momentum $k_x=0$ sector which reproduces exact dynamics at the level of a single plaquette; however, it does not capture inter-plaquette  interaction which is key to thermalization. Nevertheless, we show that several qualitative features of the dynamics obtained from such a trial wavefunction matches those obtained from exact dynamics. These include the magnetization imbalance which is a key feature obtained form our study.  
 
\begin{figure}[!htpb]
    \rotatebox{0}{\includegraphics*[width= 0.35\textwidth]{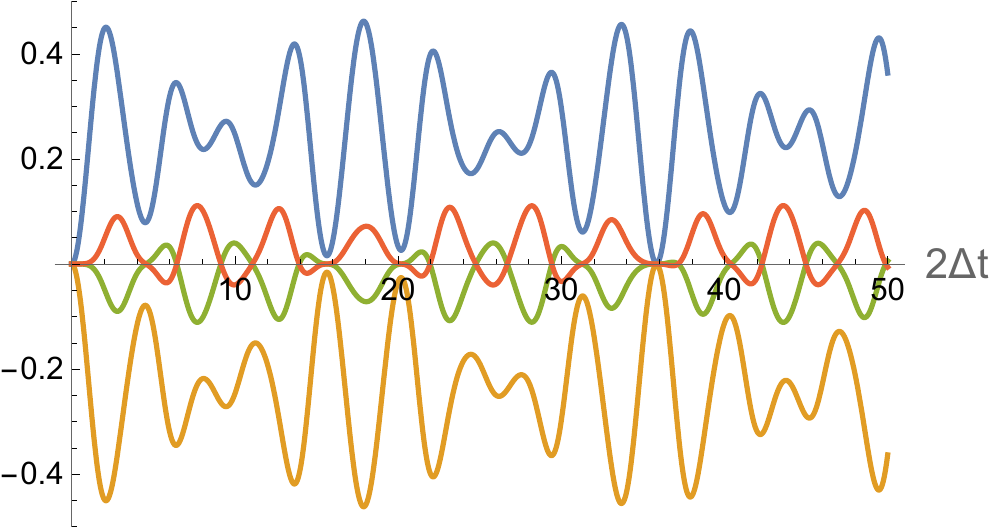}}
    \rotatebox{0}{\includegraphics*[width= 0.31\textwidth]{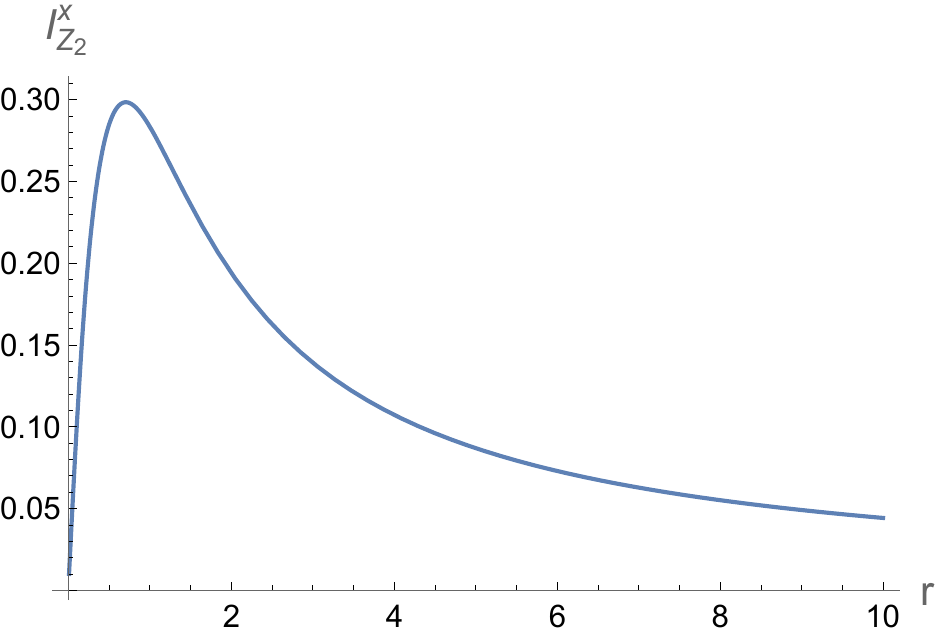}}
    \rotatebox{0}{\includegraphics*[width= 0.31\textwidth]{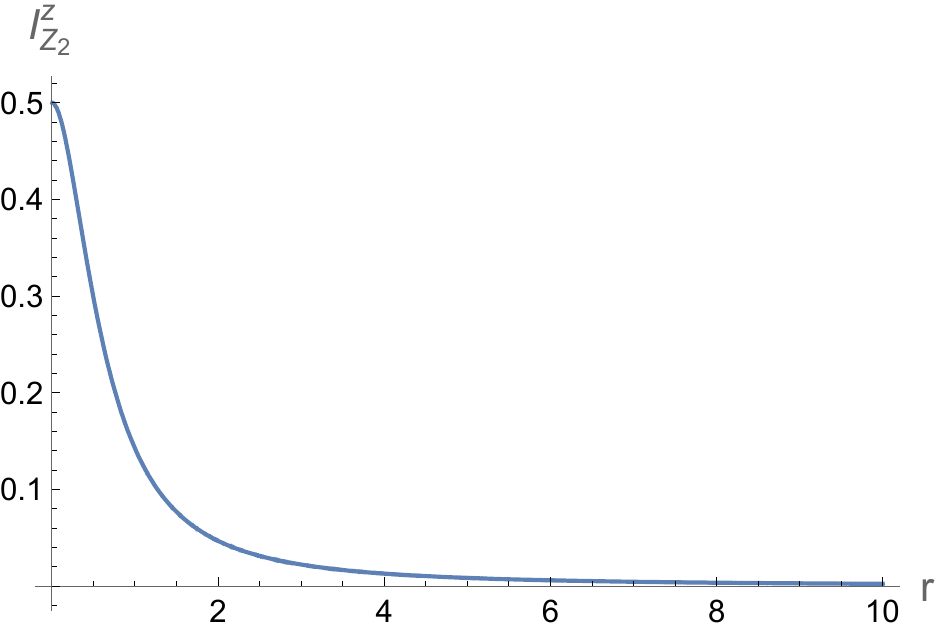}}
    \caption{Left Panel: Plot of $M_{2,2}^x(t)$ (blue), $M_{2,1}^x(t)$ (red), $M^x_{1,2}(t)$ (green) and $M^x_{1,1}(t)$ (yellow)  
    as a function of $2 \Delta t$ showing $M_{1 a}^x (t)=- M_{2 \bar a}^x (t)$ for all $t$. Middle Panel: Plot of the steady state transverse 
    imbalance ${\mathcal I}^x_{{\mathbb Z}_2}$ as a function of $r=w/(2\Delta)$. Right panel: Plot of ${\mathcal I}^z_{{\mathbb Z}_2}$ as 
    a function of $r=w/(2\Delta)$. For all plots $\hbar=1$.}
    \label{figsuppb}
\end{figure}

Using Eqs.\ \eqref{eq:variational_state} and \eqref{varw1}, one can obtain analytic expression for 
\begin{eqnarray}
M_{j,a}^{x}(t) &=& \langle \psi(t)|\hat{\tilde \sigma}^{x}_{j a}|\psi(t)\rangle/N
\end{eqnarray}
for $j=1..2$ and $a=1,2$. Note that for all other sites one obtains identical value of the magnetization within the approximation of a trial wavefunction with no inter-plaquette interaction. Using Eqs.\ \eqref{varw1}, \eqref{dyncoeff}, and \eqref{opexp1}, we find 
\begin{eqnarray} 
M^x_{2,2}(t) &=& -M^x_{1,1}(t)= \frac{r}{2} \left( \sum_{n=1,2} \cos (E_n t)/E_n^2  -2\frac{1+ 4 r^2}{E_1^2 E_2^2} \right) \left( r^2 \sum_{n=1,2} (-1)^n \cos (E_n t)/E_n^2 + \frac{1+8r^2(1+r^2)}{2E_1^2 E_2^2} \right), \nonumber\\
M^x_{2,1}(t) &=& - M^x_{1,2}(t)=  \frac{r^3}{2} \left( \sum_{n=1,2} \cos (E_n t)/E_n^2  -2\frac{1+ 4 r^2}{E_1^2 E_2^2} \right) \left( \frac{4r^2}{E_1^2 E_2^2}- \sum_{n=1,2} (-1)^n \cos (E_n t)/E_n^2  \right).  \label{transz2} 
\end{eqnarray} 
A plot of the transverse magnetization shown in the left panel of Fig.\ \ref{figsuppb} as a function of $2 \Delta t$ 
for $r=1/2$ ($\Delta=w$). We note that it reproduces the qualitative structure of the dynamics obtained from exact numerics; 
in particular $|M_{1,2}^x| \ll M_{1,1}^x$ and $M_{1,a}^x=-M_{2,\bar a}^x$ at all times. Moreover, the steady state value of the 
transverse magnetization can be obtained from Eq.\ \eqref{transz2} by time averaging over the oscillatory terms. Such a time averaging 
leads to $\overline {\cos E_n t} = \lim_{T \to \infty} (1/T) \int_0^T \cos (E_n t) dt = 0$; a straightforward calculation  yields, for the steady-state 
imbalance, 
\begin{eqnarray} 
{\mathcal I}^{x}_{{\mathbb Z}_2} &=& M^{x \,\rm steady}_{2,2} +M^{x \,\rm steady}_{2,1}-M^{x \, \rm steady}_{1,1}-M^{x \, \rm steady}_{1,2}
= \frac{r(1+4 r^2)^3}{E_1^4 E_2^4}. \label{ximbz2}
\end{eqnarray}
A plot of ${\mathcal I}^{x}_{{\mathbb Z}_2}$ is shown in the middle panel of Fig.\ \ref{figsuppb}. Note that $I^{x}_{{\mathbb Z}_2}$ vanishes as $1/r $ for large $r$ which reproduces its linear behavior around $\Delta \simeq 0$ as found in the main text. Also, it vanishes for $r \to 0$ ($\Delta/w \to \infty$) which can be understood by noting that in this limit the energy eigenstates become eigenstates of $\hat{\sigma}^z$ leading to vanishing of $M^x$ on all sites. It is worth noting that for the initial states $\ket{\psi(0)}$ that belong to the  $k_x=0$ sector (such as $\ket{\mathbb{Z}_2}$ and $\ket{{\rm vac}}$) the sign structure of the local magnetization found here agrees with that obtained from the exact numerics presented in the main text. In between, $I^x_{{\mathbb Z}_2}$ peaks at $r=1/\sqrt{2}$ or $\Delta=1/\sqrt{2}$; the position of the peak do not agree with the numerical results which we find to be at $\Delta/w > 1$. 

An exact similar calculation can be carried out for obtaining the dynamics of the longitudinal magnetization given by 
\begin{eqnarray}
M^z_{j,a}(t) = \langle \psi(t)|\hat{\sigma}^z_{j,a}|\psi(t)\rangle,
\end{eqnarray} 
starting from the ${\mathbb Z}_2$ state. The computations are identical to those charted out before and we obtain, for   
\begin{eqnarray} 
M^z_{1,1}(t) &=& M^z_{2,2}(t) = -\frac{1}{2} + \frac{r^2 (1 + 4 r^2) (5 + 40 r^2 + 48 r^4)}{4 E_1^4 E_2^4} + r^2 \Big[ \cos (2 E_1 t)/(4 E_1^4)
+\frac{\cos E_1 t}{2 E_1^4 E_2^4} (2 E_2^2 (1 + 8 r^2(1 + r^2)) \nonumber\\
&& + E_1^2 E_2^2(1+4 r^2)\cos E_2 t + \frac{4 (E_1^2 E_2^ 2-4 r^2) \cos E_2 t + r^2 E_1^2 \cos 2 E_2 t
   + 2 E_1 E_2^3
      \sin E_1 t \sin E_2 t}{4 E_1^2 E_2^4} \Big], \label{longz2} \\
M^z_{1,2}(t) &=& M^z_{2,1}(t) =  -\frac{1}{2} + \frac{r^2 (1+ 12 r^2 + 56 r^4 + 96 r^6)}{2E_1^4 E_2^4} + \frac{r^2}{4 E_1^4 E_2^4} \Big[ 
(E_2^4 \cos 2 E_1 t - 2 (E_1^2 E_2^2 + r^2 E_2^2) \nonumber\\
&& \times (4 r^2+ E_1^2 \cos E_2 t )\cos E_1 t  + E_1^2 (8 r^2(E_1^2 +r^2) \cos E_2 t
- r^2 E_1^2 \cos 2 E_2 t - 2 E_1 E_2^3
      \sin E_1 t \sin E_2 t ) )\Big]. \nonumber
\end{eqnarray}
We note that this yields the same qualitative structure of the longitudinal magnetization as that from exact numerics. The steady state imbalance can be computed in a similar manner as done for the 
transverse magnetization $M_x$ and yields
\begin{eqnarray} 
{\mathcal I}^{z}_{{\mathbb Z}_2} &=& (M^{z \, \rm steady}_{2,2} +M^{z \, \rm steady}_{1,1}-M^{z \, \rm steady}_{1,2}-M^{z \, \rm steady}_{2,1}) = 2 \frac{1+5 r^2}{E_1^2 E_2^2}. \label{zimbz2} 
\end{eqnarray}
Note that for $r=0$, $|{\mathbb Z}_2\rangle$ is an eigenstate and hence the long-time value of the imbalance is close to its value $2$ in the initial state. In contrast, ${\mathcal I}^{z}_{{\mathbb Z}_2}$ vanishes as $1/r^2$ in the large $r$ limit which explain its quadratic behavior around $\Delta \to 0$; this also shows that the imbalance is absent in the standard PXP chain for which $\Delta = 0$. 

Finally, we consider the $\ket{{\rm vac}}$ initial state. To this end, we once again solve Eq.\ \eqref{sch1} but with initial condition $c_0(0)=1$ and 
$c_\alpha(0)=0$ for $\alpha\ne 0$. The wavefunction for the single plaquette, at any time $t$, can be written as $|\psi(t)\rangle = \sum_{\alpha=0..6} d_\alpha(t) \ket{\mathcal{F}_\alpha}$, where $d_\alpha(t)$ is given by
\begin{eqnarray}
d_0(t) &=& \frac{E_1^2 + r^2 \cos E_2 t}{E_2^2}, \quad d_1(t)= -d_2^{\ast}(t)= d_3(t)= -d_4^{\ast} (t)= \frac{r(\cos E_2 t-1 -i E_2 \sin E_2 t)}{E_2^2},\nonumber\\
d_5(t) &=& d_6(t) =  \frac{2r^2(\cos E_2 t -1)}{E_2^2}. \label{vcoeff}
\end{eqnarray} 
Using Eq.\ \eqref{vcoeff}, one can easily obtain the transverse and longitudinal magnetizations for the vacuum states. For the longitudinal magnetization, we find the longitudinal magnetization is same for all sites and is given by
\begin{eqnarray}
M^z_{\rm vac}(t) = -\frac{1}{2E_2^2} +  \frac{r^2}{E_2^4} \left( 1- \cos E_2 t \right). \label{vaclongmag}
\end{eqnarray} 
Thus there is no imbalance for $M^z$ fr the vacuum initial state; this feature agrees with exact numerics. 

\begin{figure}[!htpb]
    \rotatebox{0}{\includegraphics*[width= 0.49\textwidth]{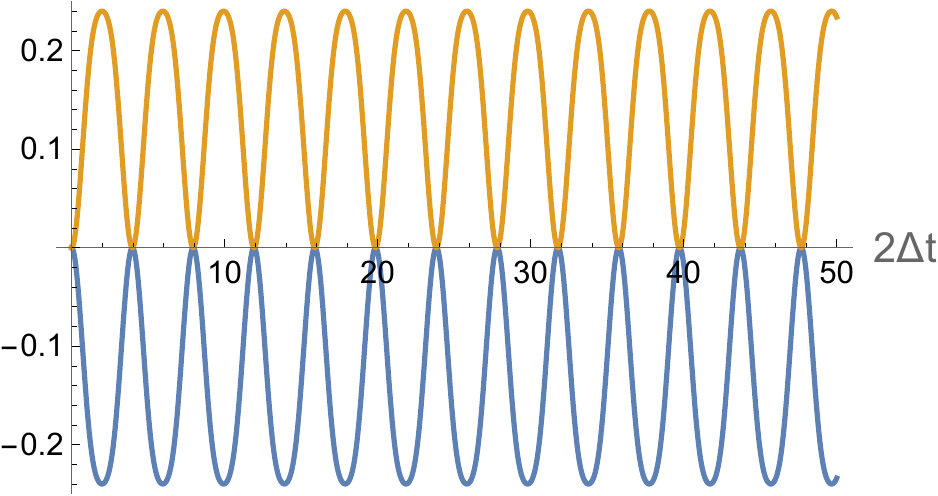}}
    \rotatebox{0}{\includegraphics*[width= 0.49\textwidth]{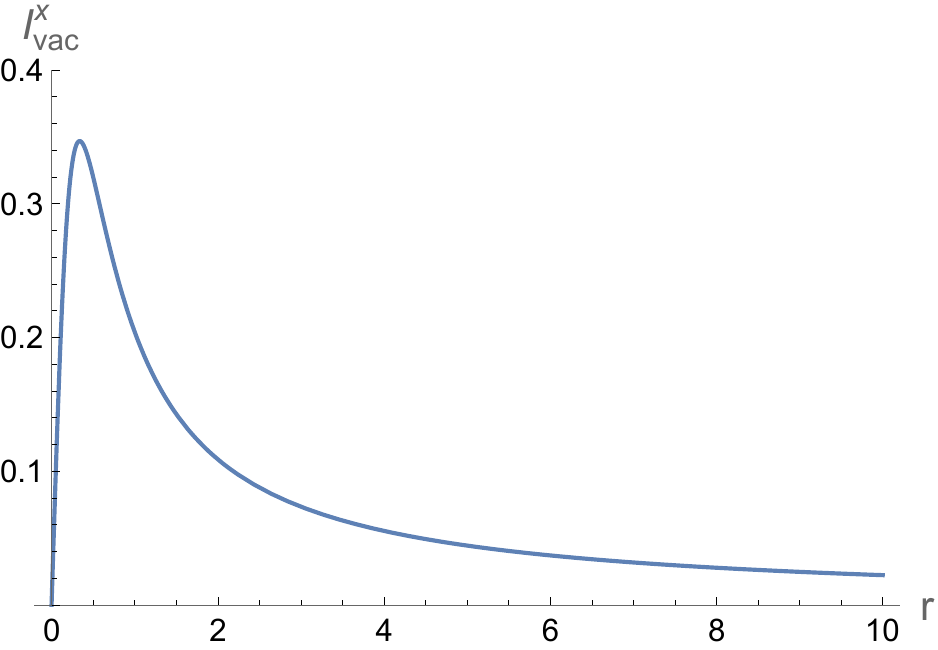}}
    \caption{Left Panel: Plot of $M_{2,2}^x(t)$ (yellow) and $M_{1,1}^x(t)$ (blue) as a function of $2 \Delta t$ starting from the $|{\rm vac}\rangle$ state. Right Panel: Plot of the steady state imbalance ${\mathcal I}^x_{\rm vac}$ as a function of $r=w/(2\Delta)$. For all plots $\hbar=1$.}
    \label{figsuppc}
\end{figure}

In contrast for the transverse magnetization, one finds $M^x_{2,1}= M^x_{2,2}= - M^x_{1,1}= -M^x_{1,2} = M_{\rm vac}^x (t)$ at all times where 
\begin{eqnarray}
M^x_{\rm vac}(t) = \frac{r(\cos E_2 t -1)(E_2^2+ 2r^2(\cos E_2 t -1)) }{E_2^4}.  \label{vaclongmag}
\end{eqnarray} 
This is shown in the left panel of Fig.\ \ref{figsuppc} and leads to a steady state imbalance, 
\begin{eqnarray} 
{\mathcal I}^{x}_{{\rm vac}} &=& \frac{2 r (1+ 3 r^2)}{(1+ 6 r^2)^2}, \label{ximbvac}
\end{eqnarray} 
as shown in the right panel of Fig.\ \ref{figsuppc}. The imbalance vanishes both at $r=0$ and $r \to \infty$; the latter shows absence of such imbalance in the pure PXP chain, However, this simplistic model does not reproduces the feature that the imbalance grows from around $\Delta \sim 0.5$; instead it shows a linear growth around $\Delta =0$. This is most likely due to the effect of inter-plaquette interaction ignored in this simplistic model and is beyond the scope of the current analysis.

\section{Chirality operators and spectral reflection symmetry}{\label{sec:supp_chirality_and_spectral_reflection}}

In this section, we demonstrate the existence of two chirality operators $\hat{C}_{1,2}$ that satisfy $\hat{C}_{1,2} |E_{\mu}\rangle= |-E_{\mu}\rangle$ for any finite $E_{\mu} \ne 0$. To this end, we consider the chirality operator $\prod_{i}\hat{\sigma}^z_i$ of the PXP chain, generalized to the case of a ladder geometry, and define the operators $\hat{\mathcal{C}}_1$ and $\hat{\mathcal{C}}_2$ using translations as follows

\begin{eqnarray}
         \hat{\mathcal{C}} &=& \prod_{j=1}^{L}\prod_{a=1}^2 \hat{\sigma}^z_{j,a}, \quad 
         \hat{\mathcal{C}}_1 = \hat{T}_x \hat{\mathcal{C}}, \quad
         \hat{\mathcal{C}}_2 = \hat{T}_x \hat{T}_y \hat{\mathcal{C}}.
    \label{eq:CC1C2-defn}
\end{eqnarray}
We note at the outset that for the present ladder geometry $\hat T_y$ is identical to $\hat{\mathcal{R}}_x$, which denotes a reflection of sites along an axis that is perpendicular to the rungs and exchanges the two sites of each rung, and set the convention that the action of translation operators $\hat{T}_{x,y}$ on a Fock state translates the spin pattern by one unit to the positive $x/y$ direction. Evidently, the operators $\hat{\mathcal{C}}_1$ and $\hat{\mathcal{C}}_2$ have the following properties
\begin{eqnarray}
    \left\{ \hat{\mathcal{C}_1} , \hat{\mathcal{H}} \right\} &=& \left\{ \hat{\mathcal{C}_2} , \hat{\mathcal{H}} \right\} = 0. \label{eq:C2-prop}
\end{eqnarray}
To see this consider $\hat{\mathcal{C}}_{2}^{-1}\hat{\mathcal{H}}\hat{\mathcal{C}}_{2}$
\begin{eqnarray}
\hat{\mathcal{C}}_{2}^{-1}\hat{\mathcal{H}}\hat{\mathcal{C}}_{2} &=&  = \hat{\mathcal{C}}^{-1}  \hat{T}_y^{-1}\hat{T}_x^{-1} \left(\hat{\mathcal{H}}_z+\hat{\mathcal{H}}_x\right)\hat{T}_x\hat{T}_y \hat{\mathcal{C}} = -\hat{\mathcal{H}_z}-\hat{\mathcal{H}}_x = -\hat{\mathcal{H}},
    \label{eq:C2-prop-proof}
\end{eqnarray}
where we have used the fact that 
\begin{eqnarray} 
\hat{T}_x^{-1}\hat{\mathcal{H}}_z\hat{T}_x &=& -\hat{\mathcal{H}}_z, \quad \hat{T}_y^{-1}\hat{\mathcal{H}}_z\hat{T}_y=\hat{\mathcal{H}}_z,\quad  \hat{\mathcal{C}}^{-1}\hat{\mathcal{H}}_z\hat{\mathcal{C}}=\hat{\mathcal{H}}_z, \nonumber\\
\hat{T}_x^{-1}\hat{\mathcal{H}}_x\hat{T}_x &=& \hat{\mathcal{H}}_x, \quad \hat{T}_y^{-1}\hat{\mathcal{H}}_x\hat{T}_y=\hat{\mathcal{H}}_x, \quad \hat{\mathcal{C}}^{-1}\hat{\mathcal{H}}_x\hat{\mathcal{C}}=-\hat{\mathcal{H}}_x. \label{oprel1} 
\end{eqnarray} 
This establishes Eq.~\eqref{eq:C2-prop} for $\hat{\mathcal{C}_2}$; a similar proof can be easily obtained for $\hat{\mathcal{C}}_1$.

Eqs.~\eqref{eq:C2-prop} implies that the operators $\hat{\mathcal{C}}_1$ and $\hat{\mathcal{C}}_2$ are indeed chirality operators. Using them we can deduced several properties which we chart below. 

First, if $\ket{E_\mu}$ is an eigenmode of $\hat{\mathcal{H}}$ with eigenvalue $E_\mu$, then $\hat{\mathcal{C}}_{1,2}\ket{E_\mu}$ is also an eigenvector of $\hat{\mathcal{H}}$ but with eigenvalue $-E_\mu$. Thus the Hamiltonian $\hat{\mathcal{H}}$ has reflection symmetry about $E=0$ for $\Delta\ne0$. 

Second, if $\hat{\mathcal{C}}_{1,2}\ket{E_\mu}$ is a normalized vector then we can identify it with $\ket{-E_\mu}$ for $E_\mu\ne0$ up to an overall sign. The situations for the zero modes have further subtleties and shall be addressed later. For $E_{\mu} \ne 0$, we find 
 \begin{equation}
     \begin{split}
         ||\; \hat{\mathcal{C}}_1\ket{E_\mu} || & = \langle E_\mu | \hat{\mathcal{C}}_1^\dagger \hat{\mathcal{C}}_1 | E_\mu \rangle =
         \langle E_\mu | \hat{\mathcal{C}}_1^{-1} \hat{\mathcal{C}}_1 | E_\mu \rangle 
         = 1.
     \end{split}
 \end{equation}
In the above derivation we have used the relation $\hat{\mathcal{C}}_1^\dagger=\hat{\mathcal{C}}_1^{-1}$ which can be shown as follows
\begin{equation}
    \begin{split}
        \hat{\mathcal{C}}_1^\dagger & = \left( \hat{T}_x\hat{\mathcal{C}} \right)^{\dagger} 
        = \hat{\mathcal{C}}^{\dagger} \hat{T}_x^{\dagger} 
     = \hat{\mathcal{C}} \hat{T}_x^{-1} 
     = \hat{\mathcal{C}}^{-1} \hat{T}_x^{-1} 
     = \left( \hat{T}_x\hat{\mathcal{C}}\right)^{-1} 
     = \hat{\mathcal{C}}_1^{-1}.
    \end{split}
\end{equation}

Third,  we further note that both $\hat{\mathcal{C}}_1$ and $\hat{\mathcal{C}}_2$ as well as the Hamiltonian $\hat{\mathcal{H}}$ commute with $\hat{\mathcal{R}}_x$. This intertwining of the chirality and reflection symmetry leads to an exponentially large number (in system size) of exact mid-spectrum zero modes with $E=0$ due to the index theorem referred to in the main text. 

To summarize, we find that $\hat{\mathcal{C}}_1\ket{E_\mu}=\ket{-E_\mu}$; similar arguments for $\hat{\mathcal{C}}_2$ shows  $\hat{\mathcal{C}}_2\ket{E_\mu}=\ket{-E_\mu}$. We shall use these properties in the next subsection to derive properties of the imbalance operators.

\section{Derivation of some exact aspects}

In this section, we provide some exact analytical results which supports the numerical observations regarding the dynamics of site resolved magnetizations and the long-time average values of the imbalance operators mentioned in the main text. Our results explain why only the zero modes of the Hamiltonian contribute to the infinite time average values of imbalances $\hat{\mathcal{I}}^z_{\ket{\mathbb{Z}_2}}$ and $\hat{\mathcal{I}}^x_{\ket{\mathbb{Z}_2}}$, while for $\hat{\mathcal{I}}^x_{\ket{{\rm vac}}}$ both zero modes and non-zero modes may contribute in general. We show that this property in turn relies on the relations amongst site-resolved magnetizations $\langle \hat{\sigma}^z_{i,a} \rangle$ and $\langle \hat{\tilde{\sigma}}^x_{i,a} \rangle$ for sites $i=1,2$ and $a=1,2$. We also provide, by using the various properties of the two chirality operators $\hat{\mathcal{C}}_{1},\hat{\mathcal{C}}_{2}$ introduced in Sec.~\ref{sec:supp_chirality_and_spectral_reflection}, an analytic derivation of these exact aspects observed in numerical analysis previously.

\subsection{Long-time average values of imbalance operators $\hat{\mathcal{I}}^z_{\ket{\mathbb{Z}_2}}$ and $\hat{\mathcal{I}}^x_{\ket{\mathbb{Z}_2}}$}{\label{sec:supp_late_time_value_of_imbalance}}

In this section, we consider the implications of the presence of the two chirality operators for long-time average imbalances for an initial $|\mathbb{Z}_2\rangle$ state. First we note that the diagonal and off diagonal matrix elements of $\hat{\sigma}^{z}_{i,a}$ operators in the space of non-zero eigenmodes of the Hamiltonian $\hat{\mathcal{H}}$ are related to each other due to these chirality operators $\hat{\mathcal{C}}_{1,2}$. This can be seen from the relations 
\begin{eqnarray}
        \langle E_\mu | \hat{\sigma}^z_{i,a} | E_\nu \rangle &=& \langle E_\mu |\hat{\mathcal{C}}_{1}^{-1} \hat{\mathcal{C}}_{1} \hat{\sigma}^z_{i,a} \hat{\mathcal{C}}_{1}^{-1} \hat{\mathcal{C}}_{1} | E_\nu \rangle 
         = \langle E_{\mu} | \hat{\mathcal{C}}_{1}^{\dagger} \left( \hat{\mathcal{C}}_{1} \hat{\sigma}^z_{i,a} \hat{\mathcal{C}}_{1}^{-1} \right) \hat{\mathcal{C}}_1| E_{\nu} \rangle 
         = \langle -E_{\mu} | \left( \hat{T}_x\hat{\mathcal{C}} \hat{\sigma}^z_{i,a} \hat{\mathcal{C}}^{-1} \hat{T}_x^{-1} \right)| -E_{\nu} \rangle \nonumber\\
         &=& \langle -E_{\mu} |\left( \hat{T}_x \hat{\sigma}^z_{i,a} \hat{T}_x^{-1} \right) | -E_{\nu} \rangle 
         = \langle -E_{\mu} | \hat{\sigma}^z_{i+1,a} | \!-\!E_{\nu} \rangle.
    \label{eq:Csigmaz}
\end{eqnarray}
Similarly using the operator ${\mathcal C}_2$, one obtains 
\begin{eqnarray}
\langle E_\mu | \hat{\sigma}^z_{i,a} | E_\nu \rangle &=&  \langle -E_{\mu} | \hat{\sigma}^z_{i+1,{\bar a}} | \!-\!E_{\nu} \rangle.  \label{eq:Csigmaz2}
\end{eqnarray} 

It is worthwhile to note that these relations do not hold in a similar manner for the zero modes for which $\hat{\mathcal{C}}_q\ket{\Phi_{\mu_0}}=\ket{\Phi_{\mu_0^\ast}}$. Since the set zero modes $\left\{ \ket{\Phi_{\mu_0}} , \mu_0=1,2,...,\mathcal{N}_0 \right\}$ form a orthonormal (albeit degenerate) subspace of eigenvectors, it is easy to see that the modified set of zero modes $\left\{ \hat{\mathcal{C}}_q\ket{\Phi_{\mu_0}} , \mu_0=1,2,...,\mathcal{N}_0 \right\}$ = $\left\{ \ket{\Phi_{\mu_0^\ast}} , \mu_0=1,2,...,\mathcal{N}_0 \right\}$ are also a set of orthonormal set of degenerate eigenvectors of $\hat{\mathcal{H}}$. This can be understood by considering the inner product between two vectors from this modified set. Since $\langle \Phi_{\mu_0} | \Phi_{\nu_0}\rangle=\delta_{\mu_0,\nu_0}$, we have
\begin{equation}
\begin{split}
     \langle \Phi_{\mu_0^\ast} | \Phi_{\nu_0^\ast} \rangle & = \langle \Phi_{\mu_0} |\hat{\mathcal{C}}_q^\dagger \hat{\mathcal{C}}_q | \Phi_{\nu_0} \rangle 
     = \langle \Phi_{\mu_0} |\hat{\mathcal{C}}_q^{-1} \hat{\mathcal{C}}_q | \Phi_{\nu_0} \rangle 
      = \delta_{\mu_0,\nu_0}.
\end{split}
\end{equation}
An analysis similar to that in Eq.~\eqref{eq:Csigmaz} then leads to 
\begin{eqnarray}
        \langle \Phi_{\mu_0} | \hat{\sigma}^z_{i,a} | \Phi_{\nu_0} \rangle &=& \langle \Phi_{\mu_0} | \hat{\mathcal{C}}_q^{-1} \hat{\mathcal{C}}_q \hat{\sigma}^z_{i,a} \hat{\mathcal{C}}_q^{-1} \hat{\mathcal{C}}_q | \Phi_{\nu_0} \rangle 
         = \langle \Phi_{\mu_0} | \hat{\mathcal{C}}_q^{\dagger} \left(\hat{\mathcal{C}}_q \hat{\sigma}^z_{i,a} \hat{\mathcal{C}}_q^{-1}\right) \hat{\mathcal{C}}_q | \Phi_{\nu_0} \rangle 
         = \langle \Phi_{\mu_0^\ast} | \left( \hat{\mathcal{C}}_q \hat{\sigma}^z_{i,a} \hat{\mathcal{C}}_q^{-1} \right) | \Phi_{\nu_0^\ast} \rangle.
    \label{eq:Cqsigmaz-zero-zero}
\end{eqnarray}
For $q=1,2$ one thus obtains 
\begin{eqnarray} 
\langle \Phi_{\mu_0} | \hat{\sigma}^z_{i,a} | \Phi_{\nu_0} \rangle &=&  \langle \Phi_{\mu_0}^{\ast} | \hat{\sigma}^z_{i+1,a} | \Phi_{\nu_0} ^{\ast}\rangle = \langle \Phi_{\mu_0}^{\ast} | \hat{\sigma}^z_{i,\bar a} | \Phi_{\nu_0}^{\ast} \rangle. \label{zeszrel} 
\end{eqnarray}

\subsubsection{Longitudinal Imbalance}

Next, we consider the dynamical consequences of Eqs.~\eqref{eq:Csigmaz} and \eqref{eq:Csigmaz2}. We note that a quench from an initial state $\ket{\psi(0)}$ leads to the instantaneous expectation value of $\hat{\sigma}^z_{i,a}$ with respect to the quantum state $\ket{\psi(t)}$ at time $t$ reads
\begin{equation}
\begin{split}
    \langle \hat{\sigma}^z_{i,a}(t) \rangle & = \sum_{\mu=1}^{\mathcal{D}_H} \sum_{\nu=1}^{\mathcal{D}_H} \langle \psi(0) | E_{\mu} \rangle (\hat{\sigma}^z_{i,a})_{\mu\nu} \langle E_\nu | \psi(0) \rangle e^{i\omega_{\mu\nu}t}. %\\
    % & = 2\sum_{\mu\notin\mathscr{H}_0}\sum_{\nu\in\mathscr{H}_0} \langle \psi(0) | E_{\mu} \rangle (\hat{\sigma}^z_{i,a})_{\mu\nu} \langle E_\nu | \psi(0) \rangle \cos\left(E_\mu t\right) \\
    % &+ \sum_{\mu\notin\mathscr{H}_0}\sum_{\nu\notin\mathscr{H}_0} \langle \psi(0) | E_{\mu} \rangle (\hat{\sigma}^z_{i,a})_{\mu\nu} \langle E_\nu | \psi(0) \rangle e^{i\omega_{\mu\nu} t} \\
    % &+ \sum_{\mu\in\mathscr{H}_0}\sum_{\nu\in\mathscr{H}_0} \langle \psi(0) | E_{\mu} \rangle (\hat{\sigma}^z_{i,a})_{\mu\nu} \langle E_\nu | \psi(0) \rangle 
\end{split}
\end{equation}
Here we have introduced the notation $\omega_{\mu\nu}=(E_\mu-E_\nu)/\hbar$ and $\left(.\right)_{\mu\nu} = \langle E_\mu | . | E_\nu \rangle$. The \textit{infinite time average} of $\langle \hat{\sigma}^z_{i,a}(t) \rangle$ is 
\begin{equation}
\begin{split}
    \overline{\hat{\sigma}^z_{i,a}} = \lim_{T\rightarrow\infty} \frac{1}{T} \int_0^T dt \; \langle \hat{\sigma}^z_{i,a}(t) \rangle =\sum_{\mu \notin \mathscr{H}_0} | \langle \psi(0) | E_\mu \rangle |^2 \; (\hat{\sigma}^z_{i,a})_{\mu\mu} + \sum_{\mu\in\mathscr{H}_0}\sum_{\nu \in \mathscr{H}_0} \langle \psi(0) | E_\mu \rangle \; \langle E_\nu | \psi(0) \rangle \; (\hat{\sigma}^z_{i,a})_{\mu\nu}.
\end{split}    
\end{equation}
For the specific initial states considered in the main text, one has 
\begin{equation}
    \langle \psi(0) | \hat{\mathcal{H}} | \psi(0) \rangle = \sum_{\mu=1}^{\mathcal{D}_H} E_\mu \big|\langle \psi(0)|E_\mu\rangle|^2 = \sum_{\mu\notin\mathscr{H}_0} E_\mu \big|\langle \psi(0)|E_\mu\rangle|^2 = 0.
\end{equation}
This implies that for these initial states, overlap of the initial state with the non-zero modes, i.e. $\left\{\ket{E_\mu},\mu\notin\mathscr{H}_0\right\}$ satisfies the relation
\begin{equation}
    |\langle \psi(0)|E_\mu\rangle|^2 = |\langle \psi(0)|-E_\mu\rangle|^2, \quad \forall E_\mu > 0.
    \label{eq:nonzero-modes-overlap-with-initial-state}
\end{equation}
From Eq.~\eqref{eq:nonzero-modes-overlap-with-initial-state} and Eq.~\eqref{eq:Csigmaz}, by setting $\mu=\nu$, we can see that
\begin{equation}
    \sum_{\mu \notin \mathscr{H}_0} | \langle \psi(0) | E_\mu \rangle |^2 \; (\hat{\sigma}^z_{i,a})_{\mu\mu} = \sum_{\mu \notin \mathscr{H}_{0}} | \langle \psi(0) | E_\mu \rangle |^2 \; (\hat{\sigma}^z_{i+1,a})_{\mu\mu}.
    \label{eq:sigmaz_C_nonzeromodes}
\end{equation}
This shows that the contribution to infinite time average from non-zero modes for operators $\hat{\sigma}^z_{i,a}$ and $\hat{\sigma}^z_{i+1,a}$ are identical. This immediately implies that the imbalance $\hat{\mathcal{I}}^z_{\ket{\mathbb{Z}_2}}$ does not have any contribution from the 
sector of non-zero energy eigenmodes as it is defined by 
\begin{equation}
    \hat{\mathcal{I}}^z_{\ket{\mathbb{Z}_2}}=\frac{1}{L}\sum_{r=1}^{L/2}\hat{T}_x^{2r}\left(\hat{\sigma}^z_{1,1}
    -\hat{\sigma}^z_{2,1}-\hat{\sigma}^z_{1,2}+\hat{\sigma}^z_{2,2}\right).
\end{equation}
Using Eq.~\eqref{eq:sigmaz_C_nonzeromodes} therefore leads to 
\begin{equation}
    \sum_{\mu\notin\mathscr{H}_0} |\langle \psi(0)|E_\mu\rangle|^2 \left(\hat{\mathcal{I}}^z_{\ket{\mathbb{Z}_2}}\right)_{\mu\mu} = 0.
\end{equation}
Note that a similar analysis for the zero modes lead to a possibility of non-zero contribution to the imbalance; this can happen 
if one or few of these modes have larger overlap with the initial state than the rest. In our numerics, we find that this is indeed the case.

\subsubsection{Transverse imbalance}

Next we carry out a similar analysis for the matrix elements of $\hat{\tilde{\sigma}}^x_{i,a}$. This yields, using ${\mathcal C}_1$, 
\begin{eqnarray}
        \langle E_\mu | \hat{\tilde\sigma}^x_{i,a} | E_\nu \rangle &=& \langle E_\mu |\hat{\mathcal{C}}_{1}^{-1} \hat{\mathcal{C}}_{1} \hat{\tilde{\sigma}}^x_{i,a} \hat{\mathcal{C}}_{1}^{-1} \hat{\mathcal{C}}_{1} | E_\nu \rangle 
        = \langle E_\mu |\hat{\mathcal{C}}_{1}^{\dagger} \left( \hat{T_x} \hat{\mathcal{C}} \hat{\tilde{\sigma}}^x_{i,a} \hat{\mathcal{C}}^{-1} \hat{T_x}^{-1} \right) \hat{\mathcal{C}}_{1} | E_\nu \rangle \nonumber\\
        &=& -\langle -E_\mu | \left( \hat{T_x} \hat{\tilde{\sigma}}^x_{i,a} \hat{T_x}^{-1} \right) |- E_\nu \rangle 
         = -\langle -E_\mu | \hat{\tilde{\sigma}}^x_{i+1,a} |- E_\nu \rangle.
    \label{eq:Csigmax}
\end{eqnarray}
Similarly using ${\mathcal C}_2$, we obtain 
\begin{eqnarray}
\langle E_\mu | \hat{\tilde\sigma}^x_{i,a} | E_\nu \rangle &=& -\langle -E_\mu | \hat{\tilde{\sigma}}^x_{i+1,{\bar a}} |- E_\nu \rangle.
    \label{eq:Csigmax2}
\end{eqnarray}
The matrix elements for the zero mode sectors can also be obtained in an identical manner and yields 
\begin{eqnarray} 
\langle \Phi_{\mu_0} | \hat{\sigma}^x_{i,a} | \Phi_{\nu_0} \rangle &=&  - \langle \Phi_{\mu_0}^{\ast} | \hat{\sigma}^x_{i+1,a} | \Phi_{\nu_0} ^{\ast}\rangle = - \langle \Phi_{\mu_0}^{\ast} | \hat{\sigma}^x_{i,\bar a} | \Phi_{\nu_0}^{\ast} \rangle. \label{zesxrel} 
\end{eqnarray} 
Using Eq.~\eqref{eq:nonzero-modes-overlap-with-initial-state} we find
\begin{equation}
    \sum_{\mu \notin \mathscr{H}_0} | \langle \psi(0) | E_\mu \rangle |^2 \; (\hat{\tilde\sigma}^x_{i,a})_{\mu\mu} = -\sum_{\mu \notin \mathscr{H}_{0}} | \langle \psi(0) | E_\mu \rangle |^2 \; (\hat{\tilde\sigma}^x_{i+1,a})_{\mu\mu}.
    \label{eq:sigmax_C_nonzeromodes}
\end{equation}
Eq.~\eqref{eq:sigmax_C_nonzeromodes} shows that the contribution to infinite time average from non-zero modes for operators $\hat{\tilde\sigma}^x_{i,a}$ and $\hat{\tilde\sigma}^x_{i+1,a}$ are exactly opposite. This immediately implies that the imbalance $\hat{\mathcal{I}}^x_{\ket{\mathbb{Z}_2}}$ does not have any contribution from the sector non-zero energy eigenmodes. To see this more clearly we note that $\hat{\mathcal{I}}^x_{\ket{\mathbb{Z}_2}}$ is defined by 
\begin{equation}
    \hat{\mathcal{I}}^x_{\ket{\mathbb{Z}_2}}=\frac{1}{L}\sum_{r=1}^{L/2}\hat{T}_x^{2r}\left(-\hat{\tilde\sigma}^x_{1,1}
    -\hat{\tilde\sigma}^x_{2,1}+\hat{\tilde\sigma}^x_{1,2}+\hat{\tilde\sigma}^x_{2,2}\right).
\end{equation}
Using Eq.~\eqref{eq:sigmax_C_nonzeromodes}, one finds  
\begin{equation}
    \sum_{\mu\notin\mathscr{H}_0} |\langle \psi(0)|E_\mu\rangle|^2 \left(\hat{\mathcal{I}}^x_{\ket{\mathbb{Z}_2}}\right)_{\mu\mu} = 0.
\end{equation}
Thus our results indicate that the long-time average imbalance found in the main text starting from the $|{\mathbb Z}_2\rangle$ state receives its contribution only from the zero-energy eigenstates. Before ending this section, we note that the imbalance operator  
$\hat{\mathcal{I}}^x_{\ket{\rm vac}}$ is defined as 
\begin{equation}
    \hat{\mathcal{I}}^x_{\ket{\rm vac}}=\frac{1}{L}\sum_{r=1}^{L/2}\hat{T}_x^{2r}\left(\hat{\tilde\sigma}^x_{1,1}+\hat{\tilde\sigma}^x_{1,2}
    -\hat{\tilde\sigma}^x_{2,1}-\hat{\tilde\sigma}^x_{2,2}\right),
\end{equation}
and therefore its contribution does not necessarily vanish from the non-zero eigenmode sector (Eqs.\ \eqref{eq:Csigmax} and \eqref{eq:sigmax_C_nonzeromodes}). In fact numerically, we find that it does receive contribution from both the zero- and non-zero mode sectors. 

\subsection{Instantaneous site-resolved magnetizations $\langle \hat{\sigma}^z_{i,a}\rangle_t$ and $\langle \hat{\tilde\sigma}^x_{i,a}\rangle_t$}

The goal in this section is to investigate the reason behind some exact relations amongst the site resolved magnetization which are observed in the numerical results of quench dynamics. From the evolution of various site resolved magnetizations in the main text, we can find that the following relations hold for the $\ket{\psi(0)}=\ket{\mathbb{Z}_2}$ initial state, for any value of $\Delta$

\begin{subequations}
    \begin{align}
        \langle \hat{\sigma}^z_{1,1} \rangle_t & = \langle \hat{\sigma}^z_{2,2} \rangle_t, \quad 
        \langle \hat{\sigma}^z_{1,2} \rangle_t = \langle \hat{\sigma}^z_{2,1} \rangle_t, \quad
        \langle \hat{\tilde\sigma}^x_{1,1} \rangle_t = -\langle \hat{\tilde\sigma}^x_{2,2} \rangle_t, \quad 
        \langle \hat{\tilde\sigma}^x_{1,2} \rangle_t = -\langle \hat{\tilde\sigma}^x_{2,1} \rangle_t,
    \end{align}
    \label{eq:site_resolved_relations_Z2}
\end{subequations}
whereas for $\ket{\psi(0)}=\ket{{\rm vac}}$ initial state we find 
\begin{subequations}
    \begin{align}
        \langle \hat{\sigma}^z_{1,1} \rangle_t & = \langle \hat{\sigma}^z_{2,2} \rangle_t = \langle \hat{\sigma}^z_{1,2} \rangle_t = \langle \hat{\sigma}^z_{2,1} \rangle_t, \quad \quad 
        \langle \hat{\tilde\sigma}^x_{1,1} \rangle_t = \langle \hat{\tilde\sigma}^x_{1,2} \rangle_t = -\langle \hat{\tilde\sigma}^x_{2,2} \rangle_t = -\langle \hat{\tilde\sigma}^x_{2,1} \rangle_t.
    \end{align}
    \label{eq:site_resolved_relations_vac}
\end{subequations}
In the following we shall explore the underlying reason for these equalities during quench dynamics. 

\subsubsection{Longitudinal Magnetization} 

We first carry out our analysis for the longitudinal magnetization. For this purpose, we first rewrite 
time dependence of magnetization $\hat{\sigma}^z_{i,a}$ as 
\begin{equation}
\begin{split}
    \langle \hat{\sigma}^z_{i,a} \rangle(t) = & \overbrace{\sum_{\mu_0,\nu_0} \langle \psi(0) | \Phi_{\mu_0} \rangle  \langle \Phi_{\nu_0} | \psi(0) \rangle \left( \hat{\sigma}^z_{i,a} \right)_{\mu_0\nu_0}}^{\text{\RomanNumeralCaps{1}}} +  \overbrace{\sum_{\mu_0,\nu} \langle \psi(0) | \Phi_{\mu_0} \rangle  \langle E_\nu | \psi(0) \rangle \left( \hat{\sigma}^z_{i,a} \right)_{\mu_0\nu} e^{-iE_\nu t}}^{\text{\RomanNumeralCaps{2}}} \\
    & + \underbrace{\sum_{\mu,\nu} \langle \psi(0) | E_\mu \rangle  \langle E_\nu | \psi(0) \rangle \left( \hat{\sigma}^z_{i,a} \right)_{\mu\nu} e^{+i\left(E_\mu-E_\nu\right) t}}_{\text{\RomanNumeralCaps{3}}}.
\end{split} 
\label{eq:dynamics_partition}
\end{equation}
Here we have used the convention that $\ket{E_\mu}$ stands for the non-zero modes of the Hamiltonian with energy eigenvalue $E_\mu$ and $\ket{\Phi_{\mu_0}}$ denotes the zero modes of the Hamiltonian.

Eq.\ \eqref{eq:dynamics_partition} consists of three distinct parts. The first part (denoted by \RomanNumeralCaps{1}) comes entirely from zero modes whereas the third part (\RomanNumeralCaps{3}) comes entirely from non-zero modes. In contrast, part \RomanNumeralCaps{2} receives contribution from both zero and non-zero modes. To analyze these contributions, we first focus on part \RomanNumeralCaps{3} of the above partition. We recall from Sec.~\ref{sec:supp_late_time_value_of_imbalance} (for example Eqs.\ \eqref{eq:Csigmaz} and \eqref{eq:Csigmaz2}) that for two non-zero modes $\ket{E_\mu}$ and $\ket{E_\nu}$ we have $\langle E_\mu | \hat{\sigma}^z_{i,a} | E_\nu \rangle = \langle -E_\mu | \hat{\sigma}^z_{i+1,a} |\!-\!E_\nu \rangle= \langle -E_\mu | \hat{\sigma}^z_{i+1,\bar a} |\!-\!E_\nu \rangle$. Moreover, the spectral reflection symmetry induced by the chirality operators $\hat{\mathcal{C}}_q$ ($q=1,2$) also allows us to relate $\langle \psi(0) | E_\mu\rangle$ and $\langle \psi(0) | \!-\!E_\mu\rangle$. To see this we note that
\begin{equation}
    \begin{split}
        \langle \psi(0) | \!-\!E_\mu \rangle & = \langle \psi(0) | \hat{\mathcal{C}}_q |E_\mu \rangle, \quad \bra{\psi(0)} \hat{\mathcal{C}}_q = \left( \hat{\mathcal{C}}_q^\dagger \ket{\psi(0)} \right)^\dagger,
    \end{split}
    \label{eq:overlap_spectral_reflection}
\end{equation}
and consider for $\ket{\psi(0)}=\ket{{\rm vac}}$ leading to 
\begin{equation}
    \begin{split}
        \bra{{\rm vac}} \hat{\mathcal{C}}_1 & = \left( \hat{\mathcal{C}}_1^\dagger \ket{{\rm vac}} \right)^\dagger 
        =  \left( \hat{\mathcal{C}}^\dagger \hat{T}_x^\dagger \ket{{\rm vac}} \right)^\dagger 
        = \left( \hat{\mathcal{C}} \hat{T}_x^{-1} \ket{{\rm vac}} \right)^\dagger 
        = \left((-1)^{N_{\text{sites}}} \ket{{\rm vac}} \right)^\dagger
        = \bra{{\rm vac}}, \nonumber\\
        \bra{{\rm vac}} \hat{\mathcal{C}}_2 & = \left( \hat{\mathcal{C}}_2^\dagger \ket{{\rm vac}} \right)^\dagger 
         =  \left( \hat{\mathcal{C}}^\dagger \hat{T}_x^\dagger \hat{T}_y^\dagger \ket{{\rm vac}} \right)^\dagger
         = \left( \hat{\mathcal{C}} \hat{T}_x^{-1} \hat{T}_y^{-1} \ket{{\rm vac}} \right)^\dagger 
         = \left((-1)^{N_{\text{sites}}} \ket{{\rm vac}} \right)^\dagger 
         = \bra{{\rm vac}}.
    \end{split}
    \label{eq:overlap_spectral_reflection_vac_q1}
\end{equation}
For $\ket{\psi(0)}=\ket{\mathbb{Z}_2}$ we have
\begin{equation}
    \begin{split}
        \bra{\mathbb{Z}_2} \hat{\mathcal{C}}_1 & = \left( \hat{\mathcal{C}}_1^\dagger \ket{\mathbb{Z}_2} \right)^\dagger 
        =  \left( \hat{\mathcal{C}}^\dagger \hat{T}_x^\dagger \ket{\mathbb{Z}_2} \right)^\dagger 
        = \left( \hat{\mathcal{C}} \hat{T}_x^{-1} \ket{\mathbb{Z}_2} \right)^\dagger 
        = \left( (-1)^{N_{\text{sites}}/2} \ket{\overline{\mathbb{Z}}_2} \right)^\dagger 
        = \ket{\overline{\mathbb{Z}}_2},  \nonumber\\
        \bra{\mathbb{Z}_2} \hat{\mathcal{C}}_2 & = \left( \hat{\mathcal{C}}_2^\dagger \ket{\mathbb{Z}_2} \right)^\dagger 
         =  \left( \hat{\mathcal{C}}^\dagger \hat{T}_x^\dagger \hat{T}_y^{\dagger} \ket{\mathbb{Z}_2} \right)^\dagger 
         = \left( \hat{\mathcal{C}} \hat{T}_x^{-1} \hat{T}_y^{-1} \ket{\mathbb{Z}_2} \right)^\dagger 
         = \left( (-1)^{N_{\text{sites}}/2} \ket{\mathbb{Z}_2} \right)^\dagger 
         = \bra{\mathbb{Z}_2}.  
    \end{split}
    \label{eq:overlap_spectral_reflection_Z2_q2}
\end{equation}
 
Using Eqs.\ \eqref{eq:Csigmaz}, \eqref{eq:Csigmaz2}, \eqref{zeszrel} along with Eq.\ \eqref{eq:overlap_spectral_reflection_Z2_q2} to obtain the properties of instantaneous magnetization starting from the $|{\mathbb Z}_2\rangle$ state.
To this end, we first consider $\langle \hat{\sigma}^z_{i,a} \rangle_{\mathbb{Z}_2}(t)$. The contribution to it from the $E=0$ sector is given by 
\begin{eqnarray}
        \langle \hat{\sigma}^z_{i,a} \rangle^{\text{I}}_{\mathbb{Z}_2}(t) &=& \sum_{\mu_0\nu_0} \langle \Phi_{\mu_0} | \hat{\sigma}^z_{i,a} | \Phi_{\nu_0} \rangle \langle \Phi_{\nu_0} | \mathbb{Z}_2 \rangle \langle \mathbb{Z}_2 |  \Phi_{\mu_0} \rangle =
        \sum_{\mu_0 \nu_0} \langle \Phi_{\mu_0^\ast} | \hat{\sigma}^z_{i+1,\overline{a}} | \Phi_{\nu_0^\ast} \rangle \langle \Phi_{\nu_0^\ast} | \mathbb{Z}_2 \rangle \langle \mathbb{Z}_2 |  \Phi_{\mu_0^\ast} \rangle 
        = \langle \hat{\sigma}^z_{i+1,\overline{a}} \rangle^{\text{I}}_{\mathbb{Z}_2}(t).
\end{eqnarray}
In addition, utilizing the $\hat{\mathcal{C}}_2$ operator ($\ket{\Phi_{\mu_0^\ast}}=\hat{\mathcal{C}}_2\ket{\Phi_{\mu_0}}$), the cross terms which receives contribution from both the 
zero and the non-zero energy sectors can be shown to satisfy 
\begin{eqnarray}
        \langle \hat{\sigma}^z_{i,a} \rangle^{\text{II}}_{\mathbb{Z}_2}(t) &=& \sum_{\mu_0\nu} \langle \Phi_{\mu_0} | \hat{\sigma}^z_{i,a} | E_\nu \rangle \langle E_\nu | \mathbb{Z}_2 \rangle \langle \mathbb{Z}_2 |  
        \Phi_{\mu_0} \rangle e^{-iE_\nu t}
        = \sum_{\mu_0 \nu} \langle \Phi_{\mu_0^\ast} | \hat{\sigma}^z_{i+1,\overline{a}} | -E_\nu \rangle \langle -E_\nu| \mathbb{Z}_2 \rangle \langle \mathbb{Z}_2 |  \Phi_{\mu_0^\ast} \rangle  e^{-iE_\nu t}
        = \langle \hat{\sigma}^z_{i+1,\overline{a}} \rangle^{\text{II}}_{\mathbb{Z}_2}(t). \nonumber\\ 
\end{eqnarray}
Moreover, the contribution from the non-zero energy sector yields 
\begin{eqnarray}
        \langle \hat{\sigma}^z_{i,a} \rangle^{\text{III}}_{\mathbb{Z}_2}(t) &=& \sum_{\mu\nu} \langle E_\mu | \hat{\sigma}^z_{i,a} | E_\nu \rangle \langle E_\nu | \mathbb{Z}_2 \rangle \langle \mathbb{Z}_2 |  \Phi_{\mu_0} \rangle e^{i\left(E_\mu-E_\nu\right) t}\nonumber\\
        &=&  \sum_{\mu \nu} \langle -E_\mu | \hat{\sigma}^z_{i+1,\overline{a}} | -E_\nu \rangle \langle -E_\nu| \mathbb{Z}_2 \rangle \langle \mathbb{Z}_2 |  -E_\mu \rangle  e^{i\left(E_\mu-E_\nu\right) t}
        = \langle \hat{\sigma}^z_{i+1,\overline{a}} \rangle^{\text{III}}_{\mathbb{Z}_2}(t).
\end{eqnarray}
From the above equations we can therefore conclude that 
\begin{eqnarray}
    \langle \hat{\sigma}^z_{i,a} \rangle_{{\rm vac}}(t) = \langle \hat{\sigma}^z_{i+1,\overline{a}}\rangle_{\mathbb{Z}_2}(t).
\end{eqnarray}

\subsubsection{Transverse Magnetization}

Next, we consider the transverse magnetization. We shall first analyze the case for initial $|Z_2\rangle$ state, {\it i.e.}, we shall study the properties of $\langle \hat{\tilde\sigma}^x_{i,a} \rangle_{\mathbb{Z}_2}(t)$. Using Eqs.\ \eqref{eq:Csigmax}, \eqref{eq:Csigmax2}, \eqref{zesxrel}, we find, for the contribution to these matrix elements from the zero-energy sector
\begin{eqnarray}
        \langle \hat{\tilde\sigma}^x_{i,a} \rangle^{\text{I}}_{\mathbb{Z}_2}(t) &=& \sum_{\mu_0\nu_0} \langle \Phi_{\mu_0} | \hat{\tilde\sigma}^x_{i,a} | \Phi_{\nu_0} \rangle \langle \Phi_{\nu_0} | \mathbb{Z}_2 \rangle \langle \mathbb{Z}_2 |  \Phi_{\mu_0} \rangle
        = -\sum_{\mu_0 \nu_0} \langle \Phi_{\mu_0^\ast} | \hat{\tilde\sigma}^x_{i+1,\overline{a}} | \Phi_{\nu_0^\ast} 
        \rangle \langle \Phi_{\nu_0^\ast} | \mathbb{Z}_2 \rangle \langle \mathbb{Z}_2 |  \Phi_{\mu_0^\ast} \rangle 
        = -\langle \hat{\tilde\sigma}^x_{i+1,\overline{a}} \rangle^{\text{I}}_{\mathbb{Z}_2}(t), \nonumber\\
\end{eqnarray}
while the cross terms yield 
\begin{eqnarray}
        \langle \hat{\tilde\sigma}^x_{i,a} \rangle^{\text{II}}_{\mathbb{Z}_2}(t) & = & \sum_{\mu_0\nu} \langle \Phi_{\mu_0} | \hat{\tilde\sigma}^x_{i,a} | E_\nu \rangle \langle E_\nu | \mathbb{Z}_2 \rangle \langle \mathbb{Z}_2 |  \Phi_{\mu_0} \rangle e^{-iE_\nu t}\nonumber\\
        & = & -\sum_{\mu_0 \nu} \langle \Phi_{\mu_0^\ast} | \hat{\tilde\sigma}^x_{i+1,\overline{a}} | -E_\nu \rangle \langle -E_\nu| \mathbb{Z}_2 \rangle \langle \mathbb{Z}_2 |  \Phi_{\mu_0^\ast} \rangle  e^{-iE_\nu t}
        = -\langle \hat{\tilde\sigma}^x_{i+1,\overline{a}} \rangle^{\text{II}}_{\mathbb{Z}_2}(t).
\end{eqnarray}
Moreover, the contribution from the non-zero sector is given by 
\begin{eqnarray}
        \langle \hat{\tilde\sigma}^x_{i,a} \rangle^{\text{III}}_{\mathbb{Z}_2}(t) &=& \sum_{\mu\nu} \langle E_\mu | \hat{\tilde\sigma}^x_{i,a} | E_\nu \rangle \langle E_\nu | \mathbb{Z}_2 \rangle \langle \mathbb{Z}_2 |  \Phi_{\mu_0} \rangle e^{i\left(E_\mu-E_\nu\right) t}\nonumber\\        
        &=& -\sum_{\mu \nu} \langle -E_\mu | \hat{\tilde\sigma}^x_{i+1,\overline{a}} | -E_\nu \rangle \langle -E_\nu| \mathbb{Z}_2 \rangle \langle \mathbb{Z}_2 |  -E_\mu \rangle  e^{i\left(E_\mu-E_\nu\right) t}
         = -\langle \hat{\tilde\sigma}^x_{i+1,\overline{a}} \rangle^{\text{III}}_{\mathbb{Z}_2}(t). 
\end{eqnarray}
Thus we finally obtain
\begin{eqnarray}
    \langle \hat{\tilde\sigma}^x_{i,a} \rangle_{{\rm vac}}(t) = -\langle \hat{\tilde\sigma}^x_{i+1,\overline{a}}\rangle_{\mathbb{Z}_2}(t).
\end{eqnarray}
This concludes the derivation of the relations suggested by the numerical quench experiments for the ${\mathbb Z}_2\rangle$ initial state.

Next, we consider the $|{\rm vac}\rangle$ initial state. The computations are exactly similar to those for the $|{\mathbb Z}_2\rangle$ initial state. 
For the $\langle \sigma_i^z\rangle(t)$, we find, using $\hat{\mathcal{C}}_1$ operator, for the zero-energy sector 
\begin{eqnarray}
        \langle \hat{\sigma}^z_{i,a} \rangle^{\text{I}}_{{\rm vac}}(t) &=& \sum_{\mu_0\nu_0} \langle \Phi_{\mu_0} | \hat{\sigma}^z_{i,a} | \Phi_{\nu_0} \rangle \langle \Phi_{\nu_0} | {\rm vac} \rangle \langle {\rm vac} |  \Phi_{\mu_0} \rangle 
        = \sum_{\mu_0 \nu_0} \langle \Phi_{\mu_0^\ast} | \hat{\sigma}^z_{i+1,a} | \Phi_{\nu_0^\ast} \rangle \langle \Phi_{\nu_0^\ast} | {\rm vac} \rangle \langle {\rm vac} |  \Phi_{\mu_0^\ast} \rangle 
        = \langle \hat{\sigma}^z_{i+1,a} \rangle^{\text{I}}_{{\rm vac}}(t). \nonumber\\
    \label{eq:relation_C1vacZ_part1}
\end{eqnarray}
The last equality in the above equation follows from the fact that $\left\{\ket{\Phi_{\mu_0}}\right\}$ and $\left\{\ket{\Phi_{\mu_0^\ast}}\right\}$ both form the zero energy degenerate subspace of $\hat{\mathcal{H}}$. Similarly, we can utilize $\hat{\mathcal{C}}_2$ operator ($\ket{\Phi_{\mu_0^\ast}}=\hat{\mathcal{C}}_2\ket{\Phi_{\mu_0}}$) to find
\begin{eqnarray}
        \langle \hat{\sigma}^z_{i,a} \rangle^{\text{I}}_{{\rm vac}}(t) & = & \sum_{\mu_0\nu_0} \langle \Phi_{\mu_0} | \hat{\sigma}^z_{i,a} | \Phi_{\nu_0} \rangle \langle \Phi_{\nu_0} | = \langle \hat{\sigma}^z_{i+1,\overline{a}} \rangle^{\text{I}}_{{\rm vac}}(t).
    \label{eq:relation_C2vacZ_part1}
\end{eqnarray}
A similar relation can be worked out for the cross terms which connects the zero- and the non-energy sectors. Utilizing $\hat{\mathcal{C}}_1$ and ${\mathcal C}_2$ operators 
a similar straightforward calculations shows 
\begin{eqnarray}
\langle \hat{\sigma}^z_{i,a} \rangle^{\text{II}}_{{\rm vac}}(t) & =& \langle \hat{\sigma}^z_{i+1,a} \rangle^{\text{II}}_{{\rm vac}}(t) =\langle \hat{\sigma}^z_{i+1,\bar a} \rangle^{\text{II}}_{{\rm vac}}(t),
    \label{eq:relation_C1vacZ_part2}
\end{eqnarray}
while for terms within the non-zero energy sectors, we find 
\begin{eqnarray}
        \langle \hat{\sigma}^z_{i,a} \rangle^{\text{III}}_{{\rm vac}}(t) &=& \langle \hat{\sigma}^z_{i+1,a} \rangle^{\text{III}}_{{\rm vac}}(t)= \langle \hat{\sigma}^z_{i+1,\bar a} \rangle^{\text{III}}_{{\rm vac}}(t).
    \label{eq:relation_C1vacZ_part3}
\end{eqnarray}
From these above relations, we can conclude that 
\begin{eqnarray}
    \langle \hat{\sigma}^z_{i,a} \rangle_{{\rm vac}}(t) =  \langle \hat{\sigma}^z_{i+1,a} \rangle_{{\rm vac}}(t) =  \langle \hat{\sigma}^z_{i+1,\overline{a}}\rangle_{{\rm vac}}(t).
\end{eqnarray}
This conforms to the numerical results and also implies that there is no imbalance for $|\sigma_{i a}^z\rangle (t)$ for the $|{\rm vac}\rangle$ initial state. 

Finally, we consider $\langle \hat{\tilde\sigma}^x_{i,a} \rangle_{{\rm vac}}(t)$. The computations are exactly identical to those elaborated in the previous 
cases and we find 
\begin{eqnarray}
    \langle \hat{\tilde\sigma}^x_{i,a} \rangle_{{\rm vac}}(t) = -\langle \hat{\tilde\sigma}^x_{i+1,a} \rangle_{{\rm vac}}(t) = -\langle \hat{\tilde\sigma}^x_{i+1,\overline{a}}\rangle_{{\rm vac}}(t),
\end{eqnarray}
which conforms to the relation between the matrix elements found in the main text. 

\section{Fidelity and Shannon entropy dynamics from $|\mathbb{Z}_2\rangle$ and
  $|\mathbb{Z}_3\rangle$ initial states in PXP chains}

\begin{figure*}[!htpb]
        \centering
        \rotatebox{0}{\includegraphics*[width= 0.49\textwidth]{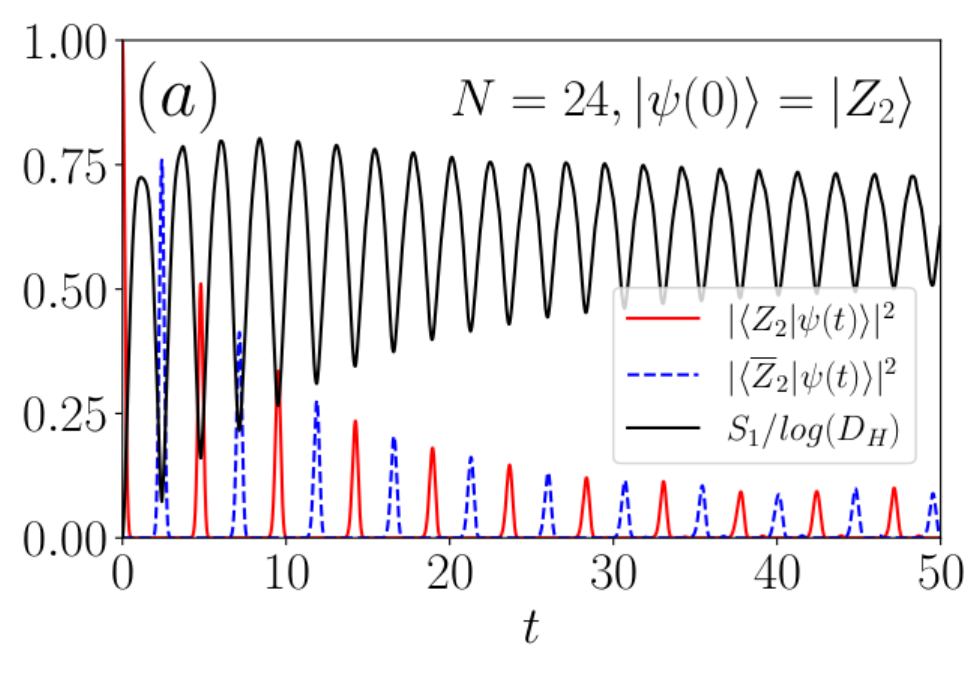}}
        \rotatebox{0}{\includegraphics*[width= 0.45\textwidth]{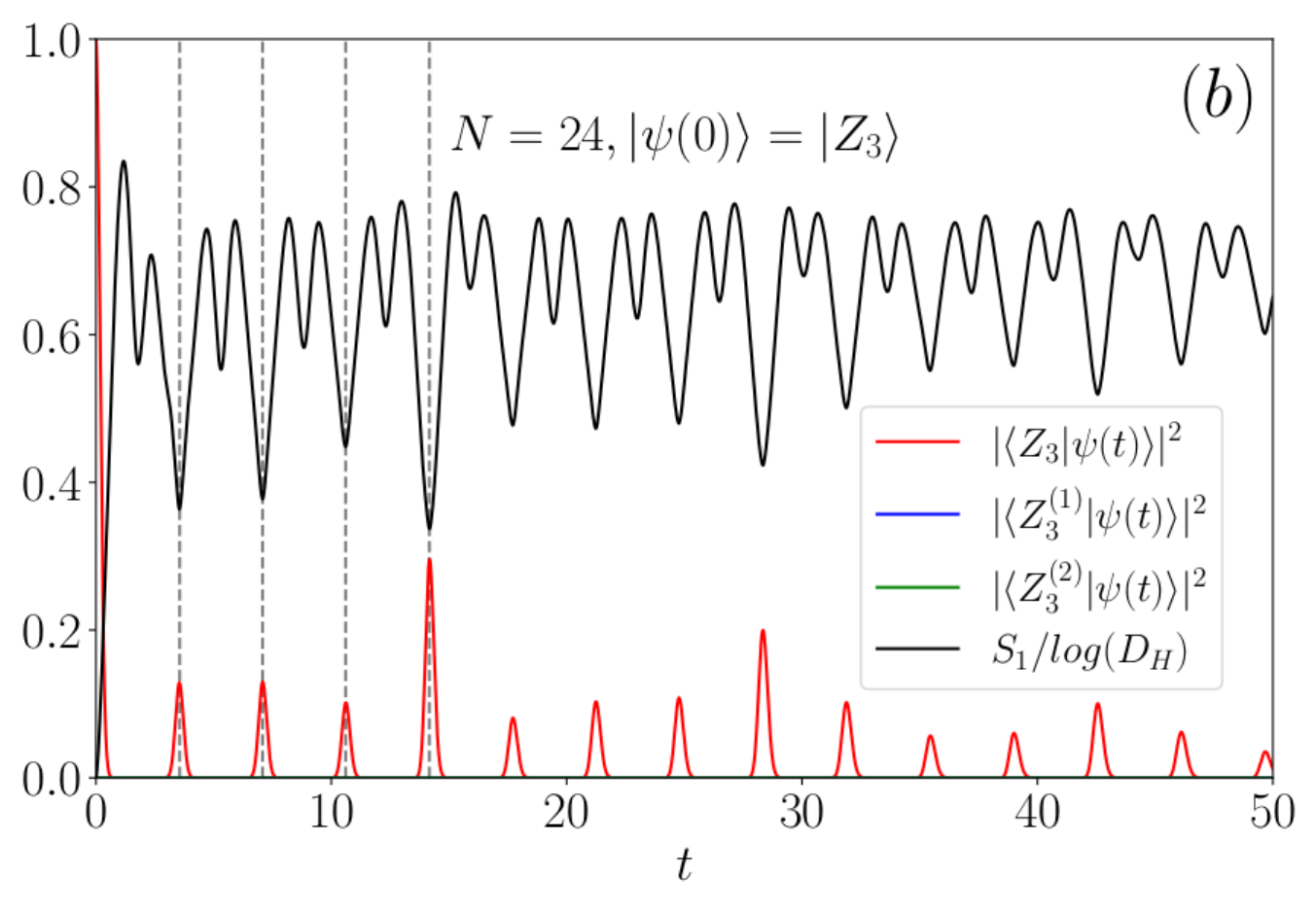}}
        \caption{(a) Plot of $|\langle \mathbb{Z}_2 | \psi(t)\rangle|^2$ (red), $|\langle \bar{\mathbb{Z}}_2 | \psi(t)\rangle|^2$ (dotted blue) and $S_1(t)$ (black) as a function of $t$ for $\ket{\psi(0)}=\ket{\mathbb{Z}_2}$ in the PXP chain with $N=24$ sites. (b) Plot of $|\langle \mathbb{Z}_3 | \psi(t)\rangle|^2$ (red), $|\langle \mathbb{Z}_3^{(1)} | \psi(t)\rangle|^2$ (blue),  $|\langle \mathbb{Z}_3^{(2)} | \psi(t)\rangle|^2$ (green) and $S_1(t)$ (black) as a function of $t$ for $\ket{\psi(0)}=\ket{\mathbb{Z}_3}$ in the PXP chain with $N=24$ sites. }
        \label{fig:PXPchains}
 \end{figure*}

In PXP chains (Eq. 1 of main text with $\Delta=0$ on a 1D chain with PBC), both period-$2$ $|\mathbb{Z}_2\rangle$ as well as period-$3$ $|\mathbb{Z}_3\rangle$ initial Fock states lead to persistent many-body revivals in $\mathcal{F}(t)$ though the latter state gives weaker revivals (Fig.~\ref{fig:PXPchains}). The Shannon entropy $S_1(t)$ shows local minima at the same times as the maxima in $\mathcal{F}(t)$ in both cases. Importantly, $S_1(t)$ shows additional pronounced local minima at times which are equidistant between two consecutive fidelity revivals in both the cases. While these additonal local minima in $S_1(t)$ can be correlated to the appearance of local maxima of $|\langle \overline{\mathbb{Z}}_2|\psi(t)\rangle|^2$ where $|\overline{\mathbb{Z}}_2 \rangle$ refers to the translated partner of $|\mathbb{Z}_2\rangle$ (Fig.~\ref{fig:PXPchains} (a)), the case with $|\mathbb{Z}_3\rangle$ is more complicated  (Fig.~\ref{fig:PXPchains} (b)). These additional local minima in $S_1(t)$ cannot be explained as revivals in either $|\langle \mathbb{Z}_3^{(1)}|\psi(t)\rangle|^2$ or $|\langle \mathbb{Z}_3^{(2)}|\psi(t)\rangle|^2$, where $| \mathbb{Z}_3^{(1)} \rangle$ and $| \mathbb{Z}_3^{(2)} \rangle$ refer to the two translated partners of $|\mathbb{Z}_3\rangle$ (Fig.~\ref{fig:PXPchains} (b)). Nonetheless, the additional minima in $S_1(t)$ still suggests that the other end state has a natural representation in terms of the superposition of a small number of Fock states, when starting from an initial $|\mathbb{Z}_3\rangle$ Fock state. 

\section{PXP and non-PXP regimes of many-body revivals in ladders}
\begin{figure*}[!htpb]
        \centering
        \rotatebox{0}{\includegraphics*[width= 0.49\textwidth]{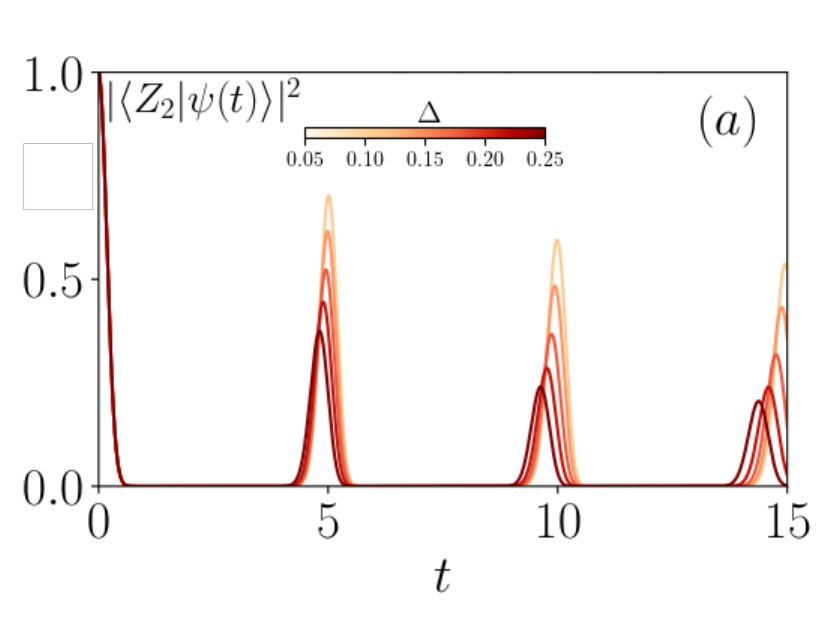}}
        \rotatebox{0}{\includegraphics*[width= 0.49\textwidth]{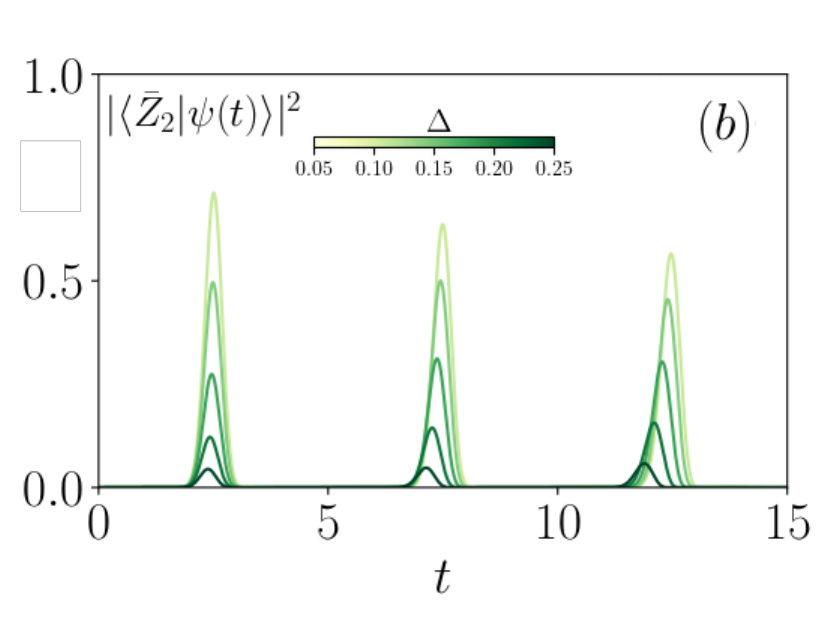}}
        \rotatebox{0}{\includegraphics*[width= 0.49\textwidth]{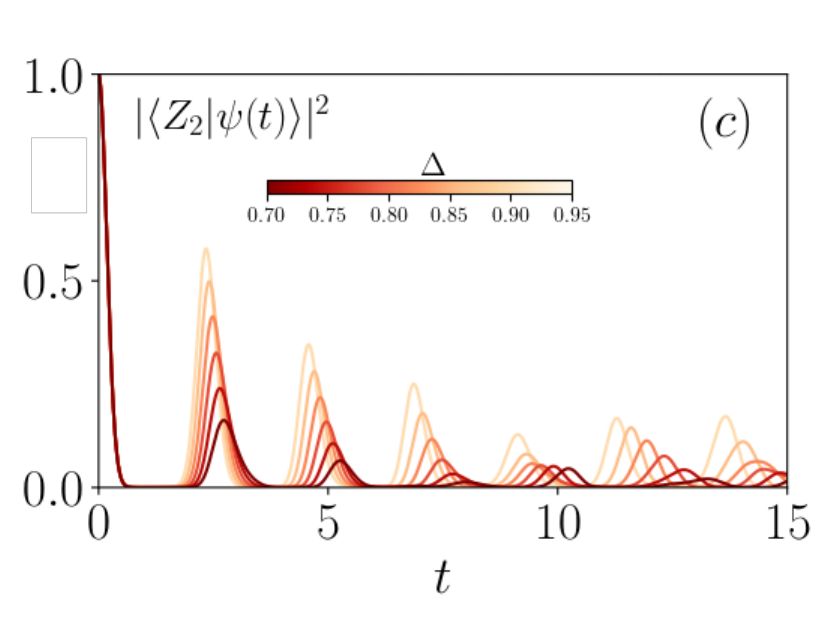}}
        \rotatebox{0}{\includegraphics*[width= 0.49\textwidth]{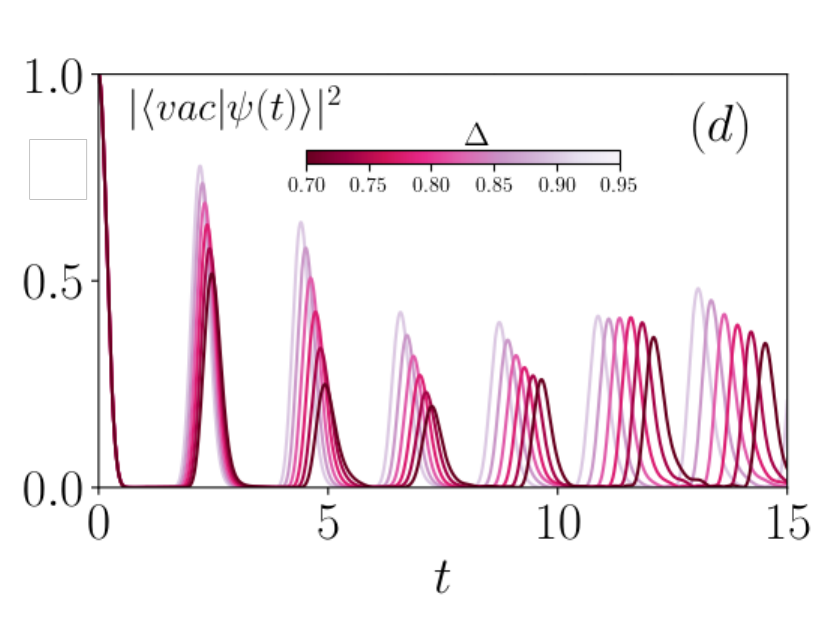}}
        \caption{Plot of (a) $|\langle \mathbb{Z}_2 | \psi(t)\rangle|^2$ and (b) $|\langle \bar{\mathbb{Z}}_2 | \psi(t)\rangle|^2$ as a function of $t$ for $\ket{\psi(0)}=\ket{\mathbb{Z}_2}$ in the PXP regime. (c) Plot of $|\langle \mathbb{Z}_2 | \psi(t)\rangle|^2$ as a function of $t$ 
        for initial state $\ket{\psi(0)}=\ket{\mathbb{Z}_2}$ in the non-PXP regime. (d) Plot of $|\langle {\rm vac} | \psi(t)\rangle|^2$ as a function of $t$ for initial state $\ket{\psi(0)}=\ket{{\rm vac}}$ in the non-PXP regime. In all panels, various values of $\Delta$ are displayed using a color bar in the inset of each corresponding panel. For all plots $N=32$ and $w=\hbar=1$. }
        \label{fig:PXP_nonPXP}
 \end{figure*}

In Fig.~\ref{fig:PXP_nonPXP} (a) and Fig.~\ref{fig:PXP_nonPXP} (b), we show the time evolution of the following fidelities, $|\langle \mathbb{Z}_2 | \psi(t)\rangle|^2$ and $|\langle \bar{\mathbb{Z}}_2 | \psi(t)\rangle|^2$, respectively, where $\ket{\psi(0)}=\ket{\mathbb{Z}_2}$ and $N=32$ for various values of $\Delta \leq 0.25$. We dub this as the ``PXP regime'' where only the $\ket{\mathbb{Z}_2}$ and its particle-hole symmetric partner $\ket{\bar{\mathbb{Z}}_2}$ show periodic revivals. While the Shannon entropy $S_1$ shows local minima at the same times as the local maxima in  $|\langle \mathbb{Z}_2 | \psi(t)\rangle|^2$ (see Fig. 4(a) in main text), $S_1$ also displays additional local minima (see Fig. 4(a) in main text) which can be correlated with $|\langle \bar{\mathbb{Z}}_2 | \psi(t)\rangle|^2$ attaining local maxima.

On the other hand, these additional local minima in $S_1$ are entirely absent (see Fig. 4(b) in main text) in the non-PXP like scarring regime from $\ket{\psi(0)}=\ket{\mathbb{Z}_2}$ where the dips in $S_1$ correspond only to the local maxima in the fidelity $|\langle \mathbb{Z}_2 | \psi(t)\rangle|^2$ where $\ket{\psi(0)}=\ket{\mathbb{Z}_2}$ and $N=32$ for various values of $\Delta \geq 0.70$ (Fig.~\ref{fig:PXP_nonPXP} (c)). Similar non-PXP like scarring is also observed starting from the $\ket{{\rm vac}}$ state (see Fig. 4(c) in main text) where the dips in $S_1$ correspond only to the local maxima in the fidelity $|\langle {\rm vac} | \psi(t)\rangle|^2$ for $N=32$ for various values of $\Delta \geq 0.70$ (Fig.~\ref{fig:PXP_nonPXP} (d)).

 \begin{figure*}[!htpb]
   \centering
   %\rotatebox{0}{\includegraphics*[width= 0.49\textwidth]{UDcut.pdf}}
   %\rotatebox{0}{\includegraphics*[width= 0.49\textwidth]{LRcut.pdf}}
   \rotatebox{0}{\includegraphics*[width= 0.98\textwidth]{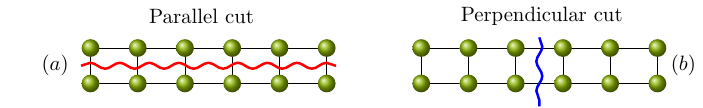}}
        \rotatebox{0}{\includegraphics*[width= 0.49\textwidth]{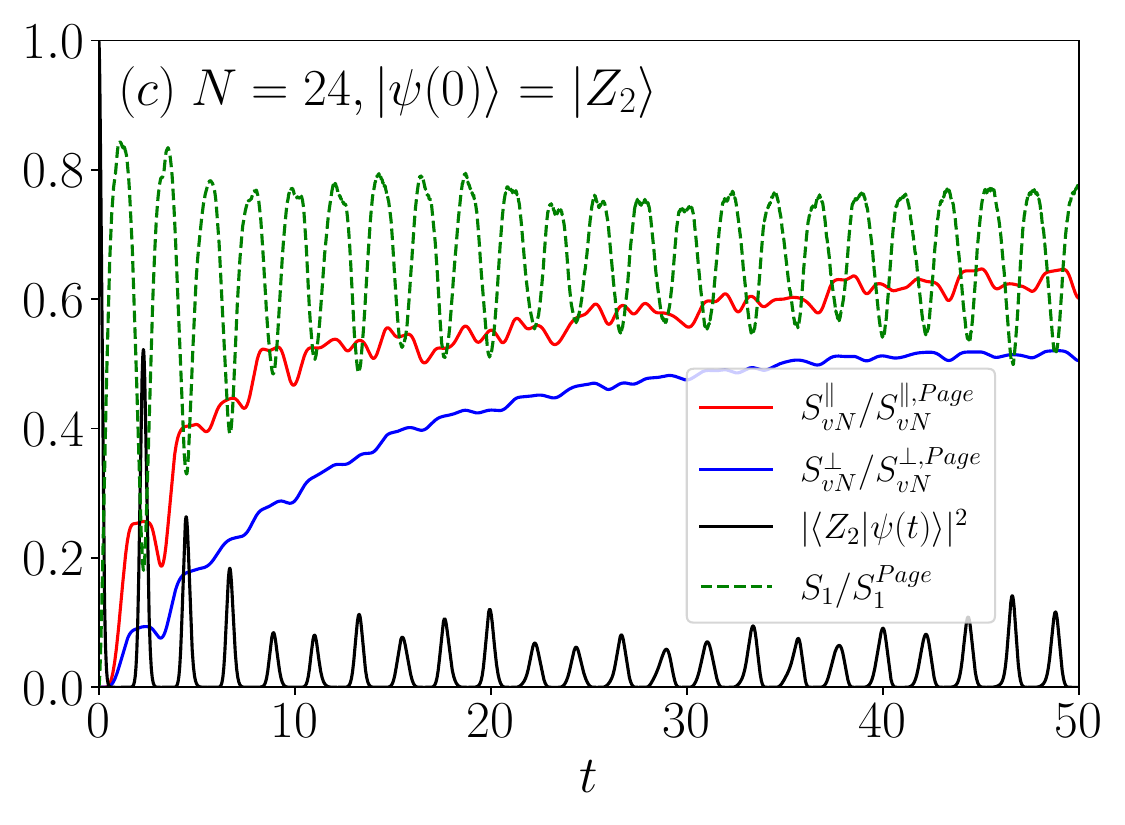}}
        \rotatebox{0}{\includegraphics*[width= 0.49\textwidth]{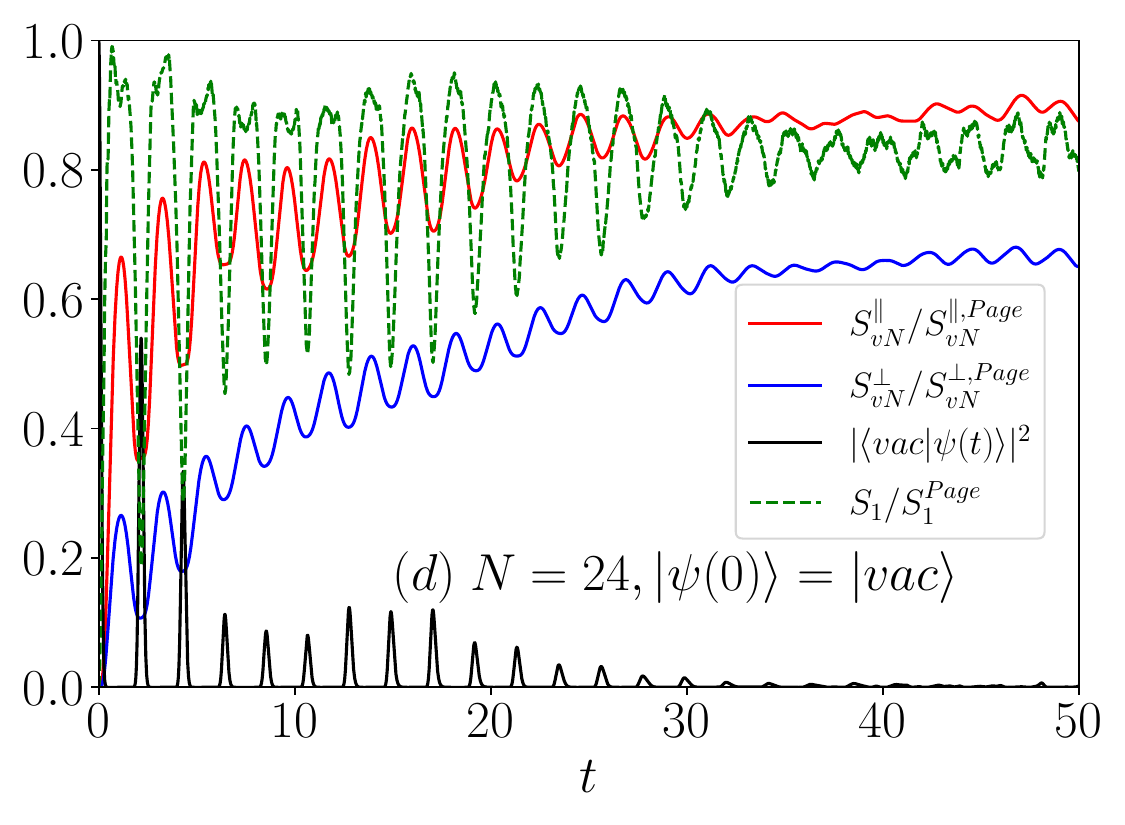}}
        \caption{The bi-partitions used to calculate $S_{{\rm vN}}^{\parallel}$ shown in (a) and $S_{{\rm vN}}^{\perp}$ shown in (b). The time evolution of the Shannon entropy $S_1/S_1^{\mathrm{Page}}$ (dotted green), fidelity $\mathcal{F}(t)$ (black), as well as the bipartite entanglement entropies $S_{{\rm vN}}^{\parallel}/S_{{\rm vN}}^{\parallel,\text{Page}}$ (red), $S_{{\rm vN}}^{\perp}/S_{{\rm vN}}^{\perp,\text{Page}}$ (blue) shown for $|\mathbb{Z}_2\rangle$ (c) and $|\rm{vac}\rangle$ (d) initial states for a ladder with $N=24$ sites and $\Delta=1$. $S_{{\rm vN}}^{\parallel,\text{Page}} \approx 5.205$, $S_{{\rm vN}}^{\perp,\text{Page}} \approx 4.935$ and $S_1^{\text{Page}} \approx 10.141$ refer to the average value of these quantities obtained by sampling $50$ random states with real coefficients from the $k_x=0$ sector of the Hilbert space. }
        \label{fig:PXPladdersEnt}
 \end{figure*}

 It is clarifying to monitor the dynamics of the bipartite entanglement entropy as a function of time in the non-PXP regime. We can define two distinct partitions which we refer to as $\parallel$ [Fig.~\ref{fig:PXPladdersEnt} (a)] ($\perp$) [Fig.~\ref{fig:PXPladdersEnt} (b)] where the partition is chosen to be parallel (perpendicular) to the legs of the ladder which divides the system into two equal halves. Next, we calculate the reduced density matrix of any of the subsystems by integrating out the degrees of freedom in the other subsystem, which gives
 \begin{eqnarray}
   \rho_{\rm A (B)} (t) = \rm{Tr}_{\rm B (A)} \rho_{\rm AB} (t)
   \end{eqnarray}
 where $\rho_{\rm AB} (t)= |\psi(t)\rangle \langle \psi(t)|$ refers to the full density matrix, and $\rm{A}$, $\rm{B}$ refer to the two subsystems. From this, the entanglement entropy can be computed by using $S_{{\rm vN}}^{\parallel/\perp}(t) = -{\rm{Tr}} (\rho_{\parallel/\perp} (t) \log(\rho_{\parallel/\perp} (t)))$ where $\parallel/\perp$ refers to the appropriate bipartition. We show the time evolution of $S_{{\rm vN}}^{\parallel}/S_{{\rm vN}}^{\parallel,\text{Page}}$, $S_{{\rm vN}}^{\perp}/S_{{\rm vN}}^{\perp,\text{Page}}$ and $S_1/S_1^{\text{Page}}$ starting from a $|\mathbb{Z}_2\rangle$ (Fig.~\ref{fig:PXPladdersEnt} (c)) as well as $|\rm{vac}\rangle$ (Fig.~\ref{fig:PXPladdersEnt} (d)) initial state for a ladder with $N=24$ sites for $\Delta=1$ using ED. Here, $S_{{\rm vN}}^{\parallel,\text{Page}} \approx 5.205$, $S_{{\rm vN}}^{\perp,\text{Page}} \approx 4.935$ and $S_1^{\text{Page}} \approx 10.141$ refer to the value of these quantities obtained by averaging over $50$ random states with real coefficients sampled from the $k_x=0$ sector of the Hilbert space for a ladder with $N=24$ sites. 
 
 In both cases, $S_{{\rm vN}}^{\parallel}(t)/S_{{\rm vN}}^{\parallel,\text{Page}}$ is higher than $S_{{\rm vN}}^{\perp}(t)/S_{{\rm vN}}^{\perp,\text{Page}}$. In particular, both  $S_{{\rm vN}}^{\parallel}(t)$ and $S_{{\rm vN}}^{\perp}(t)$ display local maxima at times that are equidistant between the consecutive local maxima of $\mathcal{F}(t)$ for dynamics from the $|\text{vac}\rangle$ state at $\Delta=1$ (Fig.~\ref{fig:PXPladdersEnt} (d)). This suggests that the other end state, where one end state is approximated by $|\text{vac}\rangle$, may be approximated in a simpler manner in a basis composed of entangled spins along rungs rather than in an unentangled Fock basis. While $S_{{\rm vN}}^{\parallel}(t)$ and $S_{{\rm vN}}^{\perp}(t)$ are relatively featureless (unlike $S_1(t)$) in Fig.~\ref{fig:PXPladdersEnt} (c), both these quantities show oscillations that track $S_1(t)$ in Fig.~\ref{fig:PXPladdersEnt} (d), highlighting the different origin of the revivals in the non-PXP regime starting from a N\'eel and a vacuum initial state, respectively.

\section{Overlap of many-body eigenstates with $\ket{\mathbb{Z}_2}$ and $\ket{{\rm vac}}$}

 \begin{figure*}[!htpb]
        \centering
        \rotatebox{0}{\includegraphics*[width= 0.29\textwidth]{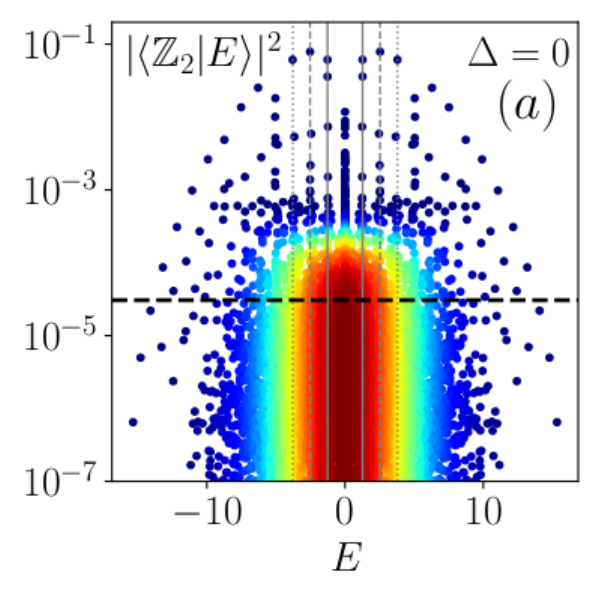}}
        \rotatebox{0}{\includegraphics*[width= 0.29\textwidth]{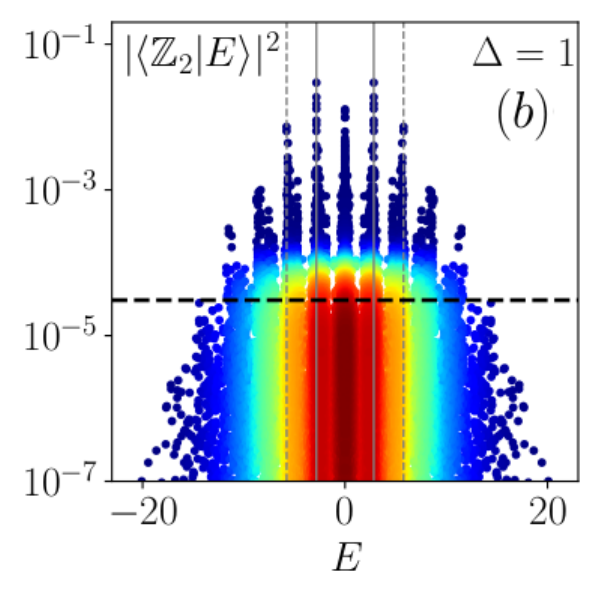}}
        \rotatebox{0}{\includegraphics*[width= 0.29\textwidth]{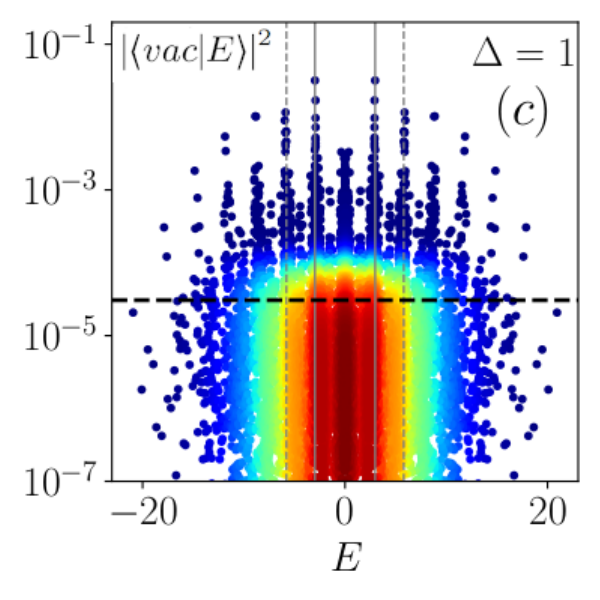}}
        \caption{Plot of overlap of many-body eigenstates of $N=28$ ladder in $k_x=0$ sector with specific Fock states (a) $|\langle \mathbb{Z}_2|E \rangle|^2$ for $\Delta=0$, (b) $|\langle \mathbb{Z}_2|E \rangle|^2$ for $\Delta=1$ and (c) $|\langle {\rm vac} |E \rangle|^2$ for $\Delta=1$.
        For all plots $w=\hbar=1$ and the color scheme indicates higher overlap for warmer colors.} 
        \label{fig:supp_overlap_figs}
    \end{figure*}

 \begin{figure*}[!h]
        \centering
        \rotatebox{0}{\includegraphics*[width= 0.29\textwidth]{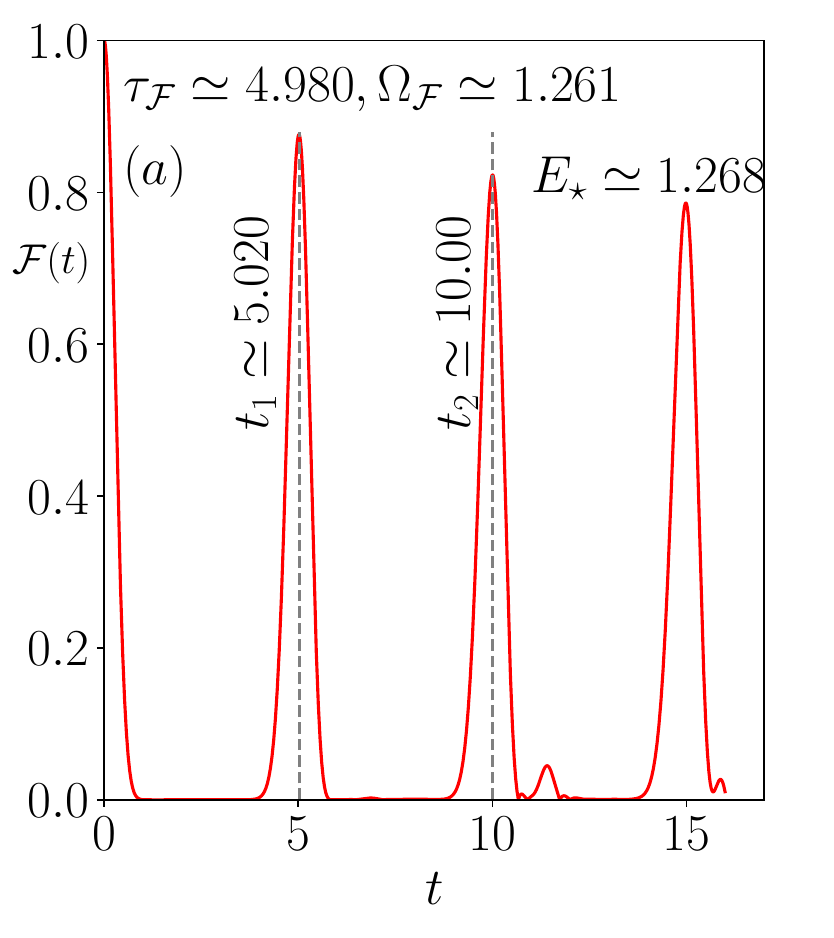}}
        \rotatebox{0}{\includegraphics*[width= 0.29\textwidth]{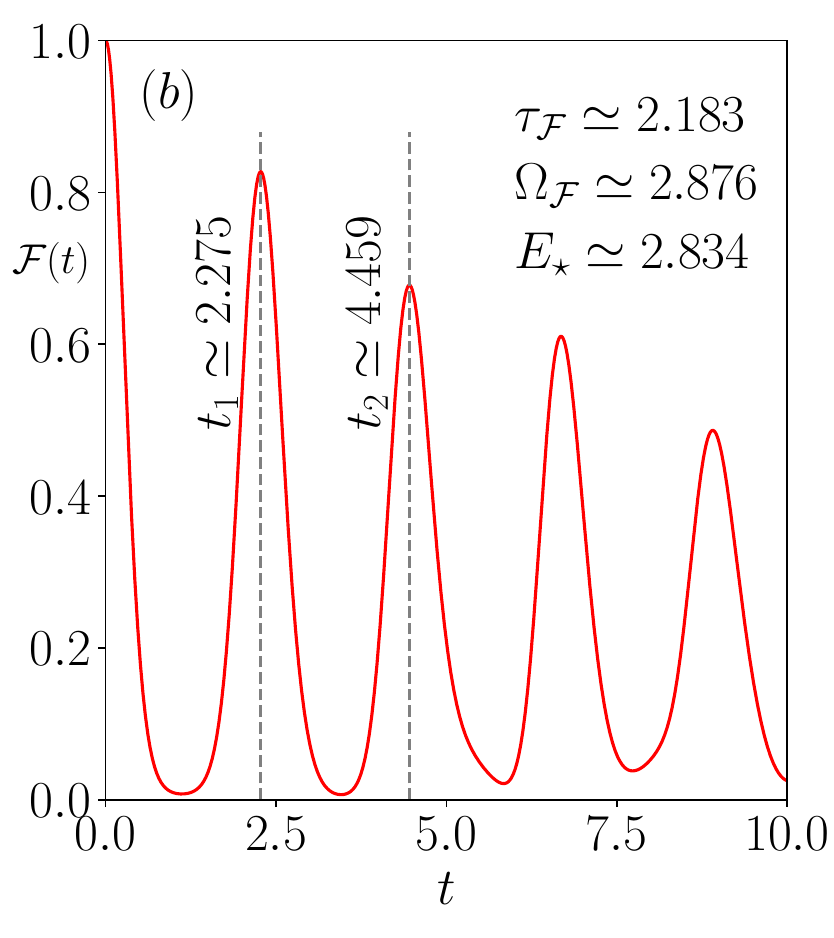}}
        \rotatebox{0}{\includegraphics*[width= 0.29\textwidth]{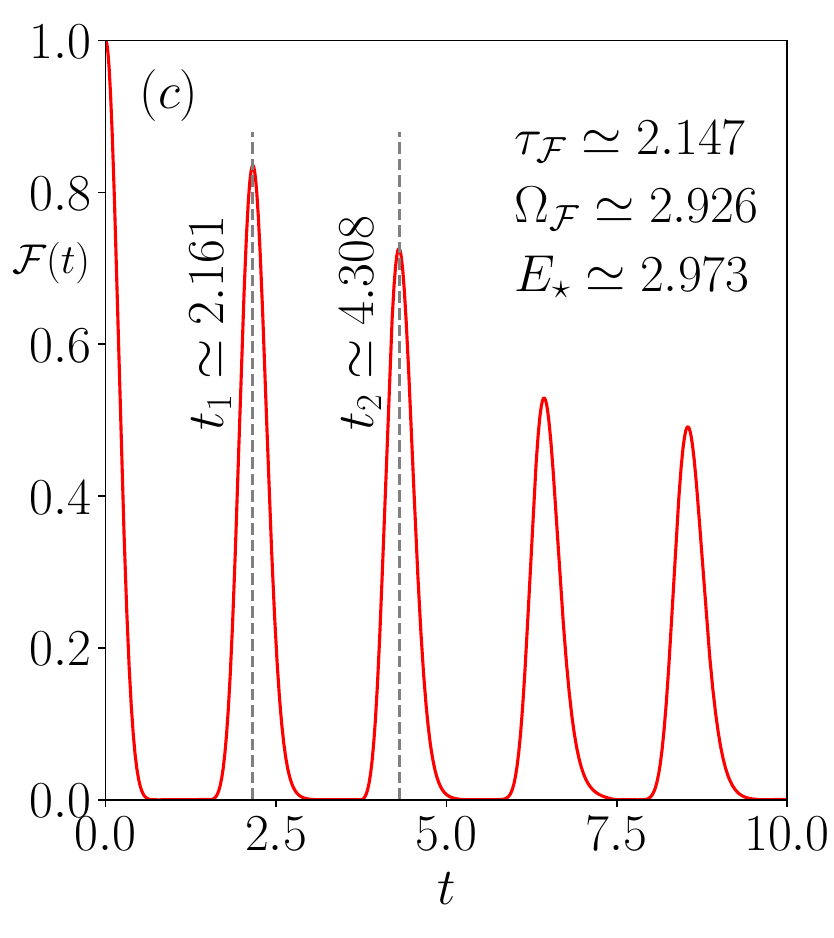}}
        \caption{Plot of the fidelity $\mathcal{F} (t)$ as a function of $t$ showing time periods of fidelity revivals for (a)\:$\Delta=0,\ket{\psi(0)}=\ket{\mathbb{Z}_2}$, (b)\:$\Delta=1,\ket{\psi(0)}=\ket{\mathbb{Z}_2}$ and (c)\:$\Delta=1,\ket{\psi(0)}=\ket{{\rm vac}}$. For all plots $N=28$ and $w=\hbar=1$.}
        \label{fig:supp_fid_revival_times}
    \end{figure*}    

    In Fig.~\ref{fig:supp_fid_revival_times}\:(a),(b),(c), we show the time period of fidelity revivals. The time period of revivals as denoted in the inset by $\displaystyle\tau_\mathcal{F}$ is computed as the difference in times of first and second revivals. i.e. $\tau_\mathcal{F}=t_2-t_1$. This time period corresponds to a frequency shown as $\Omega_\mathcal{F}=2\pi/(t_2-t_1)$, which in all cases is very close to the energy of the first non-zero energy QMBS, $E_\star$, as shown in Fig.~\ref{fig:supp_overlap_figs} by solid gray lines.

 \begin{figure*}[!h]
        \centering
        \rotatebox{0}{\includegraphics*[width= 0.29\textwidth]{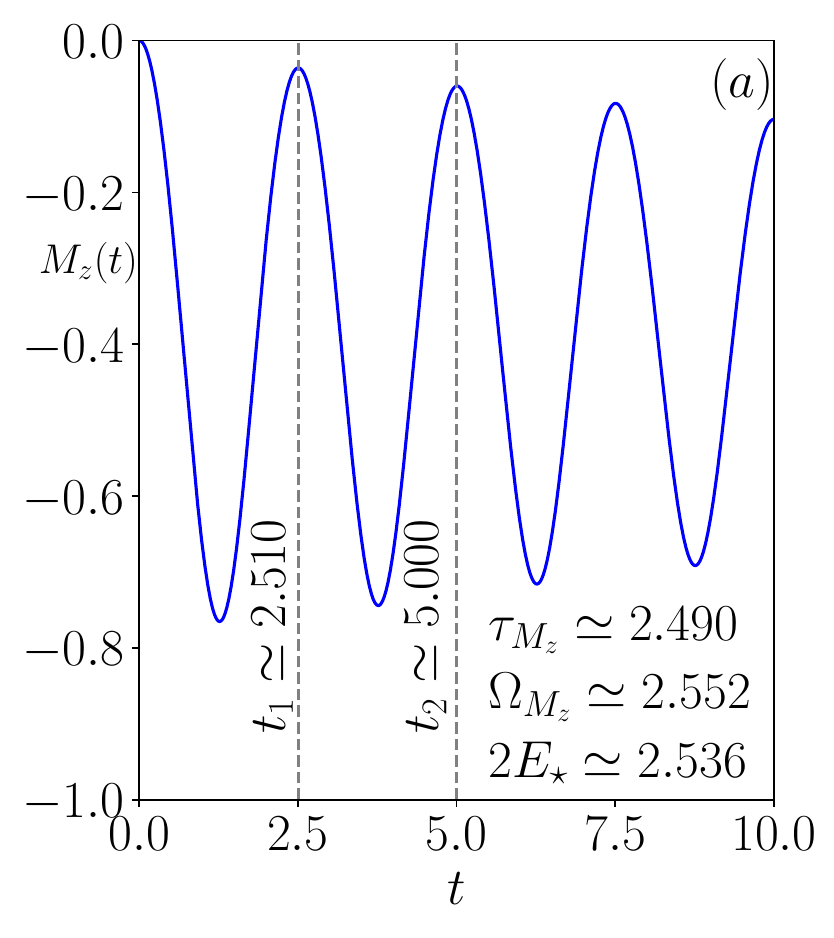}}
        \rotatebox{0}{\includegraphics*[width= 0.29\textwidth]{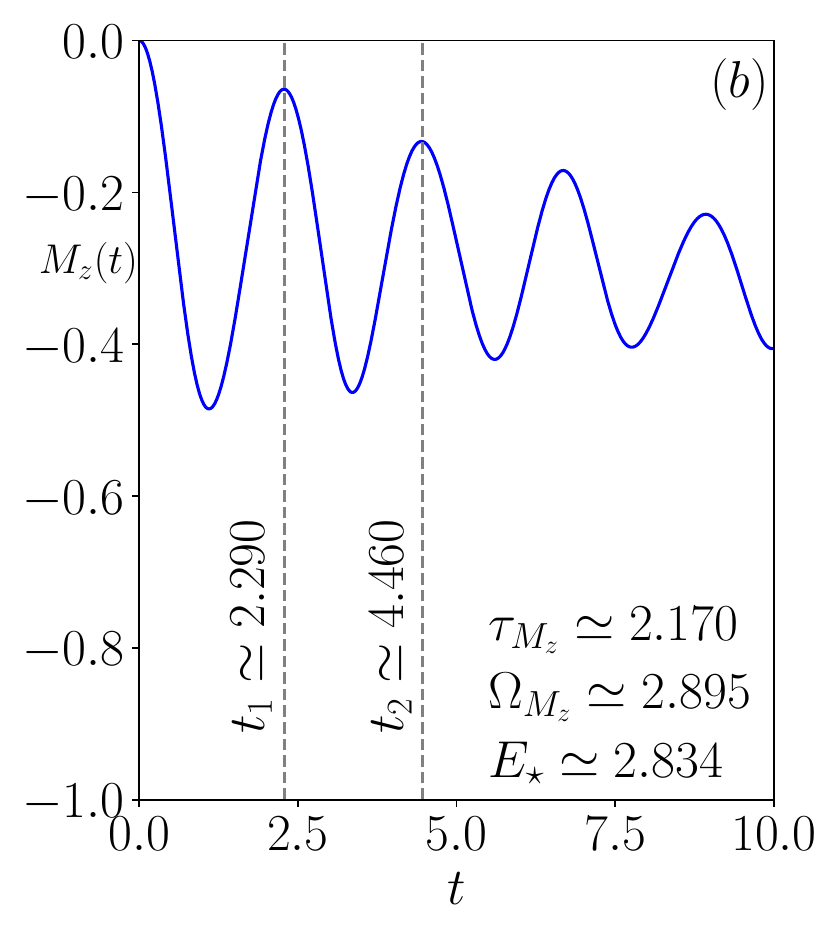}}
        \rotatebox{0}{\includegraphics*[width= 0.29\textwidth]{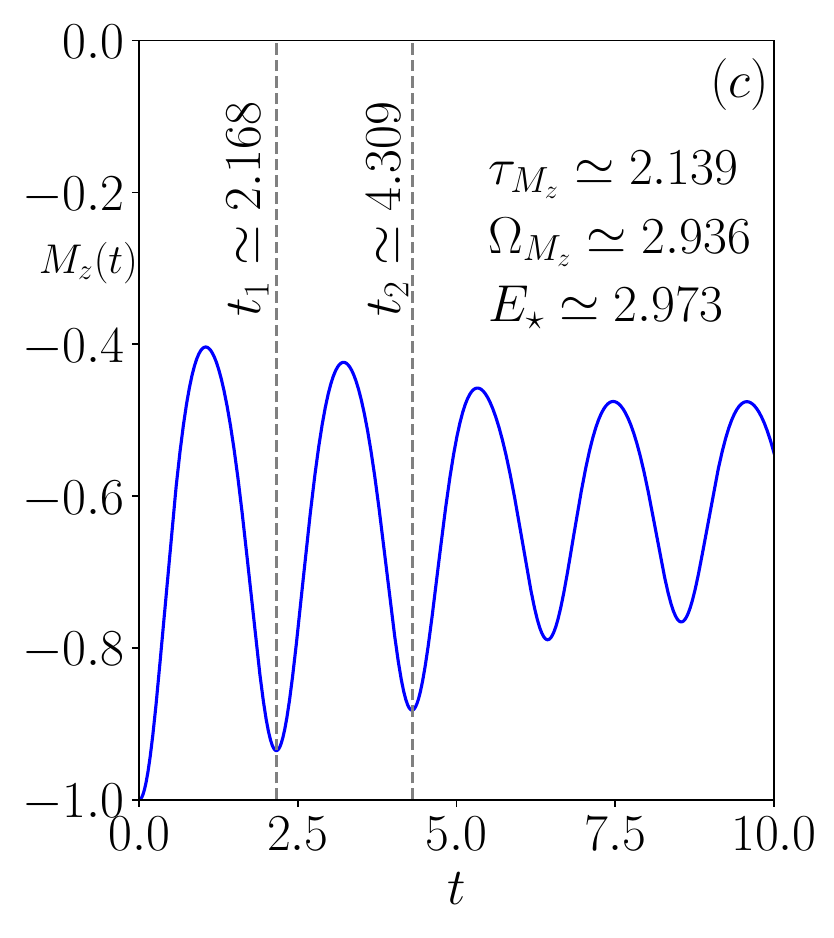}}
        \caption{Plot of the longitudinal magnetization density as a function of $t$ for (a)\:$\Delta=0,\ket{\psi(0)}=\ket{\mathbb{Z}_2}$, (b)\:$\Delta=1,\ket{\psi(0)}=\ket{\mathbb{Z}_2}$ and (c)\:$\Delta=1,\ket{\psi(0)}=\ket{{\rm vac}}$. }
        \label{fig:supp_Mz_oscll_times}
    \end{figure*}    

    In Fig.~\ref{fig:supp_Mz_oscll_times}\:(a),(b),(c), we show the time period of oscillations of longitudinal magnetization. This time period, as denoted in the inset by $\displaystyle\tau_{M_z}$ is computed as the difference in times of first and second revivals. i.e. $\tau_{M_z}=t_2-t_1$. This time period corresponds to a frequency shown as $\Omega_{M_z}=2\pi/(t_2-t_1)$, which in case (a) is very close to twice of the energy of the first non-zero energy QMBS, $E_\star$, as shown in Fig.~\ref{fig:supp_Mz_oscll_times} by solid gray lines. In other cases i.e. (b) and (c) the frequency of oscillations $\Omega_{M_z}$ is equal to $E_\star$

    \section{Bipartite entanglement entropy of the simultaneous zero modes}
\begin{figure*}[!h]
   \centering
           %\rotatebox{0}{\includegraphics*[width= 0.3\textwidth]{fig5.pdf}}
        \rotatebox{0}{\includegraphics*[width= 0.4\textwidth]{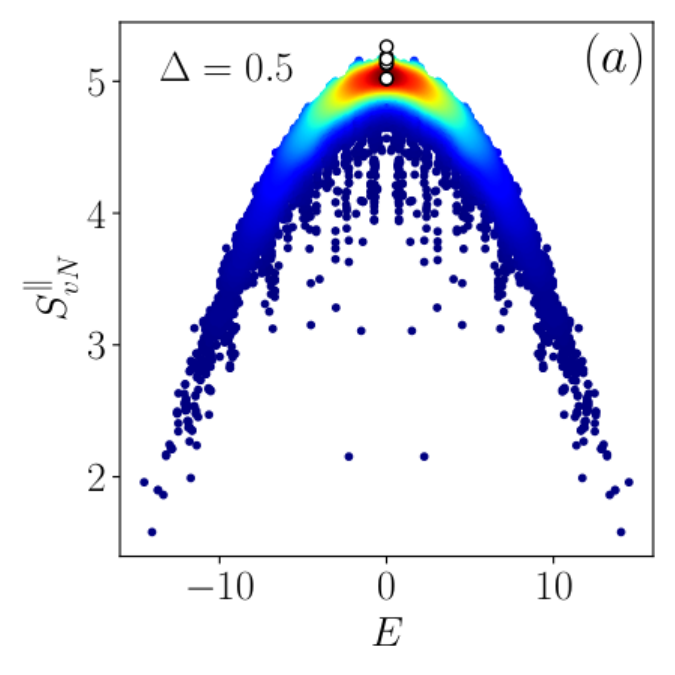}}
        \rotatebox{0}{\includegraphics*[width= 0.4\textwidth]{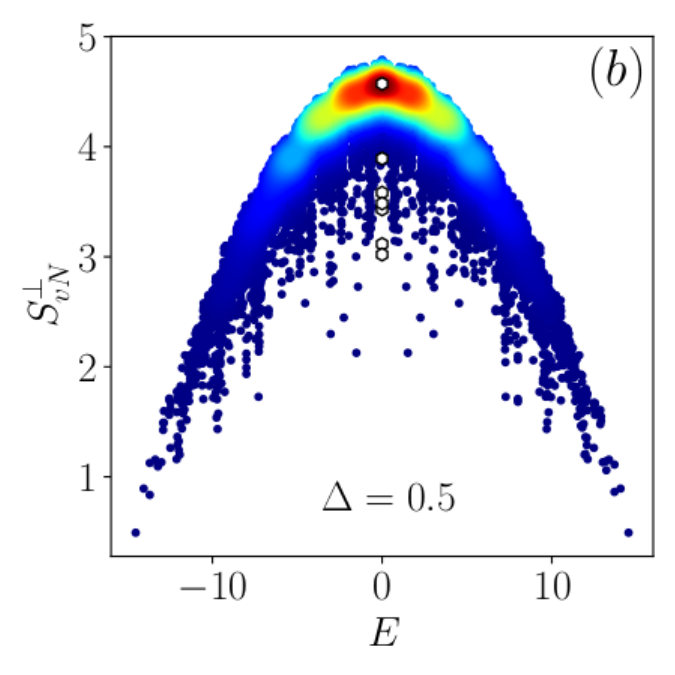}}
        \caption{Bipartite entanglement entropies (a) $S_{{\rm vN}}^{\parallel}$ and (b) $S_{{\rm vN}}^{\perp}$ of all eigenvectors (where warm colors indicate a higher density of states in the colormap) versus the anomalous zero modes (open white circles in (a) and open white hexagons in (b)) that stay unchanged as a function of $\Delta$ for $N=24$ and $\Delta=0.5$. The two cuts used to calculate the entanglement entropies are shown in Fig.~\ref{fig:PXPladdersEnt} (a), (b).}
        \label{fig:EEsimulmodes}
    \end{figure*}

    While Fig. 5 in main text showed the Shannon entropy $S_1$ of all eigenstates versus the simultaneous zero modes that stay unchanged as a function of $\Delta$ for $N=24$, $\Delta=0.5$ demonstrating that these anomalous modes are much more localized in the Hilbert space compared to neighbouring eigenstates around $E=0$, here we do the same comparison using bipartitite entanglement entropy of all the eigenstates along two cuts (Fig.~\ref{fig:PXPladdersEnt} (a), (b)) to compute $S_{{\rm vN}}^{\parallel}$ and $S_{{\rm vN}}^{\perp}$, respectively. The results in Fig.~\ref{fig:EEsimulmodes} clearly show that not all these anomalous zero modes have low entanglement entropy. In fact, $S_{{\rm vN}}^{\parallel}$ is uniformly high for all the anomalous zero modes as can be seen from Fig.~\ref{fig:EEsimulmodes} (a).  

%\bibliography{references}